\documentclass[12pt]{article}

\usepackage{natbib}
\usepackage{enumitem}
\usepackage{makecell}
\usepackage{fullpage}
\usepackage{amsthm}
\usepackage{amsmath}
\usepackage{amssymb}
\usepackage{subfigure}
\usepackage{footnote}
\usepackage{tabu}
\usepackage{tikz}
\usepackage{booktabs}

\providecommand{\keywords}[1]
{
  \small	
  \textbf{\textit{Keywords---}} #1
}
\usepackage{etoolbox}

\makeatletter
\patchcmd{\@makecaption}
  {\parbox}
  {\advance\@tempdima-\fontdimen2} 
  {}{}
\makeatother 
\usepackage{url}

\usepackage{multicol}
\usepackage{xcolor}
\usepackage{algorithm}
\usepackage[ruled,vlined,algo2e]{algorithm2e}

\usepackage{xr-hyper}
\usepackage[colorlinks, linkcolor=blue, anchorcolor=blue, citecolor=blue]{hyperref}

\usepackage{titling} 

\usepackage{amsmath,bbm,amssymb, amsfonts}
\usepackage{mathtools}
\usepackage{graphicx,psfrag,epsf}
\usepackage{enumitem}
\usepackage{natbib}
\usepackage{url} 
\usepackage{color}
\usepackage{algorithmicx}
\usepackage{booktabs}
\usepackage{siunitx}
\usepackage{array}
\usepackage{rotating}

\usepackage{algorithm}
\usepackage{algpseudocode}

\usepackage{tikz}

\makeatletter
\def\breve{\mathpalette\wide@breve}
\def\wide@breve#1#2{\sbox\z@{$#1#2$}%
	\mathop{\vbox{\m@th\ialign{##\crcr
				\kern0.08em\brevefill#1{0.8\wd\z@}\crcr\noalign{\nointerlineskip}%
				$\hss#1#2\hss$\crcr}}}\limits}
\def\brevefill#1#2{$\m@th\sbox\tw@{$#1($}%
	\hss\resizebox{#2}{\wd\tw@}{\rotatebox[origin=c]{90}{\upshape(}}\hss$}
\makeatletter

\DeclareMathOperator*{\argmin}{argmin} 
\DeclareMathOperator*{\argmax}{argmax} 

\newtheorem{proposition}{Proposition}

\newcommand{\indep}{\raisebox{0.05em}{\rotatebox[origin=c]{90}{$\models$}}}

\def\gobblestop#1#2{#1}
\def\killstop{%
	\aftergroup\gobblestop
}

\def\ba{{\mbox{\boldmath$a$}}}

\def\bm{{\bf m}}

\def\bt{{\bf t}}

\def\bw{{\bf w}}
\def\bx{{\bf x}}
\def\by{{\bf y}}
\def\bz{{\bf z}}

\def\bV{{\bf V}}

\def\bX{{\bf X}}

\def\bZ{{\bf Z}}

\def\thick#1{\hbox{\rlap{$#1$}\kern0.25pt\rlap{$#1$}\kern0.25pt$#1$}}

\def\bbeta{\boldsymbol{\beta}}

\def\smbalpha{\boldsymbol{{\scriptstyle{\alpha}}}}

\def\bXtilde{{\widetilde \bX}}

\def\smbalpha{\widehat{\smbalpha}}

\def\hbar{\bar{ h}}

\def\mybox#1{\vskip1mm \begin{center}
        \hspace{.0\textwidth}\vbox{\hrule\hbox{\vrule\kern6pt
\parbox{.9\textwidth}{\kern6pt#1\vskip6pt}\kern6pt\vrule}\hrule}
        \end{center} \vskip-5mm}
\def\lboxit#1{\vbox{\hrule\hbox{\vrule\kern6pt
      \vbox{\kern6pt#1\vskip6pt}\kern6pt\vrule}\hrule}}

\def\thickboxit#1{\vbox{{\hrule height 1mm}\hbox{{\vrule width 1mm}\kern6pt
          \vbox{\kern6pt#1\kern6pt}\kern6pt{\vrule width 1mm}}
               {\hrule height 1mm}}}

\def\bbE{\bb_{\scriptscriptstyle E}}

\def\bbV{\mathbb{V}}

\def\bbP{\mathbb{P}}

\def\calA{\mathcal{A}}

\def\calE{\mathcal{E}}

\def\calH{\mathcal{H}}
\def\calI{\mathcal{I}}
\def\calJ{\mathcal{J}}
\def\calK{\mathcal{K}}

\def\calO{\mathcal{O}}

\def\calX{\mathcal{X}}

\def\bbE{\mathbb{E}}

\def\bbP{\mathbb{P}}

\def\bbR{\mathbb{R}}

\def\bbV{\mathbb{V}}

\def\fat#1{\hbox{\rlap{$#1$}\kern0.25pt\rlap{$#1$}\kern0.25pt$#1$}}

\newcommand*\circled[1]{\tikz[baseline=(char.base)]{
            \node[shape=circle,draw,inner sep=2pt] (char) {#1};}}

\newcolumntype{R}{@{\extracolsep{0.5cm}}r@{\extracolsep{0pt}}}%
\newcolumntype{E}{@{\extracolsep{0.25cm}}c@{\extracolsep{0pt}}}%

\makeatletter
\newcommand{\distas}[1]{\mathbin{\overset{#1}{\kern\z@\sim}}}%
\makeatother

\newtheorem{thm}{Theorem}[section]
\newtheorem{cor}[thm]{Corollary}
\newtheorem{lem}[thm]{Lemma}

\newcommand{\blind}{1}

\begin{document}

\if1\blind
{
	\title{\bf Energy Balancing of Covariate Distributions}
	\author{Jared D. Huling$^{1}$\thanks{huling@umn.edu},
		Simon Mak$^{2}$\\
		\\
		$^{1}$Division of Biostatistics, University of Minnesota, Minneapolis, Minnesota \\ [8pt]
		$^{2}$Department of Statistical Science,
		Duke University, Durham, North Carolina \\ [8pt]
	}
} \fi

\if0\blind
{
	\bigskip
	\bigskip
	\bigskip
	\begin{center}
		{\Large \bf Energy Balancing of Covariate Distributions}
	\end{center}
	\medskip
} \fi

\maketitle

\begin{abstract}
Bias in causal comparisons has a direct correspondence with distributional imbalance of covariates between treatment groups. Weighting strategies such as inverse propensity score weighting attempt to mitigate bias by either modeling the treatment assignment mechanism or balancing specified covariate moments. This paper introduces a new weighting method, called energy balancing, which instead aims to balance weighted covariate distributions. By directly targeting distributional imbalance, the proposed weighting strategy can be flexibly utilized in a wide variety of causal analyses, including the estimation of average treatment effects and individualized treatment rules. Our energy balancing weights (EBW) approach has several advantages over existing weighting techniques. First, it offers a model-free and robust approach for obtaining covariate balance that does not require tuning parameters, obviating the need for modeling decisions of secondary nature to the scientific question at hand. Second, since this approach is based on a genuine measure of distributional balance, it provides a means for assessing the balance induced by a given set of weights for a given dataset. Finally, the proposed method is computationally efficient and has desirable theoretical guarantees under mild conditions. We demonstrate the effectiveness of this EBW approach in a suite of simulation experiments, and in studies on the safety of right heart catheterization and the effect of indwelling arterial catheters.

\end{abstract}%

\def\spacingset#1{\renewcommand{\baselinestretch}%
{#1}\small\normalsize}
\spacingset{1.50} 



\section{Introduction}
\label{sec:intro}

Studying the causal effect of a treatment or intervention is a central goal in many scientific disciplines. In randomized controlled trials, estimation of causal effects is possible since randomization ensures that treated units and control units are comparable \citep{dasgupta2015causal}. However, for many pressing questions, it is impossible or impractical to randomize treatment assignments. Researchers are thus left with existing sources of observational data to answer these questions. In these observational studies, it is of prime interest to make unconfounded comparisons between treatment groups, a common example being estimation of the average treatment effect (ATE).
Yet in observational settings, natural selection processes into the treatment result in imbalance in the covariate distributions between treatment groups. This often results in substantial bias in naive comparisons of the outcome of interest between groups. 

There is a vast literature on adjustment methods for correcting this imbalance in order to reduce bias in estimating treatment effects. One approach is adjustment via regression. Regression adjustments are, however, often sensitive to model misspecification \citep{rubin1979using}, and require the use of outcome information (as opposed to methods which separate a study into design and analysis stages). When careful modeling is required to make proper adjustments, it is advisable to avoid the use of outcome information to prevent a ``garden of forking paths'' problem \citep{gelman2013garden} for an analysis. Another approach is adjustment via weighting. Weighting methods avoid the use of outcome information, and control for confounding by re-weighting the treatment and control groups. 
Inverse-probability weighting (IPW) methods \citep{robins1995semiparametric, hahn1998role, robins2000marginal, hirano2001estimation, hirano2003efficient, imbens2004nonparametric}, which have origins in survey sampling, are by far the most commonly-used weighting approaches. IPW methods model the treatment assignment mechanism (or propensity score, see \citealp{rosenbaum1983central}), and inverse weight each sample by the probability of receiving its assigned treatment. Inverse weighting by the true underlying propensity score 
controls for confounding, as it re-weights the covariate distributions of the treatment groups to that of the overall population.

In practice, IPW methods require positing and fitting a model for the propensity score. While model fitting is no unfamiliar task for a statistician, it has been noted in the literature that even mild model misspecifications for the propensity score can result in substantial bias in estimating the treatment effect \citep{kang2007demystifying}. Hearkening to George Box's maxim, ``all models are wrong, but some are useful'',  it is often quite difficult in practice to obtain a useful propensity score model, especially in the presence of many covariates.  Recent work has focused on mitigating this issue, by either i) including conditions on the estimation of the propensity model which encourage moment balance of the covariates \citep{imai2014covariate}, or by ii) altogether avoiding direct modeling of the propensity score and instead estimating weights which explicitly balance moments of the covariates either exactly \citep{hainmueller2012entropy, chan2016globally} or approximately \citep{zubizarreta2015stable}. A large class of such estimators is explored in \citet{chan2016globally} and further expanded on by \citet{zhao2019}, and have a long history dating back to \citet{deming1940least}. 

Moment balancing weights can work quite well and are often more robust than inverse propensity score weights. 
However, when important moments are missed, their performance can suffer materially; in practice, it can also be difficult to decide which moments should be balanced for a given dataset. Compounding this predicament is the lack of diagnostic tools which aid in such decisions. Recent works \citep{wong2017kernel} have attempted to mitigate this issue by focusing on non-parametric approaches to moment balancing, yet they require a careful choice of kernel function and tuning of multiple hyperparameters. Given the many possible approaches for estimating weights, a natural question to ask is which set of weights works best for a given dataset. There are few available tools which can provide reliable information on which weights are most likely to result in unconfounded causal comparisons.

To make progress in addressing these issues, we show in this paper that estimation bias for the ATE has a direct link to imbalance in the covariate \textit{distributions}. This shows that balancing the full covariate distributions (and not just lower order moments) of the treatment groups to the full population is critical for reducing bias. Although this is understood in the literature \citep{li2018balancing, zhao2019}, in this paper we establish and make use of this link in a general manner,
by i) introducing a metric which can evaluate how well a set of weights mitigates imbalance for a given dataset, and ii) developing a new method for estimating weights by minimizing this metric, thus yielding good distributional balance {for a given dataset.}
The distributional balancing property of these weights allows for robust empirical performance; they tend to perform well in practice without relying on modeling assumptions for the propensity score or which moments are imbalanced.

We introduce weights that are explicitly constructed to balance the covariate distributions of the treatment groups to a target distribution (usually the full population). We do so by leveraging the energy distance in \citet{szekely2004testing}. The energy distance is a measure based on powers of the Euclidean distance and was originally introduced as a means to replace standard nonparametric goodness-of-fit tests in high dimensions. The energy distance has an exact duality with a norm on the characteristic functions, enabling its use to compare two (or more) distributions or the distributions of two samples. We show that a \textit{weighted} energy distance still retains this duality, making it a rigorously-justified and reliable metric to compare between multiple sets of weights for a given dataset. From this, we propose the so-called \textit{energy balancing weights} (EBWs), which are defined as the weights which minimize the weighted energy distance between treatment groups and the full sample, subject to constraints that mitigate variability in the weights. 
We prove that EBWs asymptotically ensure full distributional balance and result in root-$n$ consistent estimation of the ATE. Our emphasis is on the robust performance of our approach in practice, and our asymptotic analysis serves primarily to justify the use of the proposed weights. We analyze  four challenging observational studies and use each of these datasets to conduct realistic simulations, demonstrating the effectiveness of EBWs in practice.

Although we focus primarily on a simple estimand, namely, the ATE, the proposed weights can be used for a wide variety of estimands that can be characterized as statistical functionals of the population distribution function of the covariates (since EBWs are designed to re-balance distributions). 
We show that they can be used for the estimation of a wide variety of causal quantities, such as the ATE and individualized treatment rules \citep{qian2011performance, zhao2012estimating}. With minor modifications, they can also be used for estimation of the average treatment effect on the treated (ATT) and for estimating treatment effects for multi-category treatments. In spite of the fact that EBWs are not specifically designed to match low order moments, in practice they often result in better marginal mean balance of covariates than propensity score methods even in high dimensions (50-100 covariates), as seen in Sections \ref{sec:analysis}, \ref{sec:analysis_mimic} and the Supplement. EBWs are also quite stable in practice, rarely resulting in large weights, an issue that plagues standard propensity score methods \citep{kang2007demystifying}. EBWs are constructed without using outcome information; however, as for other weighting approaches, variance can be reduced via augmented estimators that make use of outcome regression models, such as the augmented estimators in \citet{wong2017kernel}, \citet{zhao2019}, and \citet{athey2018approximate}. However, we do not explore such techniques in this paper.

We note that, while the proposed EBWs can work well across a wide variety of scenarios, there are scenarios where other weighting approaches work better. For example, if it is known (or likely) that only first order moments are imbalanced for a given dataset, moment balancing weights (see, e.g., \citealp{imai2014covariate, chan2016globally}) may work better. 
In this case, a researcher may be interested in which, among the many choices of methods which can balance first order moments, works best for their application.
For these scenarios, we provide a useful, objective criterion for deciding among weights arrived at via different modeling choices. 
The weighted energy distance can provide a principled means to assess how well a given set of weights balances the distributions of treatment groups to the full population, and thus can be used as a tool to verify the balance properties of weights estimated in any fashion, such as by propensity score modeling, empirical calibration balancing \citep{chan2016globally}, or otherwise.


The remainder of this paper is organized as follows. Section \ref{sec:methods} motivates the need for distributional balance and introduces the weighted energy distance. Section \ref{sec:energy_balancing} presents the proposed EBWs, and discusses their computation and asymptotic properties. Section \ref{sec:extensions} describes extensions and further applications of EBWs. Section \ref{sec:simulation} compares in simulation studies the performance of EBWs with other weighting methods. Section \ref{sec:analysis} discusses an application of EBWs in a study of right heart catheterization and further uses this study to conduct a realistic simulation and Section \ref{sec:analysis_mimic} investigates several datasets using the MIMIC-III Critical Care Database \citep{mimiciii}. Section \ref{sec:discussion} concludes with a discussion and future work.



\section{Distributional Balance and Weighted Energy Distance}
\label{sec:methods}

\subsection{Setup}
\label{sec:setup}

Consider a sample $\{(Y_i, A_i, \bX_i)\}_{i=1}^n$ of size $n$  from a population, where $Y_i$ is the outcome of the $i$th unit, $A_i \in \{0,1\}$ is a binary indicator of receiving a treatment, and $\bX_i \in \calX = \bbR^{p}$ is a $p$-dimensional vector of covariates. Further denote $n_1 = \sum_{i=1}^nA_i$ and $n_0 = n - n_1$.
In this paper, we are interested in estimating the average causal effect of the treatment on the outcome. A formal definition of a causal effect often involves the use of so-called potential outcomes \citep{neyman1923application, rubin1974estimating, rubin1978bayesian, hernan2010causal}. The potential outcome $Y(a)$ is the outcome that would have been observed under level $a$ of the treatment. As each individual only receives one level of the treatment at a given time, only one potential outcome, either $Y_i(0)$ or $Y_i(1)$, for each individual is observable. 
We assume the standard stable unit treatment value assumption (SUTVA), which posits that the potential outcomes for each unit are unaffected by the potential outcomes of other units and that only one version of the treatment exists. Under SUTVA, the observed outcome is consistent with the potential outcomes in that $Y_i = Y_i(A_i)$. We further assume the assignment mechanism is \textit{strongly unconfounded} in the sense that $\{Y(0), Y(1)\} \perp \!\!\! \perp A \: | \: \bX$, which requires that there are no unmeasured confounders. The $\perp \!\!\! \perp$ notation of \citet{dawid1979some} denotes (conditional) independence. We further assume \textit{positivity} (or probabilistic assignment) in that the propensity score $\pi(\bx) \equiv \bbP(A = 1 \:|\: \bX = \bx)$ \citep{rosenbaum1983central} satisfies $0 < \pi(\bx) < 1$, so that everyone has a \textit{chance} of receiving the treatment. Positivity, together with no unmeasured confounders, constitute \textit{strong ignorability} \citep{rosenbaum1983central}. 

Let us denote $\mu_{a}(\bX_i) \equiv \bbE(Y(a)\:|\:\bX_i)$ as the conditional mean function and 
$\sigma^2_a(\bX_i) \equiv \bbV(Y_i(a) \:|\:\bX_i)$ as the conditional variance function of the response for $a\in\{0,1\}$. 
We consider scenarios where data have been collected from an observational study, and thus the treatment groups are not comparable due to imbalances in the distributions of their baseline covariates. These differences can be characterized through $\pi(\bx)$ or through $F_a(\bx) \equiv \bbP(\bX \leq \bx \:|\: A = a)$, the cumulative distribution function (CDF) of covariates $\bX$ conditional on treatment level $a$.

Using the notation of potential outcomes, the (population) average treatment effect (ATE) is defined as $\tau \equiv \bbE(Y(1) - Y(0))$. This can be rewritten as 
\begin{equation}
\tau = \int_{\bx\in \calX} \left[  \mu_{1}(\bx) - \mu_{0}(\bx) \right] \mathrm{d} F(\bx),
\end{equation}
where $F(\bx) = Pr(\bX \leq \bx)$ is the CDF of covariates $\bX$ marginalized over the treatment groups. In other words, $F(\bx) = F_1(\bx)P_1 + F_0(\bx)P_0$, where $P_a \equiv \bbP(A=a) \in (0,1)$ is the probability of being assigned treatment level $a\in\{0,1\}$.

\subsection{Weighted average estimates and distributional balance}
\label{sec:wavg_est}

We restrict our focus to weighted averages as estimates of the ATE. Given a vector of weights $\bw = (w_1, \dots, w_n)$, we study estimators of the form:
\begin{equation}\label{eqn:wate}
\widehat{\tau}_\bw = \frac{1}{n_1}\sum_{i=1}^nw_iY_iA_i - \frac{1}{n_0}\sum_{i =1}^nw_iY_i(1-A_i).
\end{equation}
The most commonly-used example of \eqref{eqn:wate}, inverse propensity score weighting, uses 
${w}_i = 1 / (A_i \pi(\bX_i)/n_1 + (1-A_i)(1-\pi(\bX_i))/n_0)$. 
As in \citet{imai2014covariate} and \citet{li2018balancing}, the weights in \eqref{eqn:wate} are often normalized by treatment group, i.e., $\sum_{i=1}^nw_iI(A_i = a) = n_a$ for $a \in \{0,1\}$, to improve precision \citep{kang2007demystifying, zhao2019} at the cost of a small bias. When these weights are constructed as above and normalized by treatment group, the resulting estimator is called the H{\'a}jek estimator \citep{hajek1971comment}, and may yield reduced mean squared error over its non-normalized version \citep{ding2018}. 

Given any weight vector $\bw$, we can express the error of $\widehat{\tau}_\bw$ as
\begin{align}
\widehat{\tau}_\bw - \tau = {} & \int_{\calX}  \mu_{1}(\bx) \mathrm{d} \left[ F_{n, 1, \bw} - F_n  \right](\bx) - \int_{\calX}  \mu_{0}(\bx) \mathrm{d} \left[ F_{n, 0, \bw} - F_n \right](\bx) \label{eqn:ate_bias_int} \\
& - \int_{\calX}  \left[\mu_{1}(\bx) - \mu_{0}(\bx)\right] \mathrm{d} \left[ { F - F_n } \right](\bx) \label{eqn:sample_error}  \\ 
& + \frac{1}{n_1}\sum_{i =1}^nw_i\varepsilon_iA_i -  \frac{1}{n_0}\sum_{i=1}^nw_i\varepsilon_i(1-A_i),  \label{eqn:ate_residuals} 
\end{align}
where $\varepsilon_i\equiv Y_i(A_i) - \mu_{A_i}(\bX_i)$, $F_n(\bx) = \sum_{i=1}^nI(\bX_i \leq \bx) / n$ is the empirical CDF (ECDF) of the combined sample $\{\bX_i\}_{i=1}^n$, and $F_{n, a, \bw}(\bx) = \sum_{i=1}^nw_i\\*I(\bX_i \leq \bx, A_i = a)/n_a$ is the \textit{weighted} ECDF for treatment level $a\in\{0,1\}$. 
In observational studies, $F_n$ is not impacted by the weights, so the error term \eqref{eqn:sample_error} is irreducible. However, this term goes to 0 as long as the sample is representative of the desired population. 
Since \eqref{eqn:ate_residuals} always has mean 0, the bias of $\widehat{\tau}_\bw$ in essence depends on the properties of \eqref{eqn:ate_bias_int}: the difference of integrals with respect to $F_{n,a,\bw}-F_n$. Thus, the systematic source of error from the weighted estimator $\widehat{\tau}_\bw$ arises from the imbalance between the weighted ECDFs $F_{n,a,\bw}$ and the ECDF $F_n$. Unlike the decomposition in \citet{zhao2019}, the decomposition of $\widehat{\tau}_\bw - \tau$ above holds even if the treatment effect is not constant over $\bx$. Further, none of the methodology or results of this paper require such a constant treatment effect.

The notion that the covariate distributions should be balanced to obtain a good estimate of $\tau$ is not new (see \citet{imai2014covariate} and \citet{li2018balancing}, among many others). In fact, it is well-understood that the weights resulting from correctly specified propensity score models asymptotically balance covariate distributions. However, slight model misspecifications of the propensity score may result in poor performance of \eqref{eqn:wate} \citep{kang2007demystifying}. Further, two units with the same propensity score do not necessarily have the same covariate values, so in finite samples propensity score weights may not be optimal. Other approaches which aim to estimate weights by balancing pre-specified \textit{moments} of the covariates (e.g., \citealp{imai2014covariate,chan2016globally}) tend to be more robust than directly modeling $\pi(\bx)$. Yet, the term \eqref{eqn:ate_bias_int} makes it explicit that bias results directly from \textit{distributional} imbalance. A natural conclusion is that $\bw$ should be estimated to directly balance each $F_{n, a, \bw}$ to $F_n$. In the following, we introduce a new distance metric between $F_{n, a, \bw}$ and $F_n$, which enables one to characterize how well a set of weights re-balances the weighted distribution $F_{n, a, \bw}$ to $F_n$.

\subsection{Weighted energy distance}
\label{sec:weighted_energy_general}

We introduce next a new measure of the distributional balance induced by a set of weights. This measure is based on the \textit{energy distance}, which is a metric on distributions \citep{szekely2013energy}. Due to the link between estimation bias and distributional imbalance, our measure can be used to evaluate the degree of bias one expects from a given set of weights and a given dataset. We will later leverage this measure to construct distributional balance weights which minimize this metric for a given dataset.

The energy distance (as surveyed in \citealp{szekely2013energy}) is defined as follows. Let $G$ and $H$ be two finite-mean distribution functions on $\calX$, and let $\bZ, \bZ' \distas{i.i.d} G$ and $\bV, \bV' \distas{i.i.d} H$. The energy distance between distributions $G$ and $H$ is defined as 
\begin{equation}\label{eqn:energy_distance}
    \calE(G, H) \equiv 2 \bbE \lVert \bZ - \bV \rVert_2 - \bbE \lVert \bZ - \bZ' \rVert_2 - \bbE \lVert \bV - \bV' \rVert_2,
\end{equation}
where $\lVert \, \cdot \, \rVert_2$ is the Euclidean norm.
When both $G$ and $H$ are ECDFs, i.e. $G_n$ is the ECDF of $\{\bZ_i\}_{i=1}^n\subseteq\calX$ and $H_m$ is the ECDF of $\{\bV_i\}_{i=1}^m\subseteq\calX$, the energy distance $\calE(G_n, H_m)$ can be expressed as
\begin{equation}\label{eqn:energy_distance_empirical}
    \frac{2}{nm}\sum_{i=1}^n\sum_{j=1}^m\ \lVert \bZ_i - \bV_j \rVert_2 - \frac{1}{n^2}\sum_{i=1}^n\sum_{j=1}^n\ \lVert \bZ_i - \bZ_j \rVert_2 - \frac{1}{m^2}\sum_{i=1}^m\sum_{j=1}^m\ \lVert \bV_i - \bV_j \rVert_2.
\end{equation}
The energy distance has been used within a wide variety of statistical methods, e.g., for testing equivalence of distributions, for testing statistical independence \citep{szekely2007measuring}, and for generating samples from a target distribution \citep{mak2018support}.

We propose a \textit{weighted} modification of this energy distance, which measures the distance between a weighted distribution, i.e., the weighted covariate ECDFs $F_{n,0,\bw}$ and $F_{n,1,\bw}$ for the control and treated, and a target distribution, i.e., the combined covariate ECDF $F_n$. The \textit{weighted energy distance} between $F_{n,a,\bw}$ and $F_n$ is defined as
\begin{align*}
& \quad \calE(F_{n,a,\bw}, F_{n}) \equiv {} \frac{2}{n_an}\sum_{i = 1}^n \sum_{j = 1}^nw_iI(A_i = a)\lVert {\bX}_i  
-  {\bX}_j \rVert_2 \\ 
& - \frac{1}{n_a^2}\sum_{i = 1}^{n}\sum_{j = 1}^{n}w_i w_j I(A_i = A_j = a) \lVert  {\bX}_i - {\bX}_j\rVert_2 
- \frac{1}{n^2}\sum_{i = 1}^{n}\sum_{j = 1}^{n}\lVert  {\bX}_i - {\bX}_j \rVert_2.
\end{align*}
In other words, $\calE(F_{n,a,\bw}, F_{n})$ is the energy distance between the ECDF of a sample $\{\bX_i\}_{i=1}^n$ and a weighted ECDF of a subsample $\{\bX_i\}_{i:A_i=a}$.

We show this new weighted energy distance is indeed a distance between $F_{n,a,\bw}$ and $F_n$. Let $\langle\mathbf{t}, \mathbf{s}\rangle$ be the inner product of vectors $\mathbf{t}$ and $\mathbf{s}$, and define the empirical characteristic function (ECHF) of $\{\bX_i\}_{i=1}^n$ as \[\varphi_{n}(\bt) \equiv \frac{1}{n}\sum_{i=1}^n\exp\{i\langle\mathbf{t}, {\bX}_i  \rangle\},\]
and the weighted ECHFs of $\{\bX_i\}_{i:A_i=a}$ as
\[ \varphi_{n,a,\bw}(\bt) \equiv  \frac{1}{n_a}\sum_{i=1}^nw_iI(A_i = a)\exp\{i\langle\mathbf{t}, {\bX}_i  \rangle\}.\]
The following proposition establishes the so-called \textit{distance property} of the weighted energy distance.
\begin{proposition}\label{thm:weighted_energy_duality}
Let $\bw$ be a vector of weights such that $\sum_{i=1}^nw_iI(A_i=a)=n_a$ for $a\in\{0,1\}$ and $w_i>0$ for $i=1,\dots,n$. Then
	\begin{align}
	\calE(F_{n,a,\bw}, F_{n}) = \int_{\bbR^p}\vert \varphi_{n}(\bt) - \varphi_{n,a,\bw}(\bt) \vert^2 \omega(\mathbf{t}) \mathrm{d}\mathbf{t} \quad \text{for} \quad a \in \{0,1\}, \label{eqn:empirical_energy_duality}
	\end{align}
	where $\omega(\mathbf{t}) = 1/(C_p\lVert\bt\rVert_2|^{1+p})$, $C_p = \pi^{(1+p)/2} / \Gamma((1+p)/2)$ is a constant, and $\Gamma(\cdot)$ is the complete gamma function. Thus, $\calE(F_{n,a,\bw}, F_{n}) \geq 0$ with equality to zero if and only if $\varphi_{n,a,\bw}(\bt) = \varphi_{n}(\bt)$ for all $\bt$.
\end{proposition}

\noindent Thus, the weighted energy distance is a distance between the weighted distribution of interest and the target distribution, and is thus a \textit{bona fide} measure of distributional balance of covariates induced by a set of weights. This proportion extends the duality results in Proposition 1 of \citet{szekely2013energy} and Theorem 1 of \citet{szekely2007measuring} for the weighted energy distance at hand. A subtle point is that Proposition \ref{thm:weighted_energy_duality}, combined with the decomposition presented in Section \ref{sec:wavg_est}, makes clear that $\calE(F_{n,a,\bw}, F_{n})$ more closely aligns with evaluating how well a set of weights estimates the \textit{sample} average treatment effect (SATE) \citep{ding2018} than the population ATE $\tau$ that is our main focus. Yet, as long as $F_n$ is representative of $F$, the weighted energy distance still aligns well with the population ATE.

We now show that the weighted energy distance converges to the energy distance when the weights yield a well-defined limiting distribution. 

\begin{thm}\label{thm:weighted_energy_converges_to_energy}
Assume $\bbE\left(\lVert \bX \rVert_2 \:\vert\: A = a\right) < \infty$ and  $\bbE\lVert \bX \rVert_2 < \infty$. Further assume, for a sequence of weights $\{\bw_n\}_{n=1}^{\infty}$ with $\sum_{i=1}^nw_iI(A_i=a)=n_a$ for $a\in\{0,1\}$ and $w_i>0$ for $i=1,\dots,n$ for each $n$ (so that $F_{n,a,\bw_n}$ are well-defined as CDFs), that $\lim_{n \to \infty} \varphi_{n,a,\bw_n}(\bt) \to \widetilde{\varphi}_{a}(\bt)$ almost surely for every $\bt \in \bbR^p$, where $\widetilde{\varphi}_{a}$ is some integrable characteristic function with associated CDF $\widetilde{F}_{a}(\bx)$. Then almost surely we have
\begin{equation}\label{eqn:weighted_energy_convergence}
    \lim_{n \to \infty} \calE(F_{n,a,\bw_n}, F_{n}) = \calE(\widetilde{F}_{a}, F).
\end{equation}
\end{thm}

\noindent Theorem \ref{thm:weighted_energy_converges_to_energy} shows that the weighted energy distance converges to the limiting energy distance. 
If the limiting distribution implied by a set of weights is $\widetilde{F}_{0} = \widetilde{F}_{1} = F$, then $\calE(\widetilde{F}_{0}, F)+\calE(\widetilde{F}_{1}, F)=0$. 
Proposition \ref{thm:weighted_energy_duality} and Theorem \ref{thm:weighted_energy_converges_to_energy} together imply that weights with smaller values of the sum $\calE(F_{n,1,{\bw}}, F_{n}) + \calE(F_{n,0,{\bw}}, F_{n})$ yield better distributional balance of covariates. Due to the link between imbalance and bias (see following section), this also implies better balance will yield estimates with smaller values for the terms \eqref{eqn:ate_bias_int}.

\subsection{Bias and distributional imbalance}\label{sec:biasimb}

We now demonstrate this connection between bias and distributional imbalance (as measured by the weighted energy distance) using two illustrative examples. A more formal presentation is provided later in Section \ref{sec:asymp} when proving asymptotic properties of EBWs.

In the first illustrative example, we generate a 1-dimensional covariate of sample size 250, which impacts treatment assignment via a logistic model under each of three scenarios: 1) $\text{logit}(\pi(X)) = -1 + X$, 2) $\text{logit}(\pi(X)) = -1 + X + 2X^2/3$, and 3) $\text{logit}(\pi(X)) = -1 + X + 2X^2/3 - X^3/3$. In each scenario, the response is generated as $Y=X+X^3-1/(0.1 + 0.1X^2)+\varepsilon$, where $\varepsilon\distas{} N(0,\sqrt{2})$. For each scenario, we construct IPWs based off of 3 logistic regression models, which consider only a linear term in $X$ (denoted as ``IPW (1)''), a linear plus quadratic term (``IPW (2)''), and up to the cubic term (``IPW (3)''), respectively. Thus, for Scenarios 2 and 3, at least one of the fitted models is misspecified. For each set of weights $\bw$, we compute $\calE(F_{n,0,\bw}, F_{n}) + \calE(F_{n,1,\bw}, F_{n})$, i.e., the sum of the energy distances between each treatment group and the combined sample, and compute the error $\widehat{\tau}_\bw - \tau$ of \eqref{eqn:wate} for $\tau$.

\begin{figure}
\subfigure{
\includegraphics[width=0.55\textwidth]{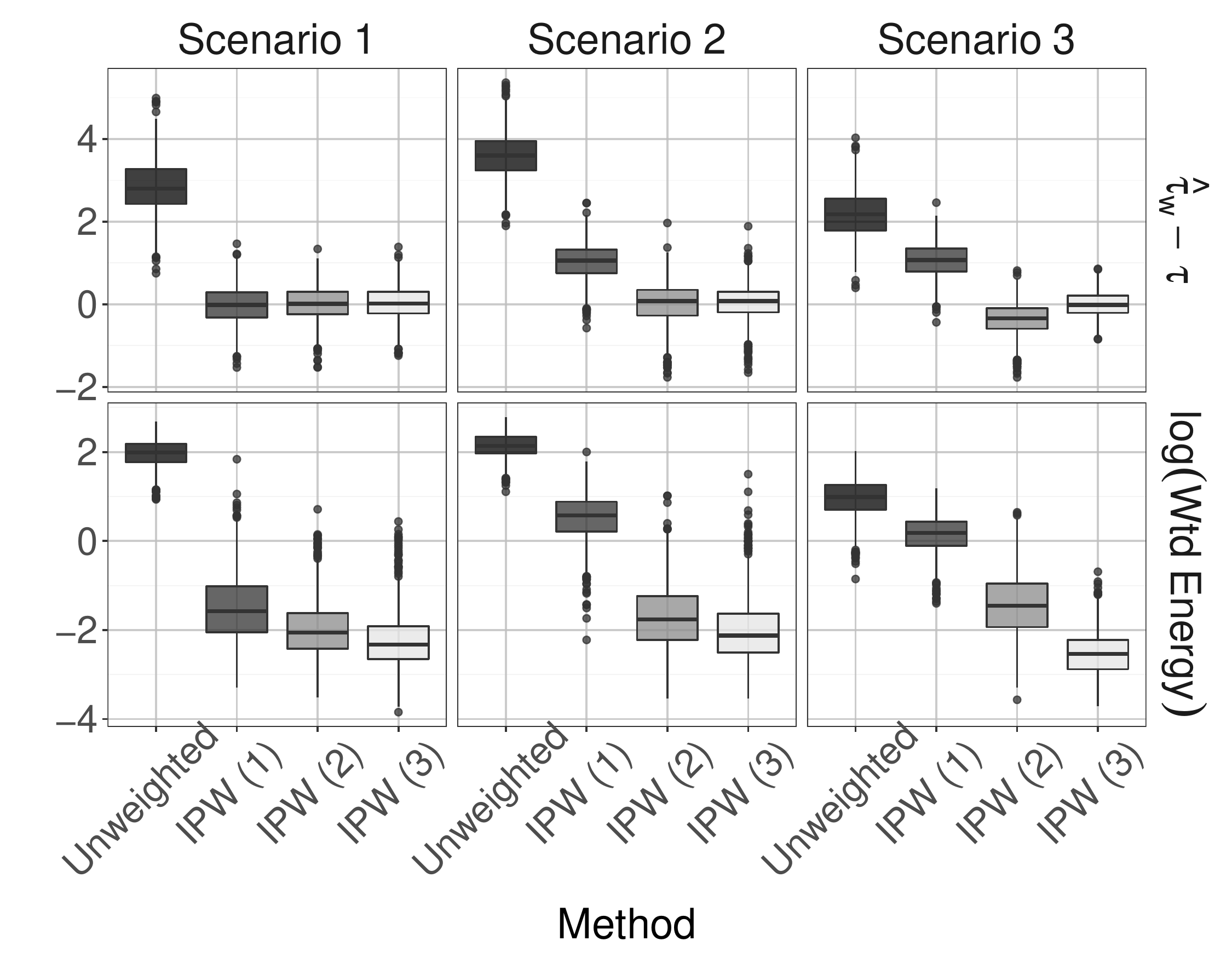}
}
\hfill
\subfigure{
\includegraphics[width=0.43\textwidth]{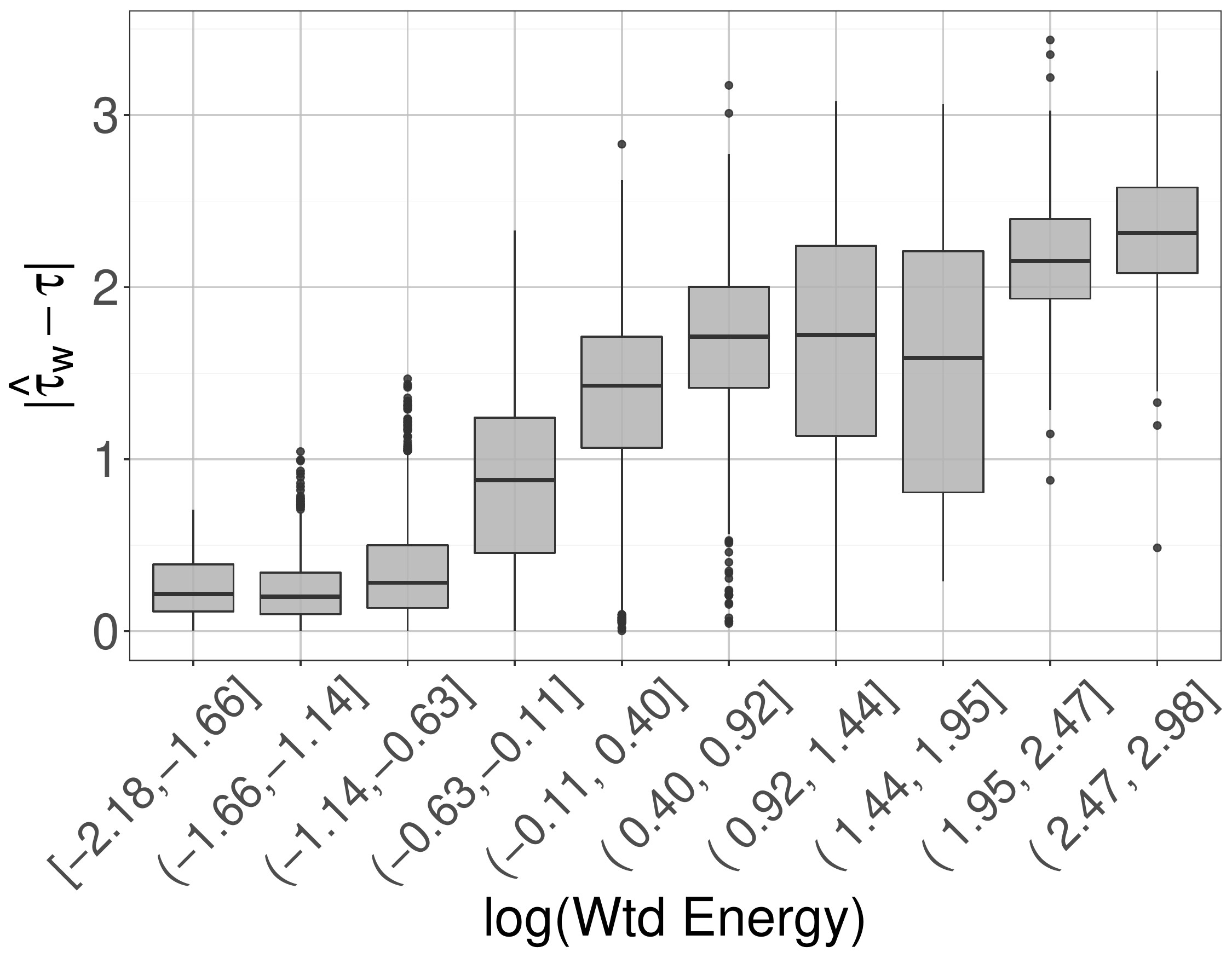}
}
\caption{\label{fig:toy_example} (a, left) Energy distances and biases for IPW estimates based on weights from the three fitted logistic regression models; (b, right) Boxplots of the biases for IPW estimates versus weighted energy distance based on weights estimated by several methods, each with different combinations of moments included for balancing or estimation.}
\end{figure}

Figure \ref{fig:toy_example} (a) displays the energy distances and biases over 1000 replications of the experiment. We see that the energy distances are the largest in all scenarios for the unweighted estimator (\eqref{eqn:wate} with all weights equal to 1). For scenarios where the weights are estimated using a misspecified model (IPW (1) in Scenario 2 and IPW (1) and (2) in Scenario 3), the energy distances are much larger than for weights based on correctly (or over-specified) models. Correspondingly, the bias is pronounced for the misspecified models. Thus, the weighted energy distance can be a useful tool to compare between different models, as weights with smaller weighted energy distances tend to yield estimates with smaller error.

In the second example, we consider a 2-dimensional example where the true assignment mechanism depends on first and second moments of the covariates. We consider several methods for estimate weights: logistic regression, the method of \citet{imai2014covariate}, and the method of \citet{chan2016globally}, each with different moments included for balancing or estimation. We then compare their weighted energy distances and absolute errors of \eqref{eqn:wate} over 1000 replications. Figure \ref{fig:toy_example} (b) displays the distances and errors for each dataset and method. We see that, in general, weights with lower weighted energy distance have a much smaller magnitude of bias in estimating the ATE.
\vspace{-20pt}
\subsection{Other measures of distributional imbalance}

There are, of course, many ways of measuring distance between distributions in the literature, including the Kolmogorov-Smirnov statistic, $f$-divergences (e.g., the Kullback-Leibler divergence and the Hellinger distance), the Wasserstein distance, and the maximum mean discrepancy. The energy distance has several advantages for characterizing distributional imbalance. First, while in general there is no uniformly most powerful nonparametric test for the difference between two distributions, the energy distance is often sensitive to differences in distributions, unlike the Kolmogorov-Smirnov statistic. Second, unlike the energy distance, $f$-divergences such as the Kullback-Leibler divergence and Hellinger distance do not metrize weak convergence; this property ensures that the distance remains stable under perturbations of the support of the distributions being measured \citep{genevay2019entropy}.  Third, the energy distance is easy and efficient to compute, unlike the Wasserstein distance. It also works reasonably well in moderately high dimensions \citep{mak2018support}, whereas the Wasserstein distance suffers from the curse of dimensionality, in the sense that empirical Wassterstein distances converge to the true Wasserstein distance at rate $\calO(n^{-1/p})$ \citep{weed2019sharp, genevay2019entropy}. Finally, while there is a direct link between maximum mean discrepancies (MMD) and the energy distance \citep{sejdinovic2013}, the energy distance does not require careful tuning of hyperparameters and tends to work well across a wide variety of scenarios. 

\section{Energy Balancing Weights}
\label{sec:energy_balancing}

\subsection{Definition}

We will now use the proposed weighted energy distance to estimate weights which i) match the distribution of covariates of the treated group to the distribution of covariates of the full population, and ii) match the distribution of covariates of the control group to the distribution of covariates of the full population. 

To achieve this, we define the \textit{energy balancing weights} (EBWs) to be
\begin{align} 
\bw^e_n \in {} &  \argmin_{\bw = (w_1, \dots, w_{n})} \left\{  \calE(F_{n,1,{\bw}}, F_{n}) + \calE(F_{n,0,{\bw}}, F_{n}) \right\} \nonumber \\
& \text{ s.t. }  \sum_{i=1}^nw_iA_i = n_1, \sum_{i=1}^nw_i(1-A_i) = n_0, w_i \ge 0 \text{ for } i=1,\dots,n. \label{eqn:energy_balancing_weights_ate}
\end{align}
Thus, the energy balancing weights $\bw^e_n$ minimize the statistical energy between the treatment and control groups to the full population. Due to the duality result of Proposition \ref{thm:weighted_energy_duality}, minimizing statistical energy is directly equivalent to balancing covariate distributions. The constraints $\sum_{i=1}^nw_iI(A_i = a) = n_a$ serve several purposes: they i) preserve the sample size of the weighted pseudo-population to that of the study population, ii) stabilize the estimated weights, and iii) ensure that $F_{n,0,{\bw}}$ and $F_{n,1,{\bw}}$ are valid distribution functions. Another important note is that, due to the bias decomposition in \eqref{eqn:ate_bias_int} and the duality result in Proposition \eqref{thm:weighted_energy_duality}, the weights $\bw^e_n$ are explicitly designed to minimize the key component of the finite-sample bias of $\widehat{\tau}_{\bw^e_n}$, in a manner agnostic to the functional forms of $\mu_0(\bx)$ and $\mu_1(\bx)$. If the functional forms of $\mu_0(\bx)$ and $\mu_1(\bx)$ are known, it is certainly possible to construct a different set of weights to better reduce bias, by emphasizing balance in regions of $\calX$ where either $\mu_0(\bX)$ or $\mu_1(\bX)$ are pronounced. However, this information is rarely (if ever) known, hence this is difficult to achieve in practice except by good fortune.

\begin{figure}[!t]
	\centering
	\includegraphics[width=0.7\textwidth]{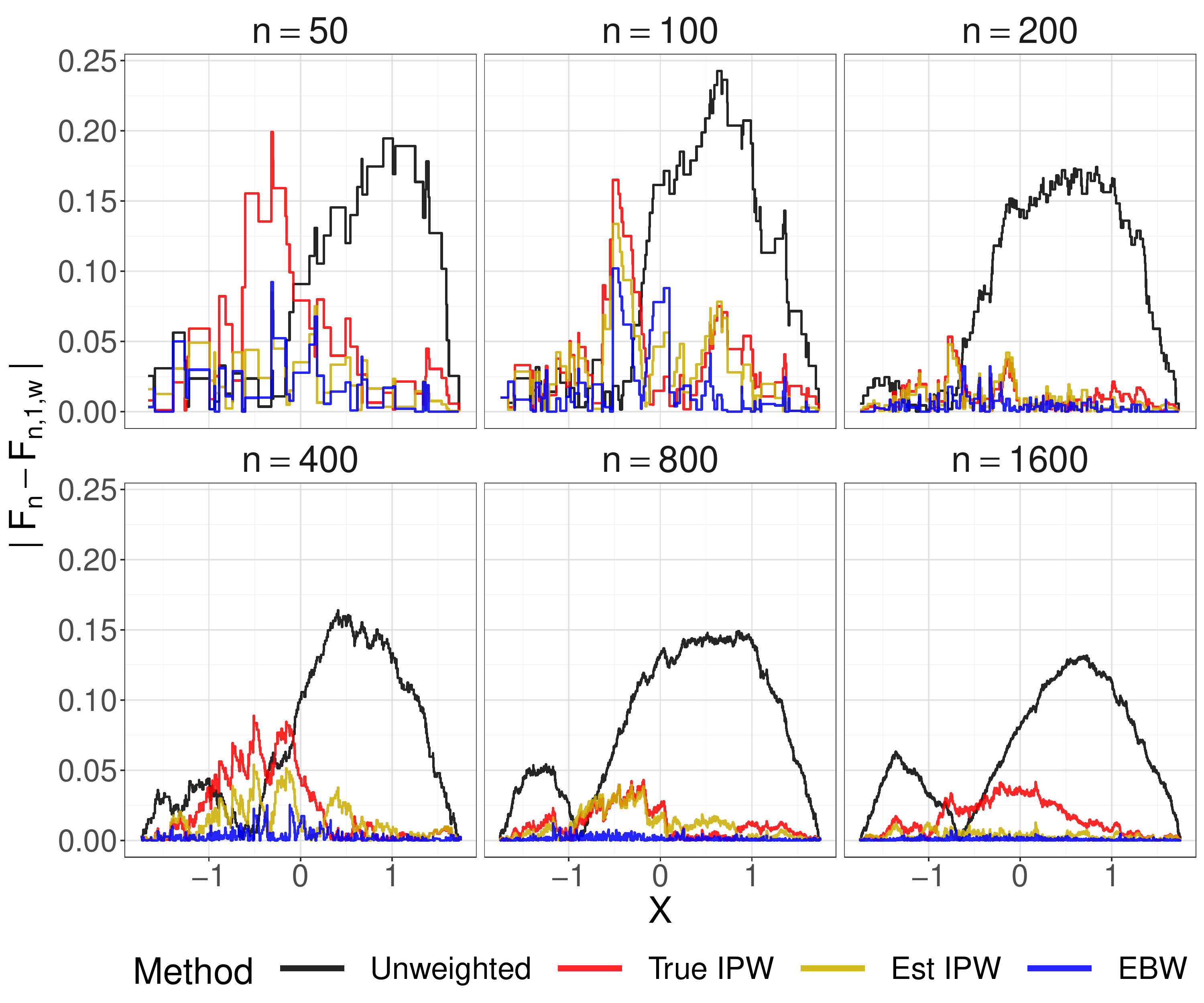}
	\caption{Displayed are the absolute differences between the ECDF of the combined sample and the weighted ECDF of the covariate in the treatment group based on energy weights. Also displayed the same for the unweighted ECDF of the treated group and the weighted ECDF based on the true and estimated propensity scores.}
	\label{fig:toy_example_3}
\end{figure}

To illustrate the effectiveness of EBWs for distributional balance, we consider data generated under Scenario 3 of the toy example in Figure \ref{fig:toy_example} (a). Figure \ref{fig:toy_example_3} shows the difference between the weighted ECDF using EBWs of the covariate in the treatment group and the ECDF of the combined (i.e., treated and untreated) sample, for varying sample sizes $n$. This is compared with the weighted ECDF using weights from the true data-generating propensity score, the estimated propensity score under the correct model, and the unweighted ECDF of the treatment group. As sample size increases, the difference vanishes for energy balancing weights, but much more slowly for both the true and estimated propensity score weights. This demonstrates the improved distributional balance provided by the proposed EBWs, which should then translate to a greater reduction of bias. 
 
\subsection{Asymptotic properties} \label{sec:asymp}

Next, we show two desirable properties of the proposed EBWs. We first show that the weighted ECDFs based on EBWs indeed converge to the population CDF $F$ of $\bX$. 
\begin{thm}\label{thm:thm1}
	Assume that $\bbE\left(\lVert \bX \rVert_2 \:\vert\: A = a\right) < \infty$ and $\bbE\lVert \bX \rVert_2 < \infty$ and that the assumptions presented in Section \ref{sec:setup} hold. 
	Let $\bw^e_n$ be as defined in \eqref{eqn:energy_balancing_weights_ate}. Then, for $a \in \{0,1\}$, 
	\begin{equation}\label{eqn:energy_weighted_cdf_convergence}
	\lim_{n \rightarrow \infty} {F}_{n,a,\bw^e_n}(\bx) \equiv \lim_{n \rightarrow \infty} \frac{1}{n_a}\sum_{i=1}^nw^e_{i}I(\bX_i \leq \bx, A_i = a)
	= F(\bx) 
	\end{equation}
	almost surely for every continuity point $\bx \in \calX$. Furthermore,\\ 	
	 \[\lim_{n \to \infty} \calE(F_{n,a,\bw^e_n}, F_{n}) = 0\]
	 holds almost surely.
\end{thm}

\noindent Thus, energy balancing weights result in the almost sure convergence of the weighted ECDFs of the treated (and untreated) group to the underlying covariate distribution $F$.   

The consistency of the energy-weighted ATE estimator $\widehat{\tau}_{\bw_n^e}$ follows immediately as a corollary of Theorem \ref{thm:thm1}:
\begin{cor}\label{thm:cor1}
	Suppose the conditions of Theorem \ref{thm:thm1} hold, and that the treatment and control mean response functions $\mu_0(\bx)$ and $\mu_1(\bx)$ are bounded and continuous on $\calX$. Then $\widehat{\tau}_{\bw_n^e}$ is a consistent estimator of $\tau$.
\end{cor}

Next, we show that the ATE estimator $\widehat{\tau}_{\bw_n^e}$ is root-$n$ consistent. To do so, we make use of the following lemma, which connects bias and distributional balance:
\begin{lem}
Let $\calH$ be the native space induced by the radial kernel $\Phi(\cdot) = -\| \cdot \|_2$ on $\calX$, and suppose $\mu_a(\cdot) \in \calH$ for $a \in \{0,1\}$. Then, for any weights $\bw$ satisfying $\sum_{i=1}^nw_iA_i = n_1, \sum_{i=1}^nw_i(1-A_i) = n_0, w_i \ge 0$, we have for $a\in\{0,1\}$:
\begin{equation}
\left[ \int \mu_{a}(\bx) \mathrm{d} \left[ F_{n, a, \bw} - F_n  \right](\bx) \right]^2 \leq C_a \mathcal{E}(F_{n,a,\bw},F_n),
\end{equation}
where $C_a \geq 0$ is a constant depending on only $\mu_a$. 
\label{lem:bd}
\end{lem}
Lemma \ref{lem:bd} provides a link between the systematic bias in \eqref{eqn:ate_bias_int} and the weighted energy distance $\mathcal{E}(F_{n,a,\bw},F_n)$. Under the mild condition that the conditional mean function $\mu_a(\bx)$ is found in $\mathcal{H}$ (this is discussed further at the end of the section), this lemma shows that the weighted energy distance is a key component in an upper bound on the systematic bias in \eqref{eqn:ate_bias_int}. We note that Lemma \ref{lem:bd} applies to any set of auto-normalized weights, justifying the use of the weighted energy distance to compare between different sets of weights. The proposed EBWs $\bw_n^e$, which minimize $\mathcal{E}(F_{n,a,\bw},F_n)$, may therefore yield lower estimation bias via this upper bound, which is in line with the empirical observations in Section \ref{sec:biasimb}.

With this, we now state the result on root-$n$ consistency:
\begin{thm}
Assume the same conditions in Theorem \ref{thm:thm1}. Let $\calH$ be the native space induced by the radial kernel $\Phi(\cdot) = -\| \cdot \|_2$ on $\calX$. Suppose the following mild conditions hold:
\begin{enumerate}[label=\textup{(A\arabic*)}]
\item $\mu_0(\cdot) \in \calH$ and $\mu_1(\cdot) \in \calH$,
\item $\textup{Var}[\mu_0(\bX)] < \infty$ and $\textup{Var}[\mu_1(\bX)] < \infty$,
\item $\sigma^2_0(\bx)$ and $\sigma^2_1(\bx)$ are bounded over $\bx \in \calX$,
\item $\mathbb{E}[h_0^2(\bX,\bX',\bX'',\bX''')] < \infty$ and $\mathbb{E}[h_1^2(\bX,\bX',\bX'',\bX''')] < \infty$, where $\bX,\bX',\bX'',\bX''' \distas{i.i.d.} F$ and, with $\pi_0(\bx) := 1-\pi(\bx)$ and $\pi_1(\bx) := \pi(\bx)$, the kernel $h_a$ is defined for $a=0,1$ as:
\begin{equation} 
\displayindent0pt
\displaywidth\textwidth
h_a(\bx,\by,\bz,\ba) = \frac{1}{\pi_a(\bx)}||\bx - \bz||_2 + \frac{1}{\pi_a(\by)}||\by - \ba||_2  - \frac{1}{\pi_a(\bx)\pi_a(\by)}||\bx - \by||_2 - ||\ba - \bz||,
\label{eq:kernh}
\end{equation}
\item The energy balancing weights $\bw_n^e = (w_{i,n}^e)_{i=1}^n$ in \eqref{eqn:energy_balancing_weights_ate} satisfy $w_{i,n}^e \leq Cn^{1/3}$ for some constant $C>0$ independent of $n$.
\end{enumerate}
Then the proposed EBW estimator $\widehat{\tau}_{\bw_n^e}$ is root-$n$ consistent, i.e.:
\begin{equation}
\sqrt{\mathbb{E}_{\bX,A,Y}[(\widehat{\tau}_{\bw_n^e}-\tau)^2]} = \mathcal{O}(n^{-1/2}).
\label{eqn:asympvar}
\end{equation}
\label{thm:rootn}
\end{thm}
\vspace{-2em} 
\noindent We give a brief discussion of Assumptions (A1)-(A5). Assumption (A1) concerns the regularity of the conditional mean functions $\mu_0$ and $\mu_1$. It can be shown that the Sobolev space $W_{\lceil (p+1)/2 \rceil,2}(\calX)$ -- the space of functions with square-integrable $r < \lceil (p+1)/2 \rceil$-th differentials -- is contained within the native space $\calH(\calX)$ (Theorem 10.42 of \citealp{wendland2004scattered}), so (A1) can be viewed as a smoothness assumption on the conditional ATE $\mu_1 - \mu_0$ (a similar assumption is made in \citealp{wong2017kernel}). Assumptions (A2) and (A3) require the conditional mean functions to have finite variance and the conditional variance function to be bounded, respectively. Assumption (A4) require the kernels $h_0$ and $h_1$ to have finite second moments; these kernels can be seen as a ``modified'' Euclidean kernel weighted by the true propensity scores. Assumption (A5) assumes all EBWs to be bounded above by $Cn^{1/3}$ for some positive constant $C>0$; this assumption has been used in several weighting-based covariate balancing methods (e.g., \citealp{wong2017kernel,athey2018approximate}). In practice, (A5) can always be checked after optimizing for the EBWs in \eqref{eqn:energy_balancing_weights_ate}. One can change the optimization procedure \eqref{eqn:energy_balancing_weights_ate} to explicitly enforce (A5), but we have never encountered ``exploding weights'' which violate (A5) in practice; the EBWs in our simulations and data analysis typically satisfy (A5) with $C=1$. 

While Theorem \ref{thm:rootn} proves the desired root-$n$ consistency of the proposed EBW estimator $\widehat{\tau}_{\bw_n^e}$, it unfortunately does not shed light on its asymptotic variance, which is useful for constructing confidence intervals on $\tau$. An exact variance expression is difficult to obtain here, since EBWs are obtained via the non-convex optimization problem \eqref{eqn:energy_balancing_weights_ate}. We instead recommend the use of bootstrapped confidence intervals for the EBW estimator.

\subsection{Connection to importance sampling}

We briefly comment on an interesting parallel between the proposed EBWs for distributional balance and the notion of importance weights for importance sampling. Importance sampling (see, e.g., \citealp{geweke1989bayesian}) is a general technique for estimating integral quantities from a desired distribution $G$, using samples of another distribution $H$. The idea is to reweigh each sample from $H$ by its importance weight $dG/dH$: the Radon-Nikodym derivative (or likelihood ratio) of $G$ with respect to $H$. Clearly, such weights perfectly balance the sample from $H$ to the desired distribution $G$. For distributional balance, $H$ is the covariate distribution for the treated or control case (which we have access to), and $G$ is the covariate distribution for the full population (which we wish to infer).

This link to importance sampling reveals two insights on the proposed distributional balance approach via EBWs. First, in order for importance weights (which are the Radon-Nikodym derivatives) to exist here, the population distribution must be absolutely continuous with respect to the treated and control distributions, which requires $f_{\bX|A=0}(\bx) > 0$ whenever $f(\bx) > 0$ and $f_{\bX|A=1}(\bx) > 0$ whenever $f(\bx) > 0$ for almost all $\bx \in \calX$, where $f$ are (conditional) densities of the covariates. But this condition is satisfied by the positivity (or probabilistic assignment) assumption in Section 2, which requires the propensity score $\pi(\bx)$ to satisfy $0 < \pi(\bx) < 1$. Hence, similar conditions are needed for distributional balance in both importance sampling and causal analysis. Second, it is known that under mild conditions, integral estimates under importance sampling are root-$n$ consistent if the underlying samples from $H$ are i.i.d. sampled \citep{robert2013monte}. The proof of Theorem \ref{thm:rootn} makes use of such results on importance weighting to establish the root-$n$ consistency of the EBW estimator.

A key difference between importance weights and EBWs is that the former depends on both the population covariate density $f(\bx)$ and the propensity score $\pi(\bx)$, both of which are unknown in practice. The proposed method offers a \textit{nonparametric} way for estimating distribution-balancing weights. It optimizes weights in the new weighted energy distance in Section 2, thereby balancing to the desired target distribution $F$ (of which $F_n$ is assumed to be a representative sample). When a distribution other than $F$ is of interest (see \citet{li2018balancing} for examples of other common target distributions), this importance sampling perspective of EBWs allows for a straight-forward modification of the criterion in Section \ref{sec:energy_balancing} to balance to the target distribution. 

\subsection{Optimization}
The optimization problem \eqref{eqn:energy_balancing_weights_ate} for computing EBWs (and the later optimization problem \eqref{eqn:improved_energy_balancing_weights_ate} for obtaining three-way EBWs) are quadratic programming problems with linear (in)equality constraints. Hence, this optimization can be efficiently solved using interior point methods for cone programming \citep{andersen2011interior}. We have implemented solutions to both  \eqref{eqn:energy_balancing_weights_ate} and \eqref{eqn:improved_energy_balancing_weights_ate} in the \texttt{R} package \texttt{ebw}, which calls the interior-point cone programming solvers available in the \texttt{R} package \texttt{cccp} \citep{cccp}. In our experience, cone programming solvers provide remarkably robust and efficient solutions. The operator splitting solver of \citet{stellato2020osqp} is another reliable option as it works for non-positive definite quadratic programs. In practice, the covariates are normalized so that each covariate has mean zero and unit variance, before being used as inputs for the optimization problem.



\section{Extensions and applications of energy balancing weights}
\label{sec:extensions}


\subsection{Three-way energy balancing weights}

The EBWs in \eqref{eqn:energy_balancing_weights_ate} are designed to balance the distributions of covariates between each treatment group to that of the combined sample $\{\bX\}_{i=1}^n$. As such, the treatment group should be asymptotically balanced to the control group. 
Yet for finite samples, EBWs do not necessarily guarantee good distributional balance between the treatment and control arms, which can be important in practice. Consider the following re-expression of the two terms in \eqref{eqn:ate_bias_int}:
\begin{align}
& \int_{\bx\in \calX}  \left[\mu_{1}(\bx) + \mu_{0}(\bx)\right] \mathrm{d} \left[ F_{n,1,\bw} - F_{n,0,\bw}\right](\bx) \label{eqn:reducible_bias_trt_ctrl} \\ 
& - \int_{\bx\in \calX}  \mu_{1}(\bx) \mathrm{d} \left[ F_n - F_{n, 0, \bw} \right](\bx) + \int_{\bx\in \calX}  \mu_{0}(\bx) \mathrm{d} \left[ F_n - F_{n, 1, \bw} \right](\bx).  \label{eqn:reducible_biases_fullsamp2}
\end{align}
The terms \eqref{eqn:reducible_bias_trt_ctrl} and \eqref{eqn:reducible_biases_fullsamp2} shed light on how the choice of $\bw$ impacts estimation error for the ATE. In particular, this error term depends not only on i) how close the weighted ECDFs $F_{n, 0, \bw}$ and $F_{n, 1, \bw}$ are to that of the combined sample $F_{n}$, but also ii) how close the weighted ECDF for the control group $F_{n, 0, \bw}$ is to the weighted ECDF of the treated group $F_{n, 1, \bw}$. The EBWs in Section \ref{sec:energy_balancing} take care of the imbalance in i), but not the imbalance in ii) between treatment and control. With finite samples, imbalance in ii) can result in \eqref{eqn:reducible_bias_trt_ctrl} dominating \eqref{eqn:reducible_biases_fullsamp2}, depending on the properties of $\mu_{0}(\bx), \mu_{1}(\bx)$, and $\mu_{0}(\bx) + \mu_{1}(\bx)$.

Indeed, the importance of this \textit{three-way balance} is recognized in the literature, and is emphasized in \citet{chan2016globally} as an important component in constructing globally efficient estimators based on moment balancing. In our case, the target is three-way \textit{distributional} balance, i.e., balance in i) and ii). 

We propose an extension of EBWs, called the \textit{improved energy balancing weights} (iEBWs), to help improve covariate balance between the treatment and control groups. These improved weights are defined as follows:
\begin{align}
	{\bw}^{ei}_n  \in {} &  \argmin_{\bw} \left\{  \calE(F_{n,1,\bw}, F_{n}) + \calE(F_{n,0,\bw}, F_{n}) + \calE(F_{n,0,\bw}, F_{n,1,\bw}) \right\} \nonumber \\
	& \text{s.t. } \sum_{i=1}^nw_iA_i = n_1, \sum_{i=1}^nw_i(1-A_i) = n_0, w_i \ge 0 \text{ for } i=1,\dots,n, \label{eqn:improved_energy_balancing_weights_ate}
\end{align}
where 
\begin{align*}
	&\calE(F_{n,1,\bw}, F_{n,0,\bw}) \equiv {}  \frac{2}{n_1n_0}\sum_{i = 1}^n \sum_{j = 1}^nw_iw_jA_i(1-A_j)\lVert {\bX}_i - {\bX}_j \rVert_2 \\
	& - \frac{1}{n_1^2}\sum_{i = 1}^{n}\sum_{j = 1}^{n}w_i w_j A_iA_j \lVert  {\bX}_i - {\bX}_j\rVert_2 - \frac{1}{n_0^2}\sum_{i = 1}^{n}\sum_{j = 1}^{n}w_i w_j (1-A_i)(1-A_j)\lVert  {\bX}_i - {\bX}_j \rVert_2
\end{align*}
is the energy distance between the weighted ECDFs for treated and control. Thus, the iEBWs ${\bw}^{ei}_n$ not only aim to minimize imbalance between treatment arms and the full population, but also the energy between the treatment arm and control arm. Note that 
$\calE(F_{n,1,\bw}, F_{n,0,\bw})$ still retains the properties of a weighted energy distance, in the sense that $\calE(F_{n,0,\bw}, F_{n,1,\bw}) = \int_{\bbR^p}\vert \varphi_{n,1,\bw}(\bt) - \varphi_{n,0,\bw}(\bt) \vert^2 \omega(\mathbf{t}) \mathrm{d}\mathbf{t}$, as Proposition \ref{thm:weighted_energy_duality} can be trivially extended to the case where both arguments of the energy distance are weighted.

\subsection{Estimation of the average treatment effect on the treated }
\label{sec:att}

A common target of estimation is the (population) average treatment effect on the treated (ATT) $\tau^{(1)} \equiv \bbE(Y(1) - Y(0) \: | \: A=1)$, which is the mean difference in potential outcomes among those who are actually treated. Due to the unconfoundedness assumption, we can write
\begin{equation}
\tau^{(1)} = \int_{\bx\in \mathcal{X}} \left[  \mu_{1}(\bx) - \mu_{0}(\bx) \right] \mathrm{d} F_1(\bx),
\end{equation}
which suggests that a plug-in estimator can be obtained by replacing $F_1(\bx)$ with a suitable energy-weighted ECDF.
To do so, we define the new weights $\bw^{e1}_n = (w^{e1}_1, \dots, w^{e1}_n)$, where $w^{e1}_i = 1$ for $\{i:A_i = 1\}$ and 
\begin{align}
	 \{ w^{e1}_i \}_{i : A_i = 0} \in {} &  \argmin_{\bw}  \calE(F_{n,0,{\bw}}, F_{n,1})   \nonumber \\
	& \text{ s.t. } \sum_{i=1}^nw_i(1-A_i) = n_0, w_i \ge 0 \text{ for } i=1,\dots,n, \label{eqn:energy_balancing_weights_att}
\end{align}
where $F_{n,1}(\bx) = \frac{1}{n_1}\sum_{i=1}^nA_iI(\bX_i\leq \bx)$. Thus, $\bw^{e1}_n$ balances the covariate distribution for the control group to the covariate distribution of the treated group. Similar to Theorem \ref{thm:thm1}, we have $\lim_{n\to\infty}F_{n, 0, \bw^{e1}_n}(\bx) = F_1(\bx)$
a.s. for every continuity point $\bx \in \mathcal{X}$.
 With the new weights $\bw^{e1}_{n}$, the natural plug-in estimate for the ATT $\tau^{(1)}$ is $\hat{\tau}_{\bw^{e1}_{n}}$ (i.e. the weighted estimator in \eqref{eqn:wate} with $\bw = \bw^{e1}_{n}$).

\subsection{Multi-category treatments}
\label{sec:multiple_treatments}

EBWs can also be constructed for multi-category treatments. Assume the treatment variable $A_i$ takes multiple values, i.e. $A_i \in \mathcal{A} = \{a_1, \dots, a_K\}$. Denote $n_a = \sum_{i=1}^nI(A_i = a)$, for $a \in \mathcal{A}$. Each unit has $K$ potential outcomes $(Y_i(a_1), \dots, Y_i(a_K))$, one for each treatment option. The strong unconfoundedness assumption in the multiple treatment setting becomes $\{Y_i(a_1), \dots, Y_i(a_K)\}\\ \indep A \: | \: \bX$. Following \citet{lopez2017estimation}, the positivity assumption required is now $0 < \pi(a, \bx) = \bbP(A = a \:|\: \bX = \bx) < 1$ for all $a \in \mathcal{A}$ and all possible $\bx \in \calX$. The standard IPW estimator for $\bbE[Y(a)]$ involves inverse weighting each sample $i$ by $\pi(A_i, \bX_i)$. Instead, we propose to estimate $\bbE[Y(a)]$ or any causal contrast $\tau^{(a - a')} \equiv \bbE[Y(a) - Y(a')]$ for $a, a' \in \calA$ with EBWs.

We define the energy balancing weights for the multiple treatment case as
\begin{align*} 
	\bw^{em}_n \in {} &  \argmin_{\bw = (w_1, \dots, w_{n})} \sum_{a \in \mathcal{A}} \calE(F_{n,a,{\bw}}, F_{n}) \nonumber \\
	& \text{ s.t. } \sum_{i=1}^nw_iI(A_i = a) = n_a \text{ for all } a \in \mathcal{A} \text{ and } w_i \ge 0 \text{ for } i=1,\dots,n. 
\end{align*}
Improved EBWs, which encourage covariate balance between all pairs of treatment options, can be defined similarly as \eqref{eqn:improved_energy_balancing_weights_ate}, where an additional weighted energy distance between each pair of treatment options is added to the objective.
Given any two treatment options $a, a' \in \mathcal{A}$, we can then estimate the causal contrast $\tau^{(a - a')}$ with
\begin{equation*}
\widehat{\tau}^{(a-a')}_{\bw^{em}_n} = \frac{1}{n_a}\sum_{i=1}^nw_iY_iI(A_i = a) - \frac{1}{n_{a'}}\sum_{i =1}^nw_iY_iI(A_i = a').
\end{equation*}

\subsection{Estimation of individualized treatment rules}
\label{sec:value}

As many treatments exhibit heterogeneous effects for different patients, there is great interest in tailoring treatment decisions to patients. A main line of work in this area is the development of statistical methods aimed at finding an optimal individualized treatment rule (ITR), which maps patient characteristics to treatment decisions. Thus, the immediate goal is to estimate a mapping $d:\mathcal{X}\mapsto \{0,1\}$ which optimizes the expected potential outcomes under the distribution induced by $d$. Following \citet{qian2011performance}, and \citet{zhao2012estimating} and assuming that larger values of the outcome are preferred, the optimal ITR is defined as 
\begin{align}
d^* &\in \argmax_{d}\bbE[Y(d)] {} \nonumber \\
&=\argmax_{d}\bbE\left[\frac{YI(A = d(\bX))}{\pi(A, \bX)}\right] = \argmin_{d}\bbE\left[\frac{YI(A \neq d(\bX))}{\pi(A, \bX)}\right], \label{eqn:optimal_itr}
\end{align}
where $\pi(a, \bX) = \bbP(A=a\:\vert\: \bX)$ and the second equality holds due to the causal assumptions outlined in Section \ref{sec:setup}. The optimal ITR $d^*$ has the property that $d^*(\bx)=a \implies \mu_a(\bx)>\mu_{1-a}(\bx)$.  The last term in \eqref{eqn:optimal_itr} appears as a weighted classification task due to the weighted 0-1 loss. With observed data, the objective becomes to minimize
\begin{equation}\label{eqn:empirical_owl}
\frac{1}{n}\sum_{i=1}^n\frac{Y_i}{\pi(A_i, \bX_i)}I(A_i \neq d(\bX_i)).
\end{equation}
Due to the non-convexity of \eqref{eqn:empirical_owl}, in practice $I(A_i \neq d(\bX_i))$ is replaced with a surrogate, such as the hinge function $(1 - (2A_i-1)d(\bX_i))^+$ (see, e.g., the outcome weighted learning (OWL) method of \citet{zhao2012estimating}). Yet, this objective still requires estimation of the propensity score $\pi(\bX)$ and involves plugging in this estimate to \eqref{eqn:empirical_owl}, which subjects it to the same issues of propensity score weighted estimates of $\tau$.
To see how EBWs can be used in place of $\pi(A, \bX)$, we can express \eqref{eqn:optimal_itr} as
\begin{align*}
 &\argmin_{d} \int_{\bx \in \mathcal{X}}\sum_{a \in \{0,1\}}{\mu_a(\bx)}I(d(\bx) \neq a)\mathrm{d} F(\bx).
\end{align*}
Thus, $d^*(\bx)$ is a functional of $F(\bx)$. This suggests a plug-in estimator by replacing $F(\bx)$ with energy-weighted ECDFs (either $F_{n,a,\bw_n^e}(\bx)$ or  $F_{n,a,\bw_n^{ei}}(\bx)$). Thus, we propose to estimate the optimal ITR as 
\begin{align}\label{eqn:energy_owl}
\hat{d}^* \in {} & \argmin_{d} \frac{1}{n}\sum_{i=1}^nY_i{w}^{e}_i\phi(A_i, d(\bX_i)),
\end{align}
where  $w_i^{e}$ could be replaced with improved weights $w_i^{ei}$ and $\phi(a, d)$ is a convex surrogate for $I(a \neq d)$, such as the hinge function or logistic loss. 
As in \citet{zhao2012estimating}, to prevent overfitting, a penalty $\lambda_n\lVert d \rVert^2$ can be added to \eqref{eqn:energy_owl} to control complexity of the estimated ITR.

To demonstrate the effectiveness of using energy balancing weights in optimal ITR estimation, we provide an illustrative example under two data-generating scenarios. For both scenarios we generate outcomes as $Y = g(\bX) + \widetilde{A}\Delta(\bX)/2 + \varepsilon$, where $g(\bX)$ are the main effects of $\bX$, $\widetilde{A} = 2A-1$, and $\Delta(\bX)=\mu_{1}(\bX) - \mu_{0}(\bX)$ is the treatment-covariate interaction, $\varepsilon \distas{}N(0,1)$, and $\bbR^{10}\ni\bX\distas{i.i.d.}\text{Unif(-1,1)}$. Both scenarios are motivated by the simulation studies of \citet{zhao2012estimating}, but generate $A$ from a logistic regression model with terms depending on up to third order polynomials in a subset of the predictors, and $g(\bX)$ contains non-linear terms in the predictors. Full details of the experiment are given in the Supplementary Material. We utilize the OWL method to obtain estimates $\hat{d}$, which uses inverse weighting via the propensity score and adds a penalty $\lambda_n\lVert d \rVert^2$ to the objective. For OWL, the propensity score is misspecified to only include linear terms in the covariates. We also estimate $d^*$ by minimizing \eqref{eqn:energy_owl} plus $\lambda_n\lVert d \rVert^2$. We denote this as OWL (EBW) for weights given by \eqref{eqn:energy_balancing_weights_ate} and OWL (iEBW) for weights given by \eqref{eqn:improved_energy_balancing_weights_ate}. We simulate 1000 independent datasets and compute the average value function $\widehat{\bbE}[Y(\hat{d})]$ evaluated on a large independent dataset in addition to the missclassification rate in estimating $d^*(\bX)$ on the independent dataset. The results are given in Table \ref{tab:optITRexample}, which indicate that energy balancing weights can yield better performing ITRs.

\begin{table}
\caption{ \label{tab:optITRexample} Displayed are the value functions and misclassification rates for the optimal ITR estimation example averaged over 1000 independent simulated datasets. In Scenario 1, the optimal value is 4.74 with $56\%$ of units with optimal assignments of $A=1$; in Scenario 2, the optimal value is 3.19 with $50\%$ of units with optimal assignments of $A=1$.}
\centering
\begin{tabular}{lcccRcc}
\toprule
 & \multicolumn{3}{c}{Scenario 1} & \multicolumn{3}{c}{Scenario 2} \\ 
 \cmidrule{2-4} \cmidrule{5-7}
 \multicolumn{1}{c}{Method}  & \multicolumn{2}{c}{Value (SD)} & Misclass & \multicolumn{2}{c}{Value (SD)} & \multicolumn{1}{c}{Misclass} \\ 
\midrule
\multicolumn{1}{l}{OWL (EBW)} & \multicolumn{1}{c}{$3.168$} & \multicolumn{1}{c}{$(0.253)$} & \multicolumn{1}{c}{$0.283$} & \multicolumn{1}{c}{$2.921$} & \multicolumn{1}{c}{$(0.166)$} & \multicolumn{1}{c}{$0.228$}  \\
\multicolumn{1}{l}{OWL (iEBW)} & \multicolumn{1}{c}{$\mathbf{3.198}$} & \multicolumn{1}{c}{$(\mathbf{0.204})$} & \multicolumn{1}{c}{$\mathbf{0.277}$} & \multicolumn{1}{c}{$\mathbf{2.931}$} & \multicolumn{1}{c}{$(\mathbf{0.165})$} & \multicolumn{1}{c}{$\mathbf{0.223}$}   \\
\multicolumn{1}{l}{OWL (PS)} & \multicolumn{1}{c}{$2.671$} & \multicolumn{1}{c}{$(0.327)$} & \multicolumn{1}{c}{$0.344$} & \multicolumn{1}{c}{$2.706$} & \multicolumn{1}{c}{$(0.196)$} & \multicolumn{1}{c}{$0.295$} \\
\bottomrule 
\end{tabular}%
\end{table}



\section{Simulation studies}
\label{sec:simulation}

To evaluate the finite sample performance and operating characteristics of our proposed estimators, we conducted a large-scale simulation study across a wide variety of data-generating scenarios. Since existing techniques, such as empirical calibration balancing and the covariate balancing propensity score, work exceedingly well when the correct moments are specified to be balanced, we primarily consider simulation settings where the relationships between covariates and both the treatment status and outcome regression model are non-linear. We consider a wide range of scenarios for the propensity score and outcome regression models, several of which are examples taken from existing works. Outcome models B and E are taken from \cite{wong2017kernel}, outcome model D is taken from \citet{cannas2019comparison}, and outcome model A is a slight modification of an outcome model from \citet{kang2007demystifying}. Outcome model C is designed to be linear in $\bX$ for the untreated group, but nonlinear and with interaction terms for the treated group. 
Propensity model III is from \citet{kang2007demystifying} but with the main effects having twice the dimension as in the \citet{kang2007demystifying} example. The remainder of the propensity models have either interactions, smooth nonlinear terms, or non-smooth nonlinear terms. We generate a $p$-dimensional covariate vector $\bZ=(Z_i, \dots, Z_p)$ from a $p$-dimensional mean zero multivariate Gaussian distribution with $\text{Cov}(Z_i,Z_j) = -0.75^{|i-j|}$, which results in positive and negative correlations between predictors. In covariate setup 1, we let the observed covariates be $\bX = \bZ$, however, similar to the widely-tested setup in \citet{kang2007demystifying}, in covariate setup 2 we define $\bX = (X_1, \dots, X_p)$ where $X_1 = \exp(Z_1/2)$, $X_2 = Z_2 / (1 + \exp(Z_1)) + 10$, $X_3 = (Z_1Z_3/25 + 0.6)^3$, $X_4 = 20 + (Z_2 + Z_4)^2$, $X_5 = \exp(Z_5/2)$, $X_6 = Z_6 / (1 + \exp(Z_5)) + 10$, $X_7 = (Z_1Z_7/25 + 0.6)^3$, $X_8 = 5 + (Z_6 + Z_8)^2$. To align with the setup of \citet{kang2007demystifying}, we utilize covariate setup 2 for any setting with propensity model III. To align with the setup of \citet{wong2017kernel}, we utilize covariate setup 2 for any setting with outcome model B or E. All other settings use covariate setup 1. 

We compare our proposed energy balancing weights (denoted ``EBW'') and the improved energy balancing weights \eqref{eqn:improved_energy_balancing_weights_ate} (denoted ``iEBW'') with several widely-used alternatives. First, we compare with inverse propensity score weights (denoted ``IPW''). In addition, we compare with the covariate balancing propensity score weights (denoted ``CBPS''), the empirical calibration balancing weights (denoted ``Cal'') with exponential tilting weights. For all methods which require specification of a model for the treatment assignment or moments to balance, only first order terms in $\bX$ were used, as in \citet{wong2017kernel}. This reflects a setting where the analyst misspecifies that only first moments should be balanced. We also utilize the kernel-based functional covariate balancing approach of \citet{wong2017kernel}, denoted as ``KCB'' for kernel covariate balancing with the second-order Sobolev kernel. As a baseline for comparison, we investigate a na\"ive unweighted estimator which simply compares the means between treated groups and denote this as ``Unweighted''. For all estimators, we use weights normalized by treatment group. Thus, the IPW approach is the H\'ajek estimator as opposed to the more unstable, non-normalized Horvitz-Thompson estimator. For each setting, we generate 1000 independent datasets and evaluate each method based on the square root of the mean squared error (RMSE) and bias in estimating the ATE. Simulations for IPW, Cal, and CBPS are conducted with the \texttt{WeightIt} \texttt{R} package \citet{WeightIt} and KCB with the \texttt{ATE.ncb} package.

For the sake of clarity of presentation, we present a subset of the results in Table \ref{tab:simulation_results_n250_p10_subset} for propensity models I-V, outcome models A and B, and dimension $p=10$. We additionally present a summary of the results across all outcome models and across all propensity models. More detailed results are presented in the Supplementary Material. From Table \ref{tab:simulation_results_n250_p10_subset}, iEBW tends to perform better than EBW for most settings both in terms of RMSE and bias. Compared with KCB, the other non-parametric balancing approach, iEBW tends to result in a smaller RMSE and bias, except for under propensity model II with outcome model A and propensity model III with outcome model B. For all other settings, iEBW and EBW outperform IPW, CBPS, and Cal, except under propensity model V with outcome model B, where Cal yields the smallest RMSE and bias and propensity model III with outcome model B, where IPW yields the smallest RMSE and bias. Although EBW and iEBW do not always result in the best performance, for no setting does either approach result in a significantly worse RMSE than the best-performing method. Thus, EBW and iEBW are robust to the specific data-generating mechanism. In contrast, there are some settings where KCB, Cal, CBPS, and IPW perform significantly worse than the best performing method.
\begin{table}
\caption{\label{tab:simulation_results_n250_p10_subset}Displayed are results for $n=250$ and $p=10$ averaged over 1000 independent simulated datasets. For each setting, the best average is bolded. }
\centering
\resizebox{1\textwidth}{!}{
		\begin{tabular}{crccRcRcRcRc}
			\toprule
			& Propensity Model: & \multicolumn{2}{c}{I} & \multicolumn{2}{c}{II} & \multicolumn{2}{c}{III} & \multicolumn{2}{c}{IV} & \multicolumn{2}{c}{V}  \\ 
			\cmidrule{3-4} \cmidrule{5-6} \cmidrule{7-8} \cmidrule{9-10} \cmidrule{11-12} 
			& Method & RMSE & Bias & RMSE & Bias & RMSE & Bias & RMSE & Bias & RMSE & Bias \\ 
			\midrule
			\multicolumn{1}{r}{$Y$ Model: A} & \multicolumn{1}{r}{Unweighted}  & \multicolumn{1}{r}{$\phantom{-}21.375$} & \multicolumn{1}{r}{$-21.110$} & \multicolumn{1}{r}{$\phantom{-}23.362$} & \multicolumn{1}{r}{$\phantom{-}22.893$} & \multicolumn{1}{r}{$\phantom{0}\phantom{-}3.851$} & \multicolumn{1}{r}{$\phantom{0}\phantom{-}0.032$} & \multicolumn{1}{r}{$\phantom{0}\phantom{-}7.772$} & \multicolumn{1}{r}{$\phantom{0}\phantom{-}6.752$} & \multicolumn{1}{r}{$\phantom{-}35.122$} & \multicolumn{1}{r}{$\phantom{-}34.983$} \\
			 & \multicolumn{1}{r}{EBW}  & \multicolumn{1}{r}{$\phantom{0}\phantom{-}8.450$} & \multicolumn{1}{r}{$\phantom{0}-8.173$} & \multicolumn{1}{r}{$\phantom{0}\phantom{-}6.908$} & \multicolumn{1}{r}{$\phantom{0}\phantom{-}6.714$} & \multicolumn{1}{r}{$\phantom{0}\phantom{-}2.133$} & \multicolumn{1}{r}{$\phantom{0}-0.094$} & \multicolumn{1}{r}{$\phantom{0}\phantom{-}1.521$} & \multicolumn{1}{r}{$\phantom{0}\mathbf{-0.200}$} & \multicolumn{1}{r}{$\phantom{-}13.542$} & \multicolumn{1}{r}{$\phantom{-}13.387$} \\
			 & \multicolumn{1}{r}{iEBW}  & \multicolumn{1}{r}{$\phantom{0}\phantom{-}\mathbf{5.239}$} & \multicolumn{1}{r}{$\phantom{0}\mathbf{-5.010}$} & \multicolumn{1}{r}{$\phantom{0}\phantom{-}5.372$} & \multicolumn{1}{r}{$\phantom{0}\phantom{-}5.204$} & \multicolumn{1}{r}{$\phantom{0}\phantom{-}\mathbf{1.790}$} & \multicolumn{1}{r}{$\phantom{0}\phantom{-}\mathbf{0.005}$} & \multicolumn{1}{r}{$\phantom{0}\phantom{-}\mathbf{1.338}$} & \multicolumn{1}{r}{$\phantom{0}-0.496$} & \multicolumn{1}{r}{$\phantom{0}\phantom{-}\mathbf{9.607}$} & \multicolumn{1}{r}{$\phantom{0}\phantom{-}\mathbf{9.475}$}\\
			 & \multicolumn{1}{r}{KCB}  & \multicolumn{1}{r}{$\phantom{0}\phantom{-}7.305$} & \multicolumn{1}{r}{$\phantom{0}-6.380$} & \multicolumn{1}{r}{$\phantom{0}\phantom{-}\mathbf{4.015}$} & \multicolumn{1}{r}{$\phantom{0}\phantom{-}\mathbf{3.413}$} & \multicolumn{1}{r}{$\phantom{0}\phantom{-}3.025$} & \multicolumn{1}{r}{$\phantom{0}\phantom{-}0.789$} & \multicolumn{1}{r}{$\phantom{0}\phantom{-}1.463$} & \multicolumn{1}{r}{$\phantom{0}-0.446$} & \multicolumn{1}{r}{$\phantom{-}15.210$} & \multicolumn{1}{r}{$\phantom{-}14.643$}\\
			 & \multicolumn{1}{r}{IPW}  & \multicolumn{1}{r}{$\phantom{-}21.462$} & \multicolumn{1}{r}{$-21.188$} & \multicolumn{1}{r}{$\phantom{-}23.371$} & \multicolumn{1}{r}{$\phantom{-}22.827$} & \multicolumn{1}{r}{$\phantom{0}\phantom{-}8.083$} & \multicolumn{1}{r}{$\phantom{0}\phantom{-}0.930$} & \multicolumn{1}{r}{$\phantom{0}\phantom{-}7.864$} & \multicolumn{1}{r}{$\phantom{0}\phantom{-}6.766$} & \multicolumn{1}{r}{$\phantom{-}35.114$} & \multicolumn{1}{r}{$\phantom{-}34.971$}\\
			 & \multicolumn{1}{r}{CBPS}  & \multicolumn{1}{r}{$\phantom{-}21.442$} & \multicolumn{1}{r}{$-21.174$} & \multicolumn{1}{r}{$\phantom{-}23.194$} & \multicolumn{1}{r}{$\phantom{-}22.688$} & \multicolumn{1}{r}{$\phantom{0}\phantom{-}4.324$} & \multicolumn{1}{r}{$\phantom{0}\phantom{-}0.612$} & \multicolumn{1}{r}{$\phantom{0}\phantom{-}7.822$} & \multicolumn{1}{r}{$\phantom{0}\phantom{-}6.759$} & \multicolumn{1}{r}{$\phantom{-}35.083$} & \multicolumn{1}{r}{$\phantom{-}34.942$}\\
			 & \multicolumn{1}{r}{Cal}  & \multicolumn{1}{r}{$\phantom{-}21.459$} & \multicolumn{1}{r}{$-21.184$} & \multicolumn{1}{r}{$\phantom{-}23.129$} & \multicolumn{1}{r}{$\phantom{-}22.597$} & \multicolumn{1}{r}{$\phantom{0}\phantom{-}3.567$} & \multicolumn{1}{r}{$\phantom{0}\phantom{-}0.744$} & \multicolumn{1}{r}{$\phantom{0}\phantom{-}7.854$} & \multicolumn{1}{r}{$\phantom{0}\phantom{-}6.764$} & \multicolumn{1}{r}{$\phantom{-}35.033$} & \multicolumn{1}{r}{$\phantom{-}34.892$}\\
			 \midrule
			\multicolumn{1}{r}{$Y$ Model: B} & \multicolumn{1}{r}{Unweighted}  & \multicolumn{1}{r}{$\phantom{-}13.678$} & \multicolumn{1}{r}{$\phantom{0}-0.377$} & \multicolumn{1}{r}{$\phantom{-}30.446$} & \multicolumn{1}{r}{$\phantom{0}\phantom{-}0.861$} & \multicolumn{1}{r}{$\phantom{-}29.859$} & \multicolumn{1}{r}{$-24.281$} & \multicolumn{1}{r}{$\phantom{-}17.670$} & \multicolumn{1}{r}{$\phantom{0}\phantom{-}0.870$} & \multicolumn{1}{r}{$\phantom{-}17.529$} & \multicolumn{1}{r}{$\phantom{0}\phantom{-}0.474$}\\
			 & \multicolumn{1}{r}{EBW}  & \multicolumn{1}{r}{$\phantom{0}\phantom{-}8.892$} & \multicolumn{1}{r}{$\phantom{0}-0.263$} & \multicolumn{1}{r}{$\phantom{0}\phantom{-}9.596$} & \multicolumn{1}{r}{$\phantom{0}\phantom{-}0.520$} & \multicolumn{1}{r}{$\phantom{-}28.707$} & \multicolumn{1}{r}{$-22.923$} & \multicolumn{1}{r}{$\phantom{-}11.779$} & \multicolumn{1}{r}{$\phantom{0}\phantom{-}0.996$} & \multicolumn{1}{r}{$\phantom{0}\phantom{-}8.679$} & \multicolumn{1}{r}{$\phantom{0}\phantom{-}0.237$} \\
			 & \multicolumn{1}{r}{iEBW}  & \multicolumn{1}{r}{$\phantom{0}\phantom{-}\mathbf{4.428}$} & \multicolumn{1}{r}{$\phantom{0}-0.198$} & \multicolumn{1}{r}{$\phantom{0}\phantom{-}\mathbf{5.740}$} & \multicolumn{1}{r}{$\phantom{0}\phantom{-}\mathbf{0.419}$} & \multicolumn{1}{r}{$\phantom{-}23.214$} & \multicolumn{1}{r}{$-18.442$} & \multicolumn{1}{r}{$\phantom{0}\phantom{-}\mathbf{8.824}$} & \multicolumn{1}{r}{$\phantom{0}\phantom{-}0.790$} & \multicolumn{1}{r}{$\phantom{0}\phantom{-}4.077$} & \multicolumn{1}{r}{$\phantom{0}\phantom{-}0.109$} \\
			 & \multicolumn{1}{r}{KCB}  & \multicolumn{1}{r}{$\phantom{0}\phantom{-}8.965$} & \multicolumn{1}{r}{$\phantom{0}-0.267$} & \multicolumn{1}{r}{$\phantom{0}\phantom{-}9.334$} & \multicolumn{1}{r}{$\phantom{0}\phantom{-}0.446$} & \multicolumn{1}{r}{$\phantom{-}17.595$} & \multicolumn{1}{r}{$-14.202$} & \multicolumn{1}{r}{$\phantom{0}\phantom{-}9.744$} & \multicolumn{1}{r}{$\phantom{0}\phantom{-}\mathbf{0.567}$} & \multicolumn{1}{r}{$\phantom{0}\phantom{-}8.945$} & \multicolumn{1}{r}{$\phantom{0}\phantom{-}0.278$} \\
			 & \multicolumn{1}{r}{IPW}  & \multicolumn{1}{r}{$\phantom{0}\phantom{-}8.979$} & \multicolumn{1}{r}{$\phantom{0}-0.238$} & \multicolumn{1}{r}{$\phantom{-}12.352$} & \multicolumn{1}{r}{$\phantom{0}\phantom{-}0.817$} & \multicolumn{1}{r}{$\phantom{-}\mathbf{15.395}$} & \multicolumn{1}{r}{$\mathbf{-11.546}$} & \multicolumn{1}{r}{$\phantom{-}17.420$} & \multicolumn{1}{r}{$\phantom{0}\phantom{-}0.972$} & \multicolumn{1}{r}{$\phantom{0}\phantom{-}8.962$} & \multicolumn{1}{r}{$\phantom{0}\phantom{-}0.265$}  \\
			 & \multicolumn{1}{r}{CBPS}  & \multicolumn{1}{r}{$\phantom{0}\phantom{-}8.929$} & \multicolumn{1}{r}{$\phantom{0}-0.241$} & \multicolumn{1}{r}{$\phantom{-}10.603$} & \multicolumn{1}{r}{$\phantom{0}\phantom{-}0.934$} & \multicolumn{1}{r}{$\phantom{-}23.757$} & \multicolumn{1}{r}{$-13.231$} & \multicolumn{1}{r}{$\phantom{-}13.973$} & \multicolumn{1}{r}{$\phantom{0}\phantom{-}0.830$} & \multicolumn{1}{r}{$\phantom{0}\phantom{-}9.074$} & \multicolumn{1}{r}{$\phantom{0}\phantom{-}0.281$} \\
			 & \multicolumn{1}{r}{Cal}  & \multicolumn{1}{r}{$\phantom{0}\phantom{-}4.643$} & \multicolumn{1}{r}{$\phantom{0}\mathbf{-0.191}$} & \multicolumn{1}{r}{$\phantom{0}\phantom{-}6.762$} & \multicolumn{1}{r}{$\phantom{0}\phantom{-}1.027$} & \multicolumn{1}{r}{$\phantom{-}44.189$} & \multicolumn{1}{r}{$-31.440$} & \multicolumn{1}{r}{$\phantom{-}18.345$} & \multicolumn{1}{r}{$\phantom{0}\phantom{-}1.138$} & \multicolumn{1}{r}{$\phantom{0}\phantom{-}\mathbf{1.702}$} & \multicolumn{1}{r}{$\phantom{0}\phantom{-}\mathbf{0.100}$} \\
			\bottomrule
		\end{tabular}%
}
\end{table}
Table \ref{tab:simulation_results_ranks_propmod_n250} contains a summary of the results averaged across outcome models (A-E) and dimension settings ($p\in\{10,25\}$). Each entry in the table is the average rank of each method in terms of RMSE and bias for each combination of outcome model and dimension; i.e. the method with the smallest RMSE for a particular setting receives a ``1'' and the method with the largest RMSE receives a ``7''. The Supplementary Material contains a similar summary, but averaged over propensity models and dimensions. In general iEBW tends to yield the best rank in terms of performance across the settings, with EBW and KCB yielding similarly low ranks. 
\begin{table}
\caption{\label{tab:simulation_results_ranks_propmod_n250}Displayed are the ranks among all methods tested of each method in terms of RMSE and bias averaged over all response models (I-VI) for $n=250$ and over the dimension settings $p=10$ and $p=25$.}
\centering
\resizebox{1\textwidth}{!}{
\begin{tabular}{cRcRcRcRcRcRc}
\toprule
 Propensity Model: & \multicolumn{2}{c}{I} & \multicolumn{2}{c}{II} & \multicolumn{2}{c}{III} & \multicolumn{2}{c}{IV} & \multicolumn{2}{c}{V} & \multicolumn{2}{c}{VI}\\ 
 \cmidrule{2-3} \cmidrule{4-5} \cmidrule{6-7} \cmidrule{8-9} \cmidrule{10-11} \cmidrule{12-13}
 & \multicolumn{2}{c}{Mean Rank} & \multicolumn{2}{c}{Mean Rank} & \multicolumn{2}{c}{Mean Rank} & \multicolumn{2}{c}{Mean Rank}
 & \multicolumn{2}{c}{Mean Rank} & \multicolumn{2}{c}{Mean Rank} \\ 
Method  & RMSE & Bias & RMSE & Bias & RMSE & Bias & RMSE & Bias & RMSE & Bias & RMSE & Bias\\ 
\midrule
\multicolumn{1}{r}{Unweighted}  & \multicolumn{1}{r}{$4.7$} & \multicolumn{1}{r}{$3.7$} & \multicolumn{1}{r}{$6.5$} & \multicolumn{1}{r}{$4.9$} & \multicolumn{1}{r}{$5.0$} & \multicolumn{1}{r}{$4.5$} & \multicolumn{1}{r}{$4.8$} & \multicolumn{1}{r}{$4.1$} & \multicolumn{1}{r}{$6.2$} & \multicolumn{1}{r}{$5.3$} & \multicolumn{1}{r}{$3.9$} & \multicolumn{1}{r}{$3.1$} \\
\multicolumn{1}{r}{EBW}  & \multicolumn{1}{r}{$2.5$} & \multicolumn{1}{r}{$2.8$} & \multicolumn{1}{r}{$2.4$} & \multicolumn{1}{r}{$2.7$} & \multicolumn{1}{r}{$3.2$} & \multicolumn{1}{r}{$4.2$} & \multicolumn{1}{r}{$2.5$} & \multicolumn{1}{r}{$3.6$} & \multicolumn{1}{r}{$\mathbf{1.9}$} & \multicolumn{1}{r}{$2.6$} & \multicolumn{1}{r}{$2.5$} & \multicolumn{1}{r}{$3.8$} \\
\multicolumn{1}{r}{iEBW}  & \multicolumn{1}{r}{$\mathbf{2.2}$} & \multicolumn{1}{r}{$\mathbf{2.5}$} & \multicolumn{1}{r}{$\mathbf{1.5}$} & \multicolumn{1}{r}{$\mathbf{1.7}$} & \multicolumn{1}{r}{$\mathbf{3.0}$} & \multicolumn{1}{r}{$3.9$} & \multicolumn{1}{r}{$\mathbf{1.7}$} & \multicolumn{1}{r}{$3.5$} & \multicolumn{1}{r}{$\mathbf{1.9}$} & \multicolumn{1}{r}{$\mathbf{2.0}$} & \multicolumn{1}{r}{$\mathbf{2.0}$} & \multicolumn{1}{r}{$3.1$} \\
\multicolumn{1}{r}{KCB}  & \multicolumn{1}{r}{$3.0$} & \multicolumn{1}{r}{$3.4$} & \multicolumn{1}{r}{$2.5$} & \multicolumn{1}{r}{$2.5$} & \multicolumn{1}{r}{$3.2$} & \multicolumn{1}{r}{$4.3$} & \multicolumn{1}{r}{$2.7$} & \multicolumn{1}{r}{$\mathbf{2.5}$} & \multicolumn{1}{r}{$3.0$} & \multicolumn{1}{r}{$3.6$} & \multicolumn{1}{r}{$2.5$} & \multicolumn{1}{r}{$\mathbf{2.3}$} \\
\multicolumn{1}{r}{IPW}  & \multicolumn{1}{r}{$5.5$} & \multicolumn{1}{r}{$5.0$} & \multicolumn{1}{r}{$6.4$} & \multicolumn{1}{r}{$5.9$} & \multicolumn{1}{r}{$5.0$} & \multicolumn{1}{r}{$\mathbf{2.9}$} & \multicolumn{1}{r}{$6.3$} & \multicolumn{1}{r}{$5.4$} & \multicolumn{1}{r}{$5.8$} & \multicolumn{1}{r}{$5.1$} & \multicolumn{1}{r}{$6.8$} & \multicolumn{1}{r}{$6.0$} \\
\multicolumn{1}{r}{CBPS}  & \multicolumn{1}{r}{$5.1$} & \multicolumn{1}{r}{$4.8$} & \multicolumn{1}{r}{$5.1$} & \multicolumn{1}{r}{$5.3$} & \multicolumn{1}{r}{$4.7$} & \multicolumn{1}{r}{$3.9$} & \multicolumn{1}{r}{$5.1$} & \multicolumn{1}{r}{$3.7$} & \multicolumn{1}{r}{$5.6$} & \multicolumn{1}{r}{$5.3$} & \multicolumn{1}{r}{$4.9$} & \multicolumn{1}{r}{$4.2$} \\
\multicolumn{1}{r}{Cal}  & \multicolumn{1}{r}{$5.0$} & \multicolumn{1}{r}{$5.8$} & \multicolumn{1}{r}{$3.6$} & \multicolumn{1}{r}{$5.0$} & \multicolumn{1}{r}{$3.9$} & \multicolumn{1}{r}{$4.3$} & \multicolumn{1}{r}{$4.9$} & \multicolumn{1}{r}{$5.2$} & \multicolumn{1}{r}{$3.6$} & \multicolumn{1}{r}{$4.1$} & \multicolumn{1}{r}{$5.4$} & \multicolumn{1}{r}{$5.5$} \\
   \bottomrule
\end{tabular}%
}
\end{table}



\section{Right Heart Catheterization Data}
\label{sec:analysis}

\subsection{Description of Data}

A study by \citet{connors1996effectiveness} was conducted to investigate the effectiveness of right heart catheterization (RHC), a diagnostic procedure for critically ill patients in intensive care units. Since RHC is more relevant for certain forms of intensive care than others, there is substantial imbalance in patient characteristics in those treated with RHC and those who did not receive RHC. The original analysis was based on propensity score matching and the data has been subsequently re-analyzed in many other works \citep{hirano2001estimation, crump2009dealing, rosenbaum2012optimal, li2018balancing}. The study consists of data on 5735 individuals, 2184 of whom received RHC, and the remaining 3551 did not receive RHC. The outcome is an indicator of survival at 30 days after admission. A panel of experts convened to discuss which variables contribute to a decision to use RHC, resulting in a large set of covariates to study (72 in total, 21 of which are continuous, 25 binary, and 26 dummy variables originating from 6 categorical covariates). The dataset is publicly available at the following site: \url{http://biostat.mc.vanderbilt.edu/wiki/pub/Main/DataSets/rhc.html}. There are substantial empirical differences in the distributions of many of these covariates between treatment groups (RHC vs. no RHC). 

In Section \ref{sec:rhc_analysis} we study the effect of RHC on 30-day survival. However, since there is no ground truth available, in Section \ref{sec:rhc_simulation} we use the RHC data to conduct a realistic simulation that demonstrates the effectiveness of EBWs.

\subsection{Analysis of Right Heart Catheterization Data}
\label{sec:rhc_analysis}

\begin{figure}[!ht]
	\centering
	\includegraphics[width=0.495\textwidth]{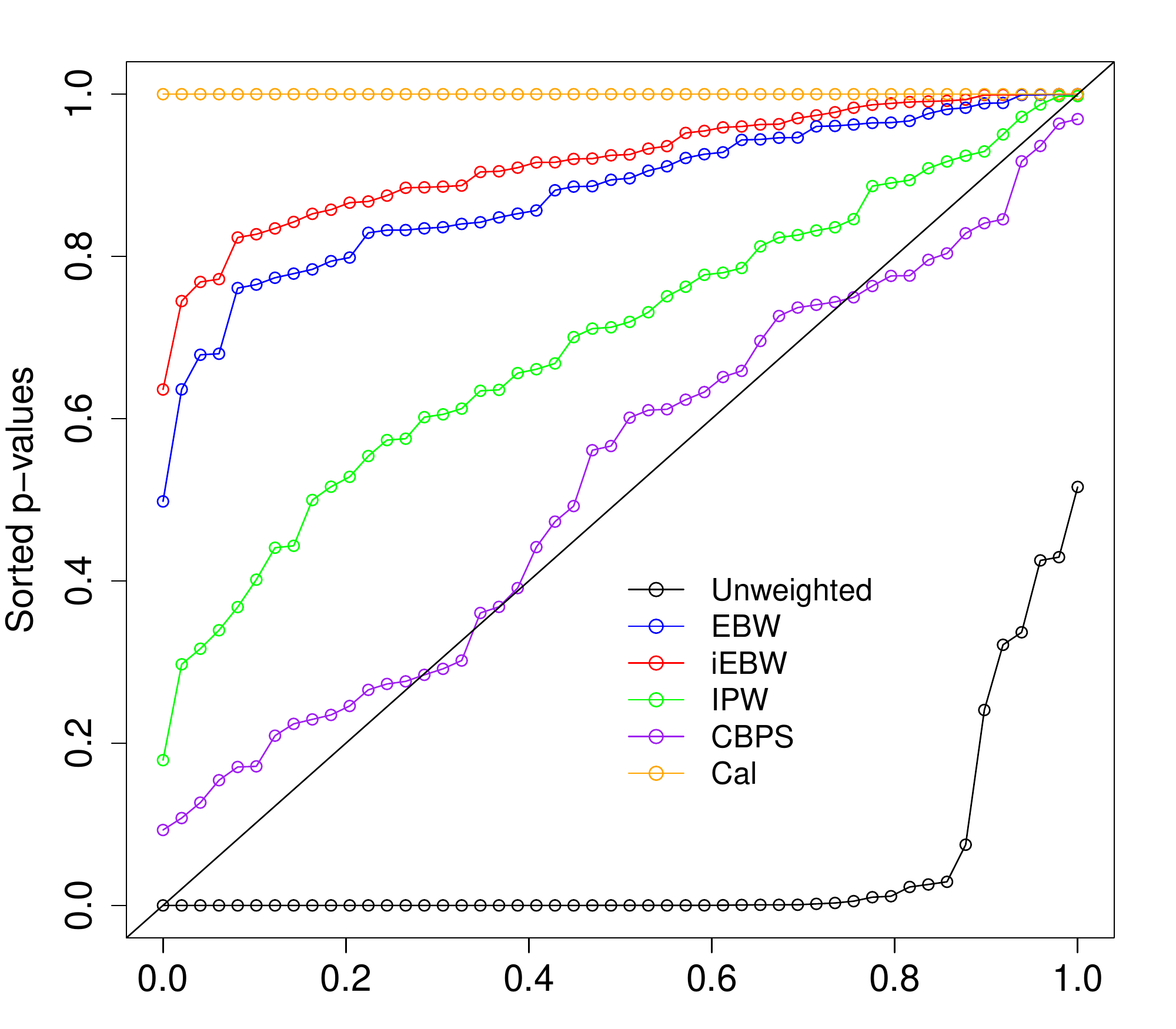}
	\includegraphics[width=0.495\textwidth]{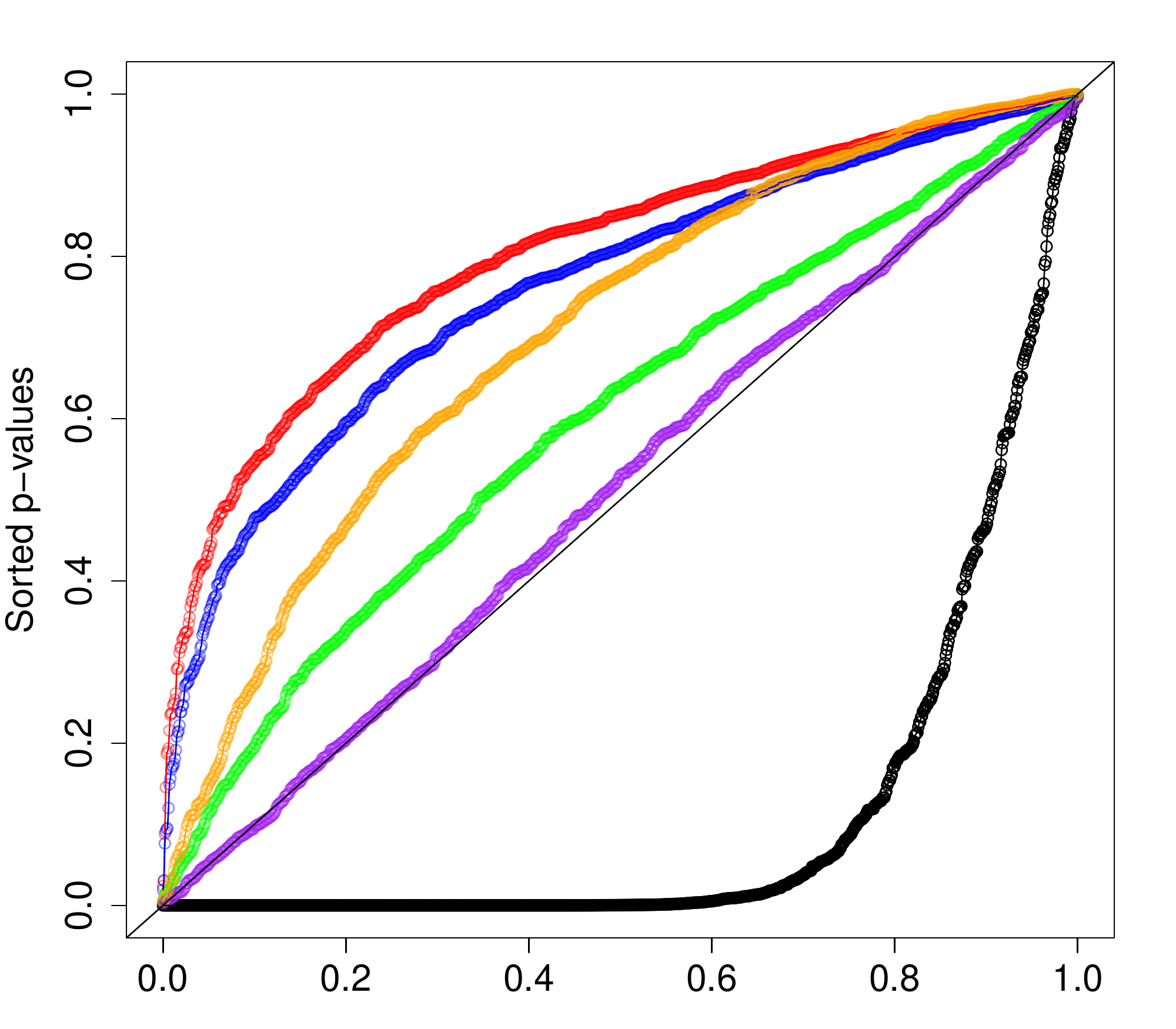}
	\caption{Displayed are $p$-values for tests of the difference in weighted means between treatment groups marginally for each covariate (left) and for each pairwise interaction of covariates (right). For continuous covariates weighted $t$-tests are used and weighted Chi-squared tests are used for discrete covariates. For an RCT, the sorted $p$-values are expected to roughly fall along the diagonal -- lines above the diagonal indicate an improvement in moment balance over random assignment. Note that by construction, the calibration balancing weights result in exact balance of the first order moments of each covariate marginally. However, exact marginal mean balance does not imply balance of higher order terms, as seen on the right plot, or of marginal distributional balance, as can be seen in Figure \ref{fig:root_imse_cdfs_rhc}.}
	\label{fig:smd_rhc_pvalues}
\end{figure}

\begin{figure}[!ht]
	\centering
	\includegraphics[width=0.495\textwidth]{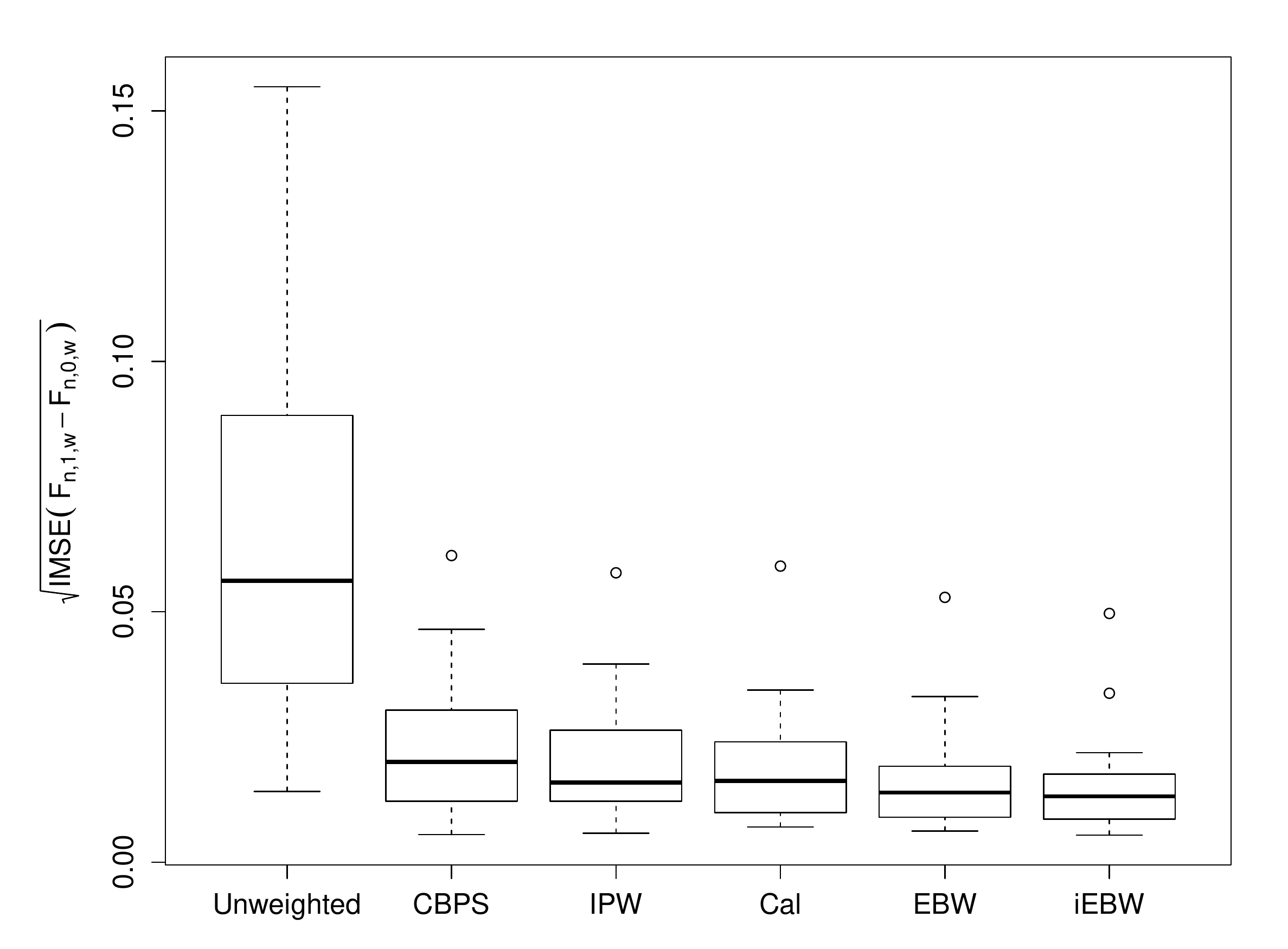}
	\includegraphics[width=0.495\textwidth]{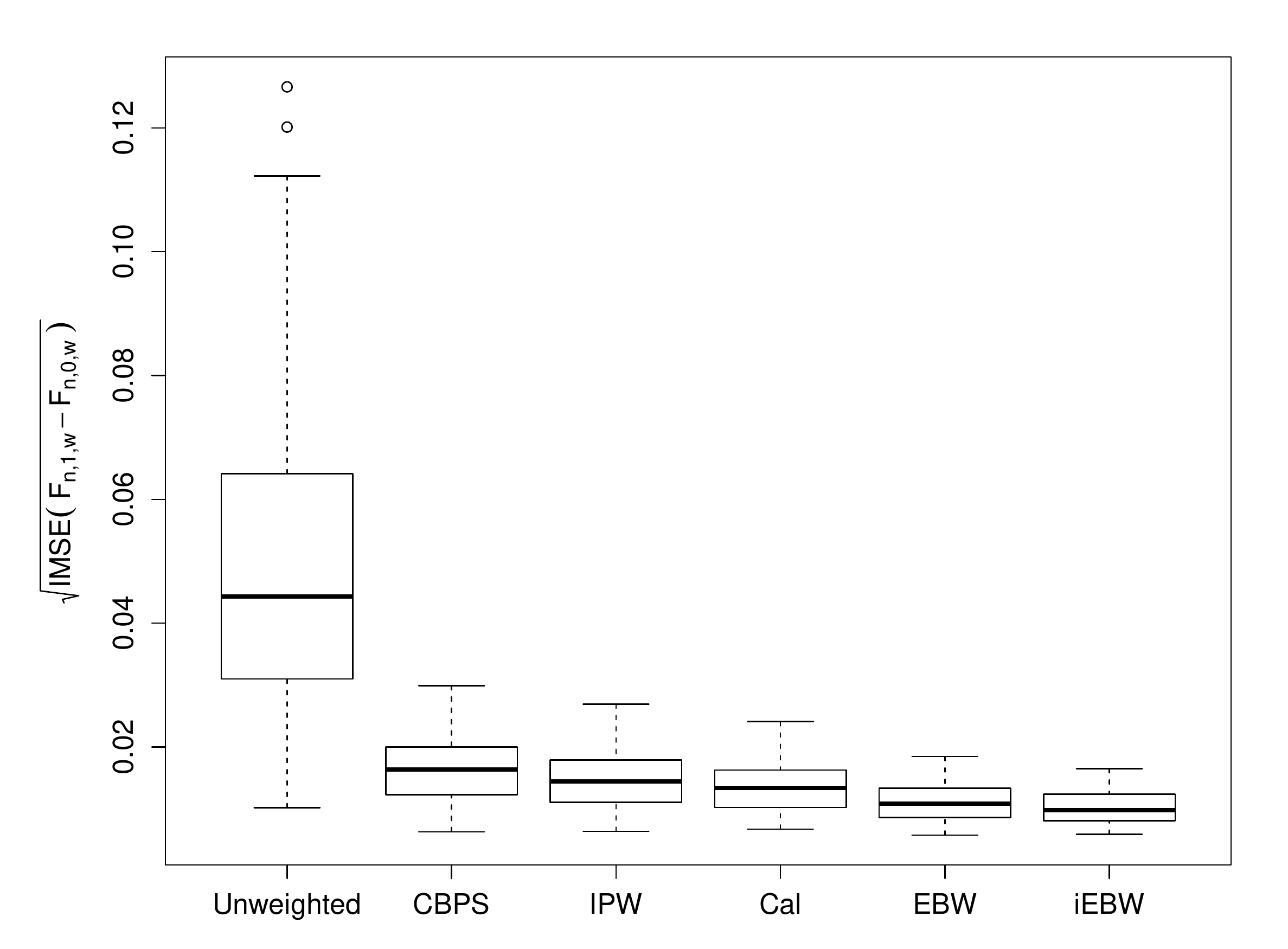}
	\caption{This figure illustrates the differences in the \textit{marginal} univariate weighted ECDFs between the treated and control populations. In particular, let $x_j \in \mathcal{X}_j$ denote the $j$ component of the covariate vector $\bx$ and let $F_{n,a,j}(x)$ denote its empirical cdf on treatment arm $a$. Similarly denote the weighted versions of this quantity. Here, we are displaying how well each method balances the marginal empirical cdfs for the treated and control arms. We do so by evaluating an estimate of $\sqrt{\int_{x_j \in \mathcal{X}_j}\left[F_{n,1,j,\boldsymbol w} - F_{n,0,j,\boldsymbol w}\right](x_j)\mathrm{d}x_j}$ obtained by integration over a grid of values for all $j=1,\dots,65$. The results across all covariates are displayed in the left two plots above. The rightmost plots similarly display the RIMSEs for all possible 65 choose 2 \textit{bivariate} cdfs. }
	\label{fig:root_imse_cdfs_rhc}
\end{figure}

We used 65 of the available covariates, as in the analysis of the same dataset in \citet{rosenbaum2012optimal}, leaving out date-related covariates. Using the 65 covariates we applied the weighting methods used in Section \ref{sec:simulation} (except the method of \citet{wong2017kernel} as the code returned constant weights of 1 regardless of the tuning parameters used) to estimate weights to balance the treated groups. To first investigate how well each method balances the marginal means of each covariate, we evaluate the absolute standardized mean differences for each covariate and $p$-values for the difference in weighted means between treated and control groups, which are displayed in Figure \ref{fig:smd_rhc_pvalues}. Empirical calibration balancing is designed to balance all specified moments exactly and since we balance the first moments, these have exact balance. Both energy balancing weights still result in tight moment balance in spite of the fact that this was not prescribed as a goal explicitly. 
In Figure \ref{fig:root_imse_cdfs_rhc} we investigate how well each method balances the distributions of all 1- and 2-dimensional projections of the covariates. To do this, we compare the weighted ECDFs for each projection by evaluating the square root of the integrated mean squared error (RIMSE) of the weighted ECDFs between the two treatment groups. Both energy balancing weights yield the smallest RIMSEs on average. However, it is important to note that lower-order projections of the distribution are not the explicit focus of energy balancing weights and thus the energy balancing weights also likely balance other aspects of the distributions of covariates between the treatment groups.
We display further measures of discrepancy of the covariates between the treatment groups in Table \ref{tab:rhc_estimates_se_and_balance}. In addition to displaying the average RIMSE values and absolute SMDs, in this table we display the weighted energy values for each of the weights, including the unweighted energy distances. By definition, energy balancing weights have the minimum weighted energy distances, however it can be seen that the empirical calibration balancing weights have a relatively small weighted energy distance, indicating its use is likely sensible for the RHC data. 

The estimates of the ATE of RHC and corresponding standard errors are displayed in Table \ref{tab:rhc_estimates_se_and_balance}. Standard errors are computed using the nonparametric bootstrap with 1000 replications. The EBW-based estimates have the smallest standard errors, and, perhaps more interestingly, yield an estimate of the effect of RHC that is slightly less deleterious than estimates in the literature, despite it still being a highly significant effect.

\begin{table}
\caption{\label{tab:rhc_estimates_se_and_balance} Estimates of the ATE and standard errors for the RHC data. Standard errors were computed for all methods using the nonparametric bootstrap with 1000 replications. Also displayed are various measures of discrepancy between the distributions of covariates for the RHC and non-RHC groups. In addition to weighted energy distances, we display mean and max ``$|$SMD$|$'', which are the average and maximum, respectively, absolute standardized marginal mean difference of covariates.
}
\centering
\begin{tabular}{crrrrrr}
  \toprule
 & Unwtd & CBPS & IPW & Cal & EBW & iEBW \\ 
  \midrule
$\widehat{\tau}_\bw$ & 0.0736 & 0.0577 & 0.0528 & 0.0547 & 0.0499 & 0.0470\\ 
  SE$(\widehat{\tau}_\bw)$ & \hphantom{-}0.0132 & \hphantom{-}0.0142 & \hphantom{-}0.0155 & \hphantom{-}0.0145 & \hphantom{-}0.0120 & \hphantom{-}0.0117 \\
  \midrule
  Energy dist \eqref{eqn:energy_balancing_weights_ate} & 8.1996 & \hphantom{9}0.8596 & \hphantom{9}0.7804 & \hphantom{9}0.4250 & \hphantom{9}{0.3236} & \hphantom{9}0.3377 \\
  Energy dist \eqref{eqn:improved_energy_balancing_weights_ate} & 23.7172 & \hphantom{9}1.8643 & \hphantom{9}1.5756 & \hphantom{9}1.0045 & \hphantom{9}0.7536 & \hphantom{9}{0.7270} \\
  Mean $|$SMD$|$ &  0.1648 & 0.0276  & 0.0182 & {0.0000}  & 0.0063  & 0.0043  \\
  Max $|$SMD$|$ & 0.5826 & 0.1158 & 0.0778 & {0.0000} & 0.0239 & 0.0168 \\
  SD $|\text{SMD}|$ &  0.1276 & 0.0207 & 0.0147 & 0.0000 & 0.0050 & 0.0036 \\
   \bottomrule
\end{tabular}
\end{table}

\subsection{Right Heart Catheterization Data Simulation}
\label{sec:rhc_simulation}

\begin{figure}[!ht]
	\centering
	\includegraphics[width=0.495\textwidth]{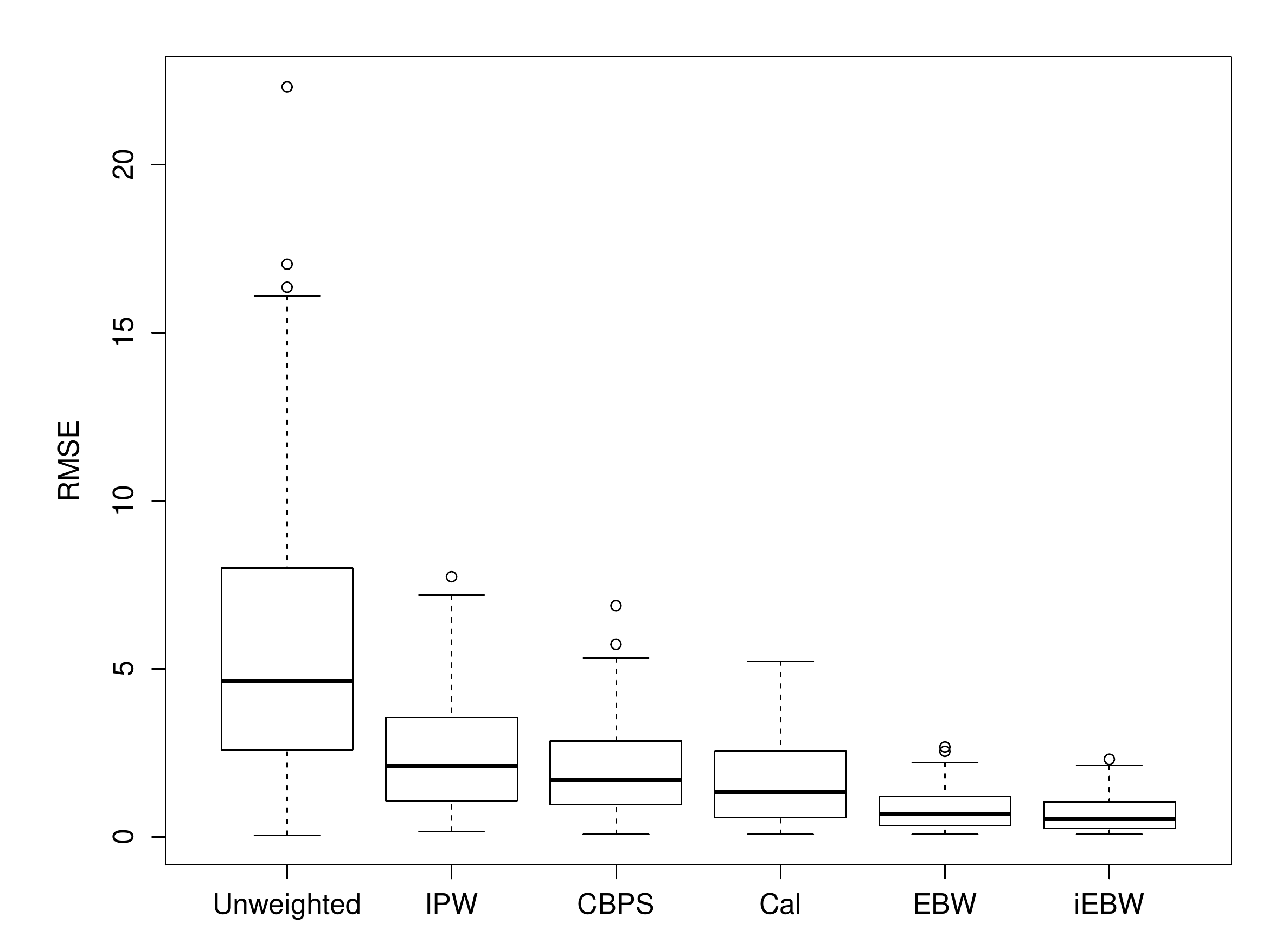}
	\caption{This figure displays the RMSEs for each method across the 100 simulation settings using the RHC data. }
	\label{fig:rmse_ate_rhc_sim}
\end{figure}

In this section we fix the covariates and treatment assignments from the RHC data and simulate responses with confounding. This produces a realistic and highly challenging simulation scenario with high dimension. We use the key functional form from outcome model D from Section \ref{sec:simulation}. Since this outcome model involves only 7 covariates, we use the functional form from this outcome model and apply it over 8 separate groups of 7 covariates in the RHC data and take the sum of all groups of 7 covariates as the mean of the outcome. We then simulate 1000 independent datasets using this procedure and each time estimate the ATE and record each method's RMSE in estimating the true ATE. Since under this model the ordering of the covariates changes the nature of the confounding, we randomly permute the columns 100 times and repeat the simulation 1000 times for each permutation and each time record the RMSE over the 1000 replications. Further details are provided in the Supplementary Material. The RMSEs for each of the 100 column permutations are displayed in Figure \ref{fig:rmse_ate_rhc_sim}. Both EBW and iEBW consistently yield among the smallest RMSE across the permutation settings. Since the outcome model in this simulation involves a constant treatment effect and thus is not impacted by potential issues of covariate overlap, in the Supplementary Material we additionally consider an outcome model with a treatment effect that depends on $\bX$; the pattern of the results is consistent with the findings using a constant treatment effect, albeit the advantage of EBWs is slightly more pronounced with the heterogeneous treatment effect scenario. 

\section{ Analysis of MIMIC-III critical care data }
\label{sec:analysis_mimic}

To further demonstrate the proposed EBWs in difficult observational studies of treatments, we analyze the effectiveness of three treatments based on separate subpopulations of the MIMIC-III v1.4 critical care database \citep{mimiciii}: a study of the effect of indwelling arterial catheters (IACs) on mortality in patients with respiratory failure \citep{hsu2015association}, a study of the effect of transthoracic echocardiography (echo) on mortality in sepsis patients \citep{feng2018transthoracic}, and a study of mechanical power of ventilation (MPV) on mortality in critically ill patients \citep{neto2018mechanical}. Each study is based on existing studies utilizing  MIMIC-III. The degree of confounding in each study varies, with the IAC and MPV studies exhibiting a great degree of confounding and the echo study with minimal confounding. For all studies, missing values were imputed using missForest \citep{stekhoven2012missforest}. We present the IAC study in this section and the remaining two studies in the Supplementary Material.

\subsection{Indwelling Arterial Catheter Data}

In this section we replicate a study originally conducted in \cite{hsu2015association} based on the MIMIC-III critical care database to study the effect of indwelling arterial catherization on 28 day mortality. The data is based on the queries provided in \url{https://github.com/MIT-LCP/mimic-code} and contains information on 2522 mechanically-ventilated patients, 1298 of whom received the treatment,  IAC. The outcome is an indicator of 28 day mortality from time of admission. Pre-treatment covariates likely to be confounders include demographics, lab values, calculated risk scores, missingness indicators, and more, totaling to a design matrix with $p=81$. We leveraged the same approach used in the RHC study to control for the confounders via various weighting approaches to estimate the ATE of IAC on 28-day mortality. The KCB approach yielded constant weights of 1 regardless of the tuning parameter. Table \ref{tab:aline_estimates_se_and_balance} contains information on the estimated effect of IAC on mortality based on each weighting method in addition to various balance statistics for the 81 covariates, including marginal mean balance, univariate and bivariate distributional balance, and weighted energy statistics. EBW and iEBW yield the best univariate and bivariate distributional balance, while Cal and CBPS result in the best marginal mean balance. Note that Cal does not achieve exact marginal mean balance due to numerical issues. 
All approaches except inverse weighting by propensity score (IPW) yield 95\% confidence intervals that contain zero, albeit with point estimates that suggest a potential benefit of IAC. IPW results in a significant estimated benefit of IAC on 28 day mortality, with a point estimate of the benefit that is substantially larger than for other methods.

\begin{table}
\caption{\label{tab:aline_estimates_se_and_balance} Estimates of the ATE and standard errors for the IAC data. Standard errors were computed for all methods using the nonparametric bootstrap with 1000 replications. Also displayed are various measures of discrepancy between the distributions of covariates for the IAC and control groups. We also display the mean and max RIMSE statistic for marginal univariate and bivariate CDF differences, as in Figure \ref{fig:root_imse_cdfs_rhc}. In addition, we display summary statistics of SMDs for marginal means and SMDs for all polynomials up to order 5 and pairwise interactions (denoted SMD(2)). 
}
\centering
\begin{tabular}{crrrrrr}
  \toprule
 & Unwtd & CBPS & IPW & Cal & EBW & iEBW \\ 
  \midrule
$\widehat{\tau}_\bw$ & \hphantom{-}0.0012 & -0.0343 & -0.0744 & -0.0150 & -0.0150 & -0.0130 \\ 
  SE$(\widehat{\tau}_\bw)$ & 0.015 & 0.0217 & 0.0446 & 0.0165 & 0.0118 & 0.0114 \\
  \midrule
  Energy dist \eqref{eqn:energy_balancing_weights_ate} & \hphantom{9}4.5625 & \hphantom{9}0.7829 & 31.336 & \hphantom{9}0.5612 & \hphantom{9}0.3869 & \hphantom{9}0.4028 \\
  Energy dist \eqref{eqn:improved_energy_balancing_weights_ate} & 13.6796 & \hphantom{9}1.6922 & 62.6704 & \hphantom{9}1.3854 & \hphantom{9}0.9614 & \hphantom{9}0.9304  \\
  Mean RIMSE, 1d & 0.0362 & 0.0130 & 0.0670 & 0.0126 & 0.0113 & \textbf{0.0111} \\
  Max RIMSE, 1d & 0.0863 & 0.0277 & 0.1646 & 0.0256 & \textbf{0.0235} & 0.0245 \\
  Mean RIMSE, 2d & 0.0416 & 0.0075 & 0.0619 & 0.0071 & 0.0067 & \textbf{0.0062} \\
  Max RIMSE, 2d & 0.3295 & 0.0283 & 0.2028 & 0.0284 & 0.0248 & \textbf{0.0189} \\
  Mean $|$SMD$|$ & 0.0732 & 0.0023 & 0.0990 & \textbf{0.0001} & 0.0060 & 0.0045 \\
  Max $|\text{SMD}|$& 0.2996 & 0.0203 & 1.2906 & \textbf{0.0028} & 0.0289 & 0.0212 \\
  Mean $|$SMD(2)$|$ & 0.0788 & 0.0081 & 0.0932 & 0.0075 & 0.0093 & \textbf{0.0073} \\
  Max $|\text{SMD}(2)|$& 0.6801 & 0.1782 & 1.2906 & 0.1412 & 0.1502 & \textbf{0.0998} \\
   \bottomrule
\end{tabular}
\end{table}

In addition, we conduct a simulation study based on the IAC data with precisely the same response-generating mechanism as for the simulation on the RHC data in Section \ref{sec:rhc_simulation}. Although the dimension of the IAC data is higher, the number of dimensions that impact the response for each data-generating setting is 63, the same as in Section \ref{sec:rhc_simulation}. The results in terms of RMSE in estimating the ATE across the 100 data-generating scenarios, each averaged over 1000 replications, are displayed in Table \ref{tab:aline_simulations}. The iEBW approach results in the smallest mean, median, and worst-case RMSE, followed by EBW, which is closely followed by Cal and CBPS. IPW in this case results in nearly worse performance than no weighting. We also conducted the same simulation study but with a treatment effect that varies in $\bX$; the results show a similar pattern and are thus displayed in the Supplementary Material.

\begin{table}
\caption{\label{tab:aline_simulations} Displayed are the (average, standard deviation, and maximum) RMSEs for each method across the 100 simulation settings using the IAC data.
}
\centering
\begin{tabular}{rrrrrrr}
  \toprule
 & Unwtd & CBPS & IPW & Cal & EBW & iEBW \\ 
  \midrule
  Median RMSE & 8.0151 & 3.2899 & 7.9435 & 3.4895 & 3.0276 & 1.8319 \\
  Mean RMSE & 9.1296 & 3.8542 & 9.5545 & 3.7609 & 3.5113 & 2.2087 \\ 
  SD RMSE & 6.7091 & 2.5588 & 7.5263 & 2.5293 & 2.5028 & 1.5785 \\
  Max RMSE & 32.3715 & 11.0479 & 39.1095 & 12.0521 & 12.4231 & 7.2066 \\
   \bottomrule
\end{tabular}
\end{table}



\section{Discussion}
\label{sec:discussion}

We have introduced a new metric, the weighted energy distance, which measures the distributional balance induced by a set of weights, and thus can be used to determine which set of weights is likely to result in low bias when estimating a causal quantity. Building on the weighted energy distance, we have introduced the energy balancing weights which minimize this distance to achieve distributional balance. The energy balancing weights are robust and reliable across many functional forms of confounding and further rarely result in large weights. Due to the distributional balancing of the energy balancing weights, they can be utilized to estimate a wide variety of causal estimands which can be represented as a statistical functional of the population distribution function of the covariates. 
While we focused entirely on the weighted energy distance, the connection between the energy distance and  distances between embeddings of probability measures into reproducing kernel Hilbert spaces \citep{sejdinovic2013} opens up the possibility of  more effective distributional balancing weights if more is known about the functional form of confounding. In particular, if the analyst believes lower order projections of the distribution should be balanced with priority over higher order aspects of the distribution, the use of a kernel which emphasizes these projections, such as the sparsity-inducing kernel in \citet{mak2017projected}, could be used.

\bigskip
\begin{center}
{\large\bf SUPPLEMENTARY MATERIAL}
\end{center}

\begin{description}

\item[Supplementary Material:] The Supplementary Material contains proofs of theoretical results, as well as additional numerical results for the MIMIC-III critical care data. (pdf)



\end{description}

\def\spacingset#1{\renewcommand{\baselinestretch}%
{#1}\small\normalsize} \spacingset{0.5}

\pagebreak

\appendix
\spacingset{1.5}


\section{Technical Proofs}
\label{sec:proofs}

\subsection{Proposition \ref{thm:weighted_energy_duality}}

\begin{proof}
	For simplicity, we focus on the case where $p=1$, but the arguments carry through for all dimensions. We further focus on the treated group (i.e. $a=1$) without loss of generality.
	We begin by noting that we can express $\vert \varphi_{n}(t) - \varphi_{n,1,\bw}( t) \vert^2$ in terms of $\varphi_{n}(t)\overline{\varphi_{n}(t)}$, $\varphi_{n}(t)\overline{\varphi_{n,1,\bw}(t)}$, $\varphi_{n, 1,\bw}(t)\overline{\varphi_{n}(t)}$, and $\varphi_{n, 1,\bw}(t)\overline{\varphi_{n, 1}(t)}$, where $\overline{\varphi_{n}(t)}$ and $\overline{\varphi_{n,1,\bw}(t)}$ are the complex conjugates of ${\varphi_{n}(t)}$ and ${\varphi_{n,1,\bw}(t)}$, respectively. For the first, we have 
	\begin{align*}
	\varphi_{n}(t)\overline{\varphi_{n}(t)} = {} & \frac{1}{n^2}\sum_{i,j}\exp\{it(X_i - X_j)\} \\
	= {} & \frac{1}{n^2}\sum_{i,j}\cos\{t(X_i - X_j)\} + V_1, 
	\end{align*}
	where $V_1$ is a term that vanishes when the integral in \eqref{eqn:empirical_energy_duality} of the main text is evaluated. Similarly, we have 
	\begin{align*}
	& \varphi_{n,1,\bw}(t)\overline{\varphi_{n,1,\bw}(t)}  = \frac{1}{n_1^2}\sum_{i,j}w_iw_jA_iA_j \cos\{t(X_i - X_j)\} + V_2  \text{ and }\\
	& \varphi_{n,1,\bw}(t)\overline{\varphi_{n}(t)}  + \varphi_{n}(t)\overline{\varphi_{n,1,\bw}(t)}  = \frac{1}{n_1n}\sum_{i,j}w_iA_i \cos\{t(X_i - X_j)\} \\
	& \;\;\;\;\;\;\;\;\; + \frac{1}{n_1n}\sum_{i,j}w_jA_j \cos\{t(X_i - X_j)\} + V_3. 
	\end{align*}
	Then combining terms, adding and subtracting 1 twice, by the constraints that the weights sum to $n_a$ for $a \in \{0,1\}$, and by Lemma 1 of \cite{szekely2013energy}, we have the desired result.
\end{proof}

\subsection{Theorem \ref{thm:weighted_energy_converges_to_energy}}
\begin{proof}

Let $\{\bXtilde_i\}_{i=1}^{n}\distas{i.i.d.} F_{n,a,\bw_n}$ and let $\widetilde{F}_{n,a,\bw_n}$ and $\widetilde{\varphi}_{n,a,\bw_n}$ be the empirical cdf and characteristic function of $\{\bXtilde_i\}_{i=1}^{n_a}$. By the Glivenko-Cantelli theorem for non-identically distributed random variables (Theorem 1 of \citet{wellner1981glivenko}, 
we have that $\lim_{n \to\infty}\sup_{\bx \in \calX}\lvert \widetilde{F}_{n,a,\bw_n}(\bx)  - \widetilde{F}_{a}(\bx)\rvert = 0$. Similar to the proof of Theorem 2 in \citet{szekely2007measuring} (with modification, since now we need a SLLN for V-statistics of triangular arrays like \citet{csorgHo2013asymptotics,patterson1989strong}), we will show that 
\begin{equation}\label{eqn:empirical_energy_convergence}
    \lim_{n \to \infty} \calE(\widetilde{F}_{n,a,\bw_n}, F_{n}) = \calE(\widetilde{F}_{a}, F)
\end{equation}
almost surely. Similar to \citet{szekely2007measuring} define $D(\delta) = \{ \bt \in \bbR^{p}: \delta \leq \vert\bt\vert_p \leq 1/\delta \}$ and $\calE_\delta(\widetilde{F}_{n,a,\bw_n} F_{n}) = \int_{D(\delta)}\vert \varphi_{n}(\bt) - \widetilde{\varphi}_{n,a,\bw_n}(\bt) \vert^2 \omega(\mathbf{t}) \mathrm{d}\mathbf{t}$. By the strong law of large numbers for V-statistics of triangular arrays \citep{csorgHo2013asymptotics,patterson1989strong}, we have that the following holds almost surely
\begin{equation*}
\lim_{n \rightarrow \infty} \calE_\delta(\widetilde{F}_{n,a,\bw_n}, F_{n}) = \calE_\delta(\widetilde{F}_{a}, F_{n}) = \int_{D(\delta)}\vert \varphi_{n}(\bt) - \widetilde{\varphi}_{a}(\bt) \vert^2 \omega(\mathbf{t}) \mathrm{d}\mathbf{t}.
\end{equation*}
We note that $\lim_{\delta\rightarrow 0}\calE_\delta(\widetilde{F}_{a}, F_{n}) = \calE(\widetilde{F}_{a}, F_{n})$, thus to verify \eqref{eqn:empirical_energy_convergence}, we must show that
\begin{equation}\label{eqn:limsup-delta}
\limsup_{\delta\rightarrow 0}\limsup_{n \rightarrow \infty} \vert \calE_\delta(\widetilde{F}_{n,a,\bw_n}, F_{n}) - \calE(\widetilde{F}_{n,a,\bw_n}, F_{n})\vert = 0.
\end{equation}

For each $\delta > 0$ we have 
\begin{align*}
\vert \calE_\delta(\widetilde{F}_{n,a,\bw_n}, F_{n}) - \calE(\widetilde{F}_{n,a,\bw_n}, F_{n})\vert \leq {} & 
\int_{\vert\bt\vert_p < \delta}\vert \varphi_{n}(\bt) - \widetilde{\varphi}_{n,a,\bw_n}(\bt) \vert^2 \omega(\mathbf{t}) \mathrm{d}\mathbf{t} \\ 
& + \int_{\vert\bt\vert_p > 1/\delta}\vert \varphi_{n}(\bt) - \widetilde{\varphi}_{n,a,\bw_n}(\bt) \vert^2 \omega(\mathbf{t}) \mathrm{d}\mathbf{t} 
\end{align*}

Note that 
\begin{align*}
\vert \varphi_{n}(\bt) - \widetilde{\varphi}_{n,a,\bw_n}(\bt) \vert^2 = {} & \left\vert  \frac{1}{n}\sum_{i=1}^n\exp\{i\langle\mathbf{t}, {\bX}_i  \rangle\} - \frac{1}{n}\sum_{i=1}^n\exp\{i\langle\mathbf{t}, \bXtilde_i  \rangle\}  \right\vert^2 \\
= {} & \left\vert  \frac{1}{n}\sum_{i=1}^n(1-\exp\{i\langle\mathbf{t},  \bXtilde_i \rangle\}) - \frac{1}{n}\sum_{i=1}^n\left(1-\exp\{i\langle\mathbf{t}, {\bX}_i  \rangle\}\right)  \right\vert^2 \\
\leq {} & \frac{1}{n}\sum_{i=1}^n\vert 1-\exp\{i\langle\mathbf{t},  \bXtilde_i \rangle\}\vert^2 + \frac{1}{n}\sum_{i=1}^n\vert 1-\exp\{i\langle\mathbf{t},  \bX_i \rangle\}\vert^2.
\end{align*}
Thus, 
\begin{align*}
\int_{\vert\bt\vert_p < \delta}\vert \varphi_{n}(\bt) - \widetilde{\varphi}_{n,a,\bw_n}(\bt) \vert^2 \omega(\mathbf{t}) \mathrm{d}\mathbf{t} \leq {} &  \frac{1}{n}\sum_{i=1}^n\int_{\vert\bt\vert_p < \delta}  \vert 1-\exp\{i\langle\mathbf{t},  \bXtilde_i \rangle\}\vert^2\omega(\mathbf{t}) \mathrm{d}\mathbf{t} \\
& + \frac{1}{n}\sum_{i=1}^n\int_{\vert\bt\vert_p < \delta} \vert 1-\exp\{i\langle\mathbf{t},  \bX_i \rangle\}\vert^2\omega(\mathbf{t}) \mathrm{d}\mathbf{t}.
\end{align*}

Similar to the arguments in the proof of Theorem 2 of \citet{szekely2007measuring}, we have that $\int_{\vert\bt\vert_p < \delta}  \vert 1-\exp\{i\langle\mathbf{t},  \bXtilde_i \rangle\}\vert^2\omega(\mathbf{t}) \mathrm{d}\mathbf{t} = |\bXtilde_i|G(\bXtilde_i\delta)$, where $G(y) = \int_{|\bt|_p < y}\frac{1-\cos(t_1)}{|\bt|^{1+p}}\mathrm{d}\bt$ where $t_1$ is the first element of $\bt$. Note that $\lim_{y\rightarrow 0}G(y)=0$ and $G(y)$ is bounded. Thus, by the strong law of large numbers, 
$\limsup_{n\rightarrow\infty}\int_{\vert\bt\vert_p < \delta}\vert \varphi_{n}(\bt) - \widetilde{\varphi}_{n,a,\bw_n}(\bt) \vert^2 \omega(\mathbf{t}) \mathrm{d}\mathbf{t} \leq \bbE\{|\bXtilde|G(|\bXtilde|\delta)\} + \bbE\{|\bX|G(|\bX|\delta)\}$. Thus, by the Lebesgue bounded convergence theorem for integrals and expectations, we have 
$\limsup_{\delta\rightarrow 0}\limsup_{n\rightarrow\infty}\int_{\vert\bt\vert_p < \delta}\vert \varphi_{n}(\bt) - \widetilde{\varphi}_{n,a,\bw_n}(\bt) \vert^2 \omega(\mathbf{t}) \mathrm{d}\mathbf{t} = 0$. By similar arguments, we have $\limsup_{\delta\rightarrow 0}\limsup_{n\rightarrow\infty}\int_{\vert\bt\vert_p > 1/\delta}\vert \varphi_{n}(\bt) - \widetilde{\varphi}_{n,a,\bw_n}(\bt) \vert^2 \omega(\mathbf{t}) \mathrm{d}\mathbf{t} = 0$.
Thus, we have shown \eqref{eqn:empirical_energy_convergence}.

Then to complete the proof it remains to show that 
\begin{equation}\label{eqn:weighted_energy_sup_convergence}
    \limsup_{n \to \infty}\; \lvert \calE(\widetilde{F}_{n,a,\bw_n} F_{n}) -\calE({F}_{n,a,\bw_n} F_{n}) \rvert = 0.
\end{equation}

We denote $\mathrm{d}\omega = \omega(\mathbf{t}) \mathrm{d}\mathbf{t}$. We begin by decomposing the above as 
\begin{align}
    & \lvert \calE(\widetilde{F}_{n,a,\bw_n}, F_{n}) -\calE({F}_{n,a,\bw_n}, F_{n}) \rvert \nonumber \\
    = {} &  \left\lvert \int_{\bbR^p} \left\{ 2\varphi_n(\bt)\left[ {\varphi}_{n,a,\bw_n}(\bt) - \widetilde{\varphi}_{n,a,\bw_n}(\bt)\right] + \widetilde{\varphi}^2_{n,a,\bw_n}(\bt) - {\varphi}^2_{n,a,\bw_n}(\bt) \right\} \mathrm{d}\omega \right\rvert \nonumber \\
    \leq {} & 2\int_{\bbR^p} \lvert\varphi_n(\bt)\rvert \left\{ \lvert\widetilde{\varphi}_a(\bt) -  {\varphi}_{n,a,\bw_n}(\bt)\rvert + \lvert \widetilde{\varphi}_a(\bt) - \widetilde{\varphi}_{n,a,\bw_n}(\bt)\rvert \right\} \mathrm{d}\omega \nonumber \\
    & + \int_{\bbR^p}  \lvert \widetilde{\varphi}_a(\bt) + \widetilde{\varphi}_{n,a,\bw_n}(\bt) \rvert \cdot \lvert \widetilde{\varphi}_a(\bt) - \widetilde{\varphi}_{n,a,\bw_n}(\bt)\rvert  \mathrm{d}\omega \nonumber \\
    & + \int_{\bbR^p}  \lvert \widetilde{\varphi}_a(\bt) + {\varphi}_{n,a,\bw_n}(\bt) \rvert \cdot \lvert \widetilde{\varphi}_a(\bt) - {\varphi}_{n,a,\bw_n}(\bt)\rvert  \mathrm{d}\omega \nonumber \\
    \leq {} & \int_{\bbR^p} \left\{ 2\lvert \varphi(\bt) \rvert + 2\lvert \varphi(\bt) - \varphi_n(\bt) \rvert +  2\lvert\widetilde{\varphi}_a(\bt)\rvert + \lvert \widetilde{\varphi}_a(\bt) - \widetilde{\varphi}_{n,a,\bw_n}(\bt) \rvert \right\} \cdot \lvert \widetilde{\varphi}_a(\bt) - \widetilde{\varphi}_{n,a,\bw_n}(\bt)\rvert  \mathrm{d}\omega \label{eqn:char_int_bound_1} \\
    & + \int_{\bbR^p} \left\{ 2\lvert \varphi(\bt) \rvert + 2\lvert \varphi(\bt) - \varphi_n(\bt) \rvert +  2\lvert\widetilde{\varphi}_a(\bt)\rvert + \lvert \widetilde{\varphi}_a(\bt) - {\varphi}_{n,a,\bw_n}(\bt) \rvert \right\} \cdot \lvert \widetilde{\varphi}_a(\bt) - {\varphi}_{n,a,\bw_n}(\bt)\rvert  \mathrm{d}\omega \label{eqn:char_int_bound_2}.
\end{align}

Note that $\varphi(\bt)$ is integrable due to the continuity of $\bX$ and that $\lvert {\varphi}_{n,a,\bw_n}(\bt) \rvert \to \vert\varphi(\bt)\rvert$ and $\lvert \widetilde{\varphi}_{n,a,\bw_n}(\bt) \rvert \to \vert\varphi(\bt)\rvert$. This ensures that the limsup of the integral converges to 0, which we will need below. 

Due to the almost sure convergence of ${\varphi}_{n,a,\bw_n}$ and $\widetilde{\varphi}_{n,a,\bw_n}$ to $\widetilde{\varphi}_a$, the terms inside the integrals \eqref{eqn:char_int_bound_1} and \eqref{eqn:char_int_bound_2} both converge almost surely to 0.
We first investigate \eqref{eqn:char_int_bound_1} and note that 
\begin{align}
    & 0 \leq  2\left\{ 2\lvert \varphi(\bt) \rvert + 2\lvert \varphi(\bt) - \varphi_n(\bt) \rvert +  2\lvert\widetilde{\varphi}_a(\bt)\rvert + \lvert \widetilde{\varphi}_a(\bt) - \widetilde{\varphi}_{n,a,\bw_n}(\bt) \rvert \right\} \cdot 
    \left\{\lvert \widetilde{\varphi}_a(\bt)\rvert + \lvert\widetilde{\varphi}_{n,a,\bw_n}(\bt)\rvert \right\} \nonumber \\
    - {} & \left\{ 2\lvert \varphi(\bt) \rvert + 2\lvert \varphi(\bt) - \varphi_n(\bt) \rvert +  2\lvert\widetilde{\varphi}_a(\bt)\rvert + \lvert \widetilde{\varphi}_a(\bt) - \widetilde{\varphi}_{n,a,\bw_n}(\bt) \rvert \right\} \cdot \lvert \widetilde{\varphi}_a(\bt) - \widetilde{\varphi}_{n,a,\bw_n}(\bt)\rvert.  \label{eqn:integral_pos_bound}
\end{align}
Note that the first term in the right hand side of \eqref{eqn:integral_pos_bound} converges to $8\{\lvert \varphi(\bt)\rvert + \lvert\widetilde{\varphi}_a(\bt)\rvert\}\lvert\widetilde{\varphi}_a(\bt)\rvert$ almost surely. Define $g_n(\bt)\equiv 2\lvert \varphi(\bt) \rvert + 2\lvert \varphi(\bt) - \varphi_n(\bt) \rvert +  2\lvert\widetilde{\varphi}_a(\bt)\rvert + \lvert \widetilde{\varphi}_a(\bt) - \widetilde{\varphi}_{n,a,\bw_n}(\bt) \rvert$ and its almost sure limit $g(\bt)\equiv 2\{\lvert\varphi(\bt)\rvert + \lvert\widetilde{\varphi}_a(\bt)\rvert\}$.
Then an application of Fatou's lemma to the right hand side of \eqref{eqn:integral_pos_bound} yields
\begin{align*}
    4\int_{\bbR^p}g(\bt)\lvert\widetilde{\varphi}_a(\bt)\rvert  \mathrm{d}\omega \leq {} & \liminf_{n\to\infty}\left\{2\int_{\bbR^p} g_n(\bt) \lvert \widetilde{\varphi}_{n,a,\bw_n}(\bt)\rvert\mathrm{d}\omega + 2\int_{\bbR^p} g(\bt) \lvert \widetilde{\varphi}_{a}(\bt)\rvert\mathrm{d}\omega \right. \\
    & \;\;\;\;\;\;\;\;\;\;\;\; \left.  -\int_{\bbR^p} g_n(\bt)\lvert \widetilde{\varphi}_a(\bt) - \widetilde{\varphi}_{n,a,\bw_n}(\bt) \rvert \mathrm{d}\omega  \right\}. 
\end{align*}
Thus we have $\limsup_{n\to\infty}\eqref{eqn:char_int_bound_1} = 0$. A similar argument holds for \eqref{eqn:char_int_bound_2}, and thus we have shown \eqref{eqn:weighted_energy_sup_convergence}, which concludes the proof.

\end{proof}

\subsection{Theorem \ref{thm:thm1}}
\begin{proof}
	Similar to \citet{amaral2017optimal}, we consider weights defined by the Radon-Nikodym derivative $h_a=f_{{\bX}}/f_{{\bX} |A = a}$ for $a \in \{0,1\}$, where $f_{{\bX}}$ is the density of $\bX$ for the full population and $f_{{\bX} |A = a}$ is the density of $\bX$ for the treated (or control) population. We then let $\mathbf{h}_a=\{h_a(\bX_1), \dots, h_a(\bX_n)\}$ be the Radon-Nikodym derivatives corresponding to the sample. We then define $\hat{h}_a(\bX_i) = {h}_a(\bX_i) / (\frac{1}{n_a}\sum_{i=1}^nI(A_i = a){h}_a(\bX_i))$ and $\hat{\boldsymbol h}_a = (\hat{h}_a(\bX_1), \dots, \hat{h}_a(\bX_n))$. By the SLLN, ${F}_{n,a,\hat{\boldsymbol h}_a}(\bx) = \frac{1}{n_a}\sum_{i=1}^n\hat{h}_a(\bX_i)I(A_i = a)I(\bX_i \leq \bx)$
	converges almost everywhere to $F(\bx)$ for every continuity point $\bx$
	\citep{tokdar2010importance, amaral2017optimal} for $a \in \{0,1\}$. Thus, as in the proof of Theorem 2 in \citet {mak2018support} by the Portmanteau and dominated convergence theorems,  we have 
	\begin{equation}\label{eqn:l2_phi_rn}
	\lim_{n\to\infty}\bbE[\vert \varphi(\mathbf{t}) - {\varphi}_{n,a,\hat{\boldsymbol h}_a}(\mathbf{t}) \vert ^ 2] = 0 \text{ for all } \mathbf{t} \text{ for } a \in \{0,1\}, 
	\end{equation}
	where ${\varphi}_{n,a,\hat{\boldsymbol h}_a}(\mathbf{t}) = \frac{1}{n_a}\sum_{i=1}^n\hat{h}_a(\bX_i)I(A_i = a) \exp\{ i, \langle \mathbf{t},\bX_i\rangle \}$ is a Radon-Nikodym derivative weighted ECHF for treatment arm $a$. Denote the expected weighted energy between the treated group and the sample population as $$\bbE[\calE(F_{n,a,{\hat{\boldsymbol h}_a}}, F_{n})] = \bbE\left[ \int_{\mathbb{R}^p}\vert \varphi(\bt) - {\varphi}_{n,a,\hat{\boldsymbol h}_a}(\mathbf{t}) \vert^2 \omega(\mathbf{t}) \mathrm{d}\mathbf{t} \right] \text{ for } a \in \{0,1\}.$$ 
	Note that although $F$ is the weighted average of two conditional distribution functions, i.e. $F(\bx) = F_1(\bx)P_1 + F_0(\bx) P_0$, due to the Theorem 2.1 and Corollary 3.1 of \citet{van1978properties}, all standard convergence properties of $F_n$ resulting from a mixture distribution such as $F$ this still hold. Specifically, a Glivenko-Cantelli theorem for empirical CDFs based on a mixture distribution as this holds. 
	Thus, by the same arguments as in \citet{mak2018support}, $\lim_{n\to\infty}\bbE[\calE(F_{n,a,{\hat{\boldsymbol h}_a}}, F_{n})] = 0$ for $a\in\{0,1\}$. Define ${\varphi}_{n,a,\bw^e_n}(\bt) = \frac{1}{n_1}\sum_{i=1}^nw^e_{i}I(A_i = a) \exp\{ i, \langle \mathbf{t},\bx_i\rangle \}$ to be the energy-weighted ECHF for treatment arm $a$. By the definition of $\mathbf{w}^e_n$, 
	\begin{align*}
	\int_{\mathbb{R}^p}\vert \varphi(\bt) - {\varphi}_{n,0,\bw^e_n}(\bt) \vert^2 \omega(\mathbf{t}) \mathrm{d}\mathbf{t} + \int_{\mathbb{R}^p}\vert \varphi(\bt) - {\varphi}_{n,1,\bw^e_n}(\bt) \vert^2 \omega(\mathbf{t}) \mathrm{d}\mathbf{t} 
	& = {} \calE(F_{n,0,{\bw^e_n}}, F_{n}) + \calE(F_{n,1,{\bw^e_n}}, F_{n}) \\
	& \leq \bbE[\calE(F_{n,0,{\hat{\boldsymbol h}_0}}, F_{n})] + \bbE[\calE(F_{n,1,{\hat{\boldsymbol h}_1}}, F_{n})].
	\end{align*} 
	Thus, $\lim_{n\to\infty} \calE(F_{n,a,{\bw^e_n}}, F_{n}) = 0$ for $a \in \{0,1\}$. If we choose any subsequence $\{n_k\}_{k=1}^\infty$ of $\mathbb{N}_+$, we have the same property that $\lim_{k\to\infty}  \calE(F_{n_k,0,{\bw^e_{n_k}}}, F_{n}) = 0$ for $a \in \{0,1\}$. By the Riesz-Fischer Theorem, a sequence of functions $f_n$ which converge to $f$ in $L_2$ has a subsequence $f_{n_k}$ which converges almost everywhere to $f$, implying the existence of a subsubsequence $\{n'_k\}_{k=1}^\infty \subseteq\{n_k\}_{k=1}^\infty$ such that ${\varphi}_{{n'_k},a,\bw^e_{n'_k}}(\bt)$ converges to ${\varphi}(\bt)$ almost everywhere as $k\to\infty$ for $a \in \{0,1\}$. Since $(n_k)$ was chosen arbitrarily, $\lim_{n\to\infty}{\varphi}_{{n},a,\bw^e_{n}}(\bt) = {\varphi}(\bt)$ almost everywhere. Thus the main convergence result of Theorem \ref{thm:thm1} holds. That $\lim_{n \to \infty} \calE(F_{n,a,\bw^e_n}, F_{n}) = 0$ holds almost surely is a consequence of \eqref{eqn:energy_weighted_cdf_convergence} of the main text and Theorem \ref{thm:weighted_energy_converges_to_energy}.
	
\end{proof}


\subsection{Corollary \ref{thm:cor1}}
\begin{proof}
	From (3) of the main text, the bias of $\hat{\tau}_{\bw_n^e}$ can be written as:
	\begin{align}
		\begin{split}
			|\mathbb{E}[\hat{\tau}_{\bw_n^e}]-\tau| & = \left| \int_{\bx\in \mathcal{X}}  \mu_{1}(\bx) \mathrm{d} \left[ F - F_{n, 1, \bw_n^e} \right](\bx) - \int_{\bx\in \mathcal{X}}  \mu_{0}(\bx) \mathrm{d} \left[ F - F_{n, 0, \bw_n^e} \right](\bx) \right|\\
			& \leq \left| \int_{\bx\in \mathcal{X}}  \mu_{1}(\bx) \mathrm{d} \left[ F - F_{n, 1, \bw_n^e} \right](\bx)\right| + \left| \int_{\bx\in \mathcal{X}}  \mu_{0}(\bx) \mathrm{d} \left[ F - F_{n, 0, \bw_n^e} \right](\bx) \right|.
		\end{split}
		\label{eqn:bias}
	\end{align}
	By Theorem \ref{thm:thm1}, we know that $F_{n, 1, \bw_n^e}(\bx)$, the \textit{weighted} treatment covariate distribution, converges to $F$, the population covariate distribution. By the Portmanteau Theorem (Theorem 2.1, \citealp{billingsley1993convergence}), it follows that:
	\[ \int_{\bx\in \mathcal{X}} \mu_{1}(\bx) \mathrm{d} F_{n,1,\bw_n^e}(\bx) \xrightarrow{n \rightarrow \infty} \int_{\bx\in \mathcal{X}} \mu_{1}(\bx) \mathrm{d} F (\bx).\]
	An analogous argument yields a similar result for the control group: 
	\[\int_{\bx\in \mathcal{X}} \mu_{0}(\bx) \mathrm{d} F_{n,0,\bw_n^e} (\bx) \xrightarrow{n \rightarrow \infty} \int_{\bx\in \mathcal{X}} \mu_{0}(\bx) \mathrm{d} F (\bx).\]
	Hence, from \eqref{eqn:bias}, we have $\lim_{n \rightarrow \infty} |\mathbb{E}[\hat{\tau}_{\bw_n^e}]-\tau| = 0$, which proves the claim.
\end{proof}

\subsection{Lemma \ref{lem:bd}}
This follows directly from Theorem 4 of \cite{mak2018support}.

\subsection{Theorem \ref{thm:rootn}}
The proof of Theorem \ref{thm:rootn} requires a few lemmas. 

The first lemma shows that, under i.i.d. sampling of the covariates $\bX_1, \cdots, \bX_n \sim F$, the expected energy distance between $F_n$ (its empirical distribution) and $F$ (the population distribution) converges at a rate of $\mathcal{O}(1/n)$:

\begin{lem}
Suppose $\bX_1, \cdots, \bX_n \distas{i.i.d.} F$. Then $\mathbb{E}[\calE(F,F_n)] = \mathcal{O}(1/n)$.
\label{lem:expen}
\end{lem}

\begin{proof}[Lemma \ref{lem:expen}] 
By Proposition 1 of \cite{szekely2013energy}, we have:
\[ \calE(F,F_n) = \int_{\mathbb{R}^p} |\varphi(\bt)-\varphi_{n}(\bt)|^2 \omega(\bt) d\bt. \]
Taking an expectation on both sides, it follows that:
\begin{align*}
\mathbb{E}[\calE(F,F_n)] &= \mathbb{E}\left[\int_{\mathbb{R}^p} |\varphi(\bt)-\varphi_{n}(\bt)|^2 \omega(\bt) d\bt\right]\notag\\
&= \int_{\mathbb{R}^p} \mathbb{E} \left[ |\varphi(\bt)-\varphi_{n}(\bt)|^2 \right] \omega(\bt) d\bt \tag{Tonelli's theorem, since the integrand is non-negative}\\
&= \int_{\mathbb{R}^p} \frac{\mathbb{V}[\text{Re}(\phi_1(\bt))]+\mathbb{V}[\text{Im}(\phi_1(\bt))]}{n} \omega(\bt) d\bt \tag{$\mathbb{E} \left[ |\varphi(\bt)-\varphi_{n}(\bt)|^2 \right]$ is a variance term, since $\mathbb{E}\varphi_{n}(\bt) = \varphi(\bt)$}\\
&= \mathcal{O}\left( \frac{1}{n} \right),
\end{align*}
where constant terms depend on $F$ and $p$.
\end{proof}

The second lemma shows that, under the additional causal assumptions of positivity and strong ignorability as well as mild distributional assumptions on $F_0$ and $F_1$, the same convergence rate of $\calO(1/n)$ holds for the energy distance between $F_{n,a,\bw_n^{e}}$ (the \textit{energy-weighted} distribution for the treated or control) and $F_n$ (the empirical covariate distribution):

\begin{lem}
Assume that the causal assumptions of positivity and strong ignorability hold. Let $\bw_n^e$ be the solution to the energy balancing objective \eqref{eqn:energy_balancing_weights_ate} of the main text. Under assumption (A4), we have $\calE(F_{n,a,\bw_n^{e}},F_n) = \calO(1/n)$ almost surely.

\label{lem:expen_rn}
\end{lem}

\begin{proof}[Lemma \ref{lem:expen_rn}] 
First consider the non-normalized Radon-Nikodym derivative weights $\bw_n^{nnrn}$. We will first show that $\calE(F_{n,a,\bw_n^{nnrn}},F_n)$ is simply a degenerate two-sample $V$-statistic to show its convergence rate. The weights $\bw_n^{nnrn}$ are functions of $\bx$ in the sense that $w_i^{nnrn} = w^{nnrn}(\bX_i) = 1/\pi(A_i, \bX_i)$, where $\pi(a, \bx) = \bbP(A=a\:|\:\bX=\bx)$. Then $\calE(F_{n,a,\bw_n^{nnrn}},F_n)$ is a two-sample $V$-statistic with kernel $h(\bx_i,\bx_j; \bx_\ell, \bx_m) = w^{nnrn}(\bx_i)||\bx_i - \bx_\ell||_2 + w^{nnrn}(\bx_j)||\bx_j - \bx_m||_2 - w^{nnrn}(\bx_i)w^{nnrn}(\bx_j)||\bx_i - \bx_j||_2 - ||\bx_\ell - \bx_m||_2$. Denote $\{\widetilde{\bX}_1, \dots, \widetilde{\bX}_{n_a}\} = \{{\bX}_i : A_i = a\}$.
Then $\calE(F_{n,a,\bw_n^{nnrn}},F_n)$ can be written as the following $V$-statistic
\[
\calE(F_{n,a,\bw_n^{nnrn}},F_n) = \frac{1}{n^2n^2_a}\sum_{i=1}^{n_a}\sum_{j=1}^{n_a}\sum_{\ell=1}^{n}\sum_{m=1}^{n}h(\widetilde{\bX}_i, \widetilde{\bX}_j;\bX_\ell, \bX_m).
\]
From positivity and strong ignorability it can be shown that $\calE(F_{n,a,\bw_n^{nnrn}},F_n)$ is first-order degenerate in the sense that $\bbE h(\widetilde{\bx}, \widetilde{\bX}_j;\bX_\ell, \bx) = 0$ for any $\widetilde{\bx}$ and $\bx$. Thus, if $\bbE h^2 < \infty$, then $\calE(F_{n,a,\bw_n^{nnrn}},F_n) = \calO(n^{-1})$ by extensions of asymptotic results for one-sample $V$-statistics \citep{serfling1980approximation, korolyuk1989theory} to multi-sample $V$-statistics as in \citet{rizzo2002test}. Note that this also implies that $\calE(F_{n,a,\bw_n^{rn}},F_n) = \calO(n^{-1})$, since $\bw_n^{nnrn}$ and $\bw_n^{rn}$ differ only by a normalizing constant such that $\frac{1}{n}\sum_{i=1}^nw_i^{rn} = \bbE\left[\frac{1}{n}\sum_{i=1}^nw_i^{nnrn}\right] = 1$. By the definition of $\bw_n^{e}$, we have $\calE(F_{n,a,\bw_n^{e}},F_n) \leq \calE(F_{n,a,\bw_n^{rn}},F_n)$ for each $n$, which proves the desired result  $\calE(F_{n,a,\bw_n^{e}},F_n) = \calO(n^{-1})$.
\end{proof}

The next lemma shows that, under a mild regularity condition on the energy balancing weights, the sum of the squared weights is upper bounded by $\calO(n)$:

\begin{lem}
Let ${\bw}^{e}_n = ({w}^e_{1,n}, \cdots, {w}^e_{n,n})$ be the solution to the energy balancing objective \eqref{eqn:energy_balancing_weights_ate} of the main text. Under assumptions (A4) and (A5), we have almost surely that:
\[ \sum_{i:A_i=0} \frac{{{}{w}^e_{i,n}}^2}{n_0} \leq B \quad \text{and} \quad \sum_{i:A_i=1} \frac{{{}{w}^e_{i,n}}^2}{n_1} \leq B \]
for all $n > n^*$ for some $n^*>1$ and some constant $B>0$ that does not depend on $n$.
\label{lem:sqwts}
\end{lem}

\begin{proof}[Lemma \ref{lem:sqwts}] 

Note that $\calE(F_{n,1,\widetilde{\bw}^{e}_n}, F_n) = \calO(n^{-1})$ by Lemma \ref{lem:expen_rn}. We consider for simplicity the univariate case $p=1$ and only focus on the treated group, i.e. those with $A=1$; however, the same results apply directly for $A=0$. For clarity of presentation, we denote $w_i\equiv{w}^e_{i,n}$.
By the weighted energy distance duality, we have 
\begin{align}
& \calE(F_{n,1,{\bw}^{e}_n}, F_n) \nonumber \\ & =  \int_\bbR|\phi_n(t) - \phi_{n,1,\widetilde{\bw}^{e}_n}(t)|^2\omega(t) \; dt \nonumber\\ 
& = \frac{1}{n^2}\int_\bbR\left|\sum_{i=1}^n(1-w_iA_i\frac{n}{n_1})\exp(itX_i)\right|^2\omega(t) \; dt \nonumber \\
& = \frac{1}{n^2}\int_\bbR\sum_{i=1}^n\sum_{j=1}^n\left\{\left(1-w_iA_i\frac{n}{n_1} -w_jA_j\frac{n}{n_1} + w_iw_jA_iA_j\frac{n^2}{n^2_1}\right)\exp(it(X_i+X_j)) \right\} \omega(t) \; dt.\label{eqn:duality_expanded}
\end{align}

Suppose that the number of weights $w_i$ that are ``near'' the maximum $Cn^{1/3}$  (i.e. are of the same order with respect to $n$) is of order $\calO(n^{1/3})$. Denote the index set of these observations as $\calI_n\equiv\{i:w_i=\calO(n^{1/3})\}$, and note that this supposition implies $|\calI_n| = \calO(n^{1/3})$. Further suppose the ``worst case'' scenario that $A_i=1$ for all $i\in \calI_n$, $\text{Re}(\exp(it(X_i+X_j))) > 0$ and $\text{Im}(\exp(it(X_i+X_j))) > 0$ for all $i,j$, and that $\text{Re}(\exp(itX_i))<0$, $ \text{Im}(\exp(itX_i))<0$, $\text{Re}(\exp(itX_j))<0$, and $\text{Im}(\exp(itX_j))<0$, so that every term in the double sum inside the integral in \eqref{eqn:duality_expanded} is positive. Then the double sum in \eqref{eqn:duality_expanded} is larger than 
\begin{align*}
& \sum_{i\in \calI_n}\sum_{j\in \calI_n}\left\{\left(1-w_iA_i\frac{n}{n_1} -w_jA_j\frac{n}{n_1} + w_iw_jA_iA_j\frac{n^2}{n^2_1}\right)\exp(it(X_i+X_j)) \right\} \\
& = \sum_{i\in \calI_n}\sum_{j\in \calI_n}\left\{\left(1+\calO(n^{1/3}) +\calO(n^{1/3}) + \calO(n^{2/3})\right)\exp(it(X_i+X_j)) \right\} \\
& = \sum_{i\in \calI_n}\left(\calO(n^{1/3})+\calO(n^{2/3}) +\calO(n^{2/3}) + \calO(n)\right) \\
& = \calO(n^{4/3}),
\end{align*}
which implies that $\calE(F_{n,1,\bw^e}, F_n) = \calO(n^{-2/3})$, which is a contradiction to Lemma \ref{lem:expen_rn}. 
Thus, we cannot have $|\calI_n|$ as large as $\calO(n^{1/3})$. Using a similar argument, one can then show that the maximum size $\calI_n$ can take to avoid such a contradiction is $|\calI_n| = \calO(n^{1/6})$.

Assume, therefore, the worst case scenario that $|\calI_n| = \calO(n^{1/6})$. To study the behavior of $\sum_{i:A_i=1} {w_{i,n}^2}/{n_1}$, we consider the set $\calJ_n = \{i: i\not\in\calI_n,w_i=O(r(n)) \text{ where } \lim_{n\to\infty}r(n)=\infty \text{ and } \lim_{n\to\infty}r(n)/n^{1/3}=0\}$. Thus, if we define $\calK_n = \{i:w_i = O(1)\}$, then $\{i:A_i=1\} = \calI_n\cup\calJ_n\cup\calK_n$. We now seek to find how large $|\calJ_n|$ can be to avoid a contradiction like the above. Consider the cross terms of $\calJ_n$ and $\calI_n$ in \eqref{eqn:duality_expanded}, which are 
\begin{align*}
& \sum_{i\in \calI_n}\sum_{j\in \calJ_n}\left\{\left(1-w_iA_i\frac{n}{n_1} -w_jA_j\frac{n}{n_1} + w_iw_jA_iA_j\frac{n^2}{n^2_1}\right)\exp(it(X_i+X_j)) \right\}  \\
& = \sum_{i\in \calI_n}\sum_{j\in \calJ_n}\left\{\left(1+\calO(n^{1/3}) +\calO(r(n)) + \calO(r(n)n^{1/3})\right)\exp(it(X_i+X_j)) \right\} \\
& = \sum_{j\in \calJ_n}\left(\calO(n^{1/6})+\calO(n^{1/2}) +\calO(r(n)n^{1/6}) + \calO(r(n)n^{1/2}) \right) \\
& = \calO(|\calJ_n|r(n)n^{1/2}).
\end{align*}
Thus, to avoid a contradiction to Lemma \ref{lem:expen_rn}, we need $\calO(|\calJ_n|r(n)n^{1/2}) = \calO(n)$, i.e., $|\calJ_n|r(n) = \calO(n^{1/2})$. With this, the sum $\sum_{i:A_i=1} {w_{i,n}^2}$ then becomes:
\begin{align*}
\sum_{i:A_i=1} w_{i,n}^2 = {} & \sum_{i\in\calK_n} w_{i,n}^2+\sum_{i\in\calJ_n} w_{i,n}^2+\sum_{i\in\calI_n} w_{i,n}^2\\ 
= {} & \sum_{i\in\calK_n} \calO(1)+\sum_{i\in\calJ_n} \calO(r^2(n))+\sum_{i\in\calI_n} \calO(n^{2/3}) \\
= {} & \calO(n) + \calO(n^{5/6}) + \calO(|\calJ_n|r^2(n)) = \calO(n),
\end{align*}
where the last equality holds since $|\calJ_n|r(n)$ is at most of order $\calO(n^{1/2})$, $\lim_{n\to\infty}r(n)/n^{1/2}=0$, and $|\calK_n|=\calO(n)$ because $n=|\calK_n|+|\calJ_n|+|\calI_n|$ .

From this (and the symmetry of the argument for $A=0$), it follows that
\[ \sum_{i:A_i=0} \frac{w_{i,n}^2}{n_0} \leq B \quad \text{and} \quad \sum_{i:A_i=1} \frac{w_{i,n}^2}{n_1} \leq B \]
for all $n>n^*$ for some $n^*>1$, which proves the lemma.
\end{proof}

\begin{proof}[Proof of Theorem \ref{thm:rootn}]
With these lemmas in hand, we can now tackle the main theorem. Let us condition on both $\bX$ and $A$. From \eqref{eqn:ate_bias_int}-\eqref{eqn:ate_residuals}  of the main text, we can rewrite the mean squared error of $\hat{\tau}_{\bw_n^e}$ as:
\allowdisplaybreaks
\begin{align*}
\mathbb{E}_{Y|\bX,A}[(\hat{\tau}_{\bw_n^e}-\tau)^2] = {} &  \mathbb{V}_{Y|\bX,A}\left[\frac{1}{n_0}\sum_{i:A_i=0} w_i^e \epsilon_i\right] + \mathbb{V}_{Y|\bX,A}\left[\frac{1}{n_1} \sum_{i:A_i=1} w_i^e \epsilon_i\right]\\
& + \left( \int  \mu_{1}(\bx) \mathrm{d} \left[ F - F_{n, 1, \bw_n^e} \right](\bx) - \int  \mu_{0}(\bx) \mathrm{d} \left[ F - F_{n, 0, \bw_n^e} \right](\bx) \right)^2\\
= {} &  \mathbb{V}_{Y|\bX,A}\left[\frac{1}{n_0}\sum_{i:A_i=0} w_i^e \epsilon_i\right] + \mathbb{V}_{Y|\bX,A}\left[\frac{1}{n_1} \sum_{i:A_i=1} w_i^e \epsilon_i\right]\\
& + \left( \int  \mu_{1}(\bx) \mathrm{d} \left[ F_n - F_{n, 1, \bw_n^e} \right](\bx) - \int \mu_{1}(\bx) \mathrm{d} \left[ F_n - F \right](\bx) \right.\\
& \quad \quad \left. - \int  \mu_{0}(\bx) \mathrm{d} \left[ F_n - F_{n, 0, \bw_n^e} \right](\bx) + \int  \mu_{0}(\bx) \mathrm{d} \left[ F_n - F \right](\bx) \right)^2\\
\leq {} &  \underbrace{\sum_{a=0}^1 \frac{1}{n_a^2} \sum_{i:A_i=a} (w_i^e)^2 \sigma^2_a(\bX_i)}_{\circled{1}} + 4 \underbrace{\sum_{a=0}^1 \left(\int \mu_a(\bx) \mathrm{d} \left[ F_n - F_{n,a,\bw_n^e} \right](\bx) \right)^2}_{\circled{2}}\\
& + 4 \underbrace{\sum_{a=0}^1 \left(\int \mu_a(\bx) \mathrm{d} \left[ F - F_n \right](\bx) \right)^2}_{\circled{3}},
\label{eqn:bigdecomp}
\end{align*}
where the last step follows from the identity $(a+b+c+d)^2 \leq 4(a^2+b^2+c^2+d^2)$.


Consider first the terms in \circled{1}. Since $\sigma^2_a(\bx)$ is assumed to be bounded over $\mathcal{X}$, define $\bar{\sigma}^2 \equiv \max_{a \in \{0,1\}}\{\sup_{\bx \in \calX} \sigma^2_a(\bx)\}$. We have:
\begin{align*}
\mathbb{E}_{\bX,A}\left[\sum_{a=0}^1 \frac{1}{n_a^2} \sum_{i:A_i=a} (w_i^e)^2 \sigma^2_a(\bX_i)\right] &\leq \bar{\sigma}^2 \mathbb{E}_{\bX,A}\left[ \sum_{a=0}^1\frac{1}{n_a^2} \sum_{i:A_i=a} (w_i^e)^2\right]\\
&\leq B \bar{\sigma}^2 \mathbb{E}_{A}\left[ Z_0 + Z_1 \right] \tag{Lemma \ref{lem:sqwts}},
\end{align*}
\noindent where $Z_a = 1/n_a$ if $n_a > 0$ and $0$ otherwise. Note that, for $a \in \{0,1\}$, $n_a \sim \text{Bin}(n,P_a)$, where $P_a = \mathbb{P}(A=a)$. It follows that:
\[\mathbb{E}_{A}\left[Z_a\right] \leq \mathbb{E}_{A}\left[\frac{2}{n_a+1}\right] = \frac{2(1-(1-P_a)^{n+1})}{P_a(n+1)} \leq \frac{2}{P_a(n+1)} = \calO\left( \frac{1}{n}\right).\]
From this, we get that $\mathbb{E}_{\bX,A}[\circled{1}]$ is also $\calO(1/n)$.

\notag Consider next the terms in \circled{2}. For each $a \in \{0,1\}$, we have:
\begin{align*}
&\mathbb{E}_{\bX,A}\left[ \left(\int \mu_a(\bx) \mathrm{d} \left[ F_n - F_{n,a,\bw_n^e} \right](\bx) \right)^2 \right]\\
& \quad \quad \quad \leq \mathbb{E}_{\bX,A} \left[ \sup_{\zeta \in \calH : ||\zeta||_{\mathcal{H}} \leq ||\mu_a||_{\mathcal{H}}}\left(\int \zeta(\bx) \mathrm{d} \left[ F_n - F_{n,a,\bw_n^e} \right](\bx) \right)^2 \right] \\
& \quad \quad \quad \leq C \mathbb{E}_{\bX,A}\left[\calE(F_{n,a,\bw_n^e},F_n)\right] \tag{Lemma \ref{lem:bd}}\\
& \quad \quad \quad = \calO\left(\frac{1}{n}\right). \tag{Lemma \ref{lem:expen_rn}}
\end{align*}
Finally, consider the terms in \circled{3}. Since $\bX_1, \cdots, \bX_n \distas{i.i.d.} F$, for each $a \in \{0,1\}$, we have:
\[ \mathbb{E}_{\bX,A}\left[\left( \int \mu_a(\bx) \mathrm{d} \left[ F - F_{n} \right](\bx) \right)^2\right] =  \frac{\text{Var}[\mu_a(\bX)]}{n} = \mathcal{O}\left(\frac{1}{n} \right).\]
\notag Using the above bounds on \circled{1}, \circled{2} and \circled{3}, the desired claim is proven:
\[\mathbb{E}_{Y,\bX,A}[(\hat{\tau}_{\bw_n^e}-\tau)^2] = \mathbb{E}_{\bX,A} \mathbb{E}_{Y|\bX,A}[(\hat{\tau}_{\bw_n^e}-\tau)^2] = \mathcal{O}\left( \frac{1}{n} \right).\]

\end{proof}



\section{Additional Simulation Results}
\label{sec:supp_details}

\subsection{Additional details for simulations}

In this section we provide specific details of all of the propensity score models and outcome models used in the simulations in Section \ref{sec:multiple_treatments} of the main text. The propensity score models are described in Table \ref{tab:simulation_settings_propensity}. The average proportion of those treated in propensity models I, II, III, IV, and V are 0.35, 0.31, 0.50, 0.51, and 0.51, respectively. The conditional mean functions of the outcome given the covariates and treatment for outcome models (A-E) are provided in Table \ref{tab:simulation_settings_response}.

\begin{table}[ht]
\centering
\resizebox{1\textwidth}{!}{
\begin{tabular}{cc}
\toprule
 Model & $\eta = \text{logit}\{\bbP(A=1|\bX)\} = $  \\ 
\midrule
I & $2X_1X_2I(\lvert X_1 \rvert >1, \lvert X_2 \rvert >1) + 2X_2X_3I(\lvert X_2 \rvert < 1, \lvert X_3 \rvert < 1)$ \\ 
& $+2X_3X_4I(\lvert X_3 \rvert >1, \lvert X_4 \rvert >1) + 2X_4X_1I(\lvert X_1 \rvert <1, \lvert X_4 \rvert < 1)$ \\ 
& $ + I(\lvert X_1\rvert > 0.5, \lvert X_2\rvert > 0.5, \lvert X_3\rvert>0.5, \lvert X_4\rvert > 0.5)$ \\
& $+ I(\lvert X_1\rvert < 0.25, \lvert X_2\rvert > 0.25 , \lvert X_3\rvert < 0.25, \lvert X_4\rvert > 0.25) $ \\
II & $-2 + \log\lvert X_1-X_2 \rvert - \log\lvert X_2-X_3 \rvert + \lvert(X_3-X_4)X_1X_2\rvert^{1/2}$ \\
III & $-X_1 + 0.5X_2 - 0.25X_3 - 0.1X_4 - X_5 + 0.5X_6 - 0.25X_7-0.1X_8$ \\
IV &  $c \sum_{i =1}^3\sum_{j=i}^{4}(-1)^{2j-i}X_iX_j$, where $c$ is chosen  such that $\text{SD}(\eta)=5$  \\
V & $-2 + 2X_1X_2 + (X_1-X_2)^2 - 2X_3X_4 - (X_3+X_5)^2$ \\
VI & $ \lvert X_1 - 2X_2 \rvert \cdot \lvert X_2 - 2X_3 \rvert - \lvert X_3 - 2X_4 \rvert \cdot \lvert X_4 - 2X_5 \rvert + X_6 - 0.5X_7 -0.25X_8$ \\
\bottomrule
\end{tabular}%
}
\caption{Propensity models used in the simulation studies. The average proportion of those treated in propensity models I, II, III, IV, and V are 0.35, 0.31, 0.50, 0.51, and 0.51, respectively.}
\label{tab:simulation_settings_propensity}
\end{table}

\begin{table}[ht]
\centering
\resizebox{1\textwidth}{!}{
\begin{tabular}{cc}
\toprule
 Model & $\mu = \bbE[Y \vert \bX, A] = $  \\ 
\midrule
A & $210 + 27.4\lvert X_1\rvert + 13.7\lvert X_2\rvert + 13.7\lvert X_3\rvert  + 13.7\lvert X_4\rvert$  \\
B & $X_1X_2^3X_3^2X_4 + X_4\lvert X_1 \rvert^{1/2}$ \\
C & $2\sum_{j=1}^4\left( 1- X_j I(X_j>0)A \right) \cdot \left( X_j - 2X_{j+1} \right)$ \\
D & $\sum_{j=1}^7X_j\beta_j + \beta_2X_2^2 + \beta_4X_4^2+\beta_7X_7^2 +0.5\beta_1X_1X_3+0.7\beta_2X_2X_4 +0.7\beta+2X_2X_4 + 0.5\beta_3X_3X_5$ \\
& $+ 0.7\beta_4X_4X_6 + 0.5\beta_5X_5X_7 + 0.5\beta_1X_1X_6 + 0.7\beta_2X_2X_3 + 0.5\beta_3X_3X_4 + 0.5\beta_4X_4X_5 + 0.5\beta_5X_5X_6$ \\
E &  $210 + (1.5A - 0.5) \left( 27.4 X_1 + 13.7 X_2 + 13.7 X_3  + 13.7 X_4 \right)$ \\
\bottomrule
\end{tabular}%
}
\caption{The coefficients in Model $D$ above are $\bbeta = (0.8, 0.25, 0.6, -0.4, -0.8, -0.5, 0.7)$. }
\label{tab:simulation_settings_response}
\end{table}

\subsection{Detailed Simulation Results}

Table \ref{tab:simulation_results_ranks_ymod_n250} contains a summary of the results averaged across propensity models (I-VI) and dimension settings ($p\in\{10,25\}$). Each entry in the table is the average rank of each method in terms of RMSE and bias for each combination of outcome model and dimension; i.e. the method with the smallest RMSE for a particular setting receives a ``1'' and the method with the largest RMSE receives a ``7''.

\begin{table}
\caption{\label{tab:simulation_results_ranks_ymod_n250} Displayed are the ranks among all methods tested of each method in terms of RMSE and bias averaged over all response models (A-E) for $n=250$ and  over the dimension settings $p=10$ and $p=25$.}
\centering
\resizebox{\textwidth}{!}{
\begin{tabular}{cRcRcRcRcRc}
\hline
$Y$ Model: & \multicolumn{2}{c}{A} & \multicolumn{2}{c}{B} & \multicolumn{2}{c}{C} & \multicolumn{2}{c}{D} & \multicolumn{2}{c}{E} \\ 
  \cmidrule{2-3} \cmidrule{4-5} \cmidrule{6-7} \cmidrule{8-9} \cmidrule{10-11}
 & \multicolumn{2}{c}{Mean Rank} & \multicolumn{2}{c}{Mean Rank} & \multicolumn{2}{c}{Mean Rank} & \multicolumn{2}{c}{Mean Rank} & \multicolumn{2}{c}{Mean Rank} \\ 
Method  & RMSE & Bias & RMSE & Bias & RMSE & Bias & RMSE & Bias & RMSE & Bias \\ 
\hline
\multicolumn{1}{r}{Unweighted}  & \multicolumn{1}{r}{$5.1$} & \multicolumn{1}{r}{$4.8$} & \multicolumn{1}{r}{$6.2$} & \multicolumn{1}{r}{$4.8$} & \multicolumn{1}{r}{$4.8$} & \multicolumn{1}{r}{$4.7$} & \multicolumn{1}{r}{$6.0$} & \multicolumn{1}{r}{$4.8$} & \multicolumn{1}{r}{$3.8$} & \multicolumn{1}{r}{$\mathbf{2.2}$} \\
\multicolumn{1}{r}{EBW}  & \multicolumn{1}{r}{$2.3$} & \multicolumn{1}{r}{$2.5$} & \multicolumn{1}{r}{$3.4$} & \multicolumn{1}{r}{$4.0$} & \multicolumn{1}{r}{$\mathbf{2.0}$} & \multicolumn{1}{r}{$\mathbf{2.2}$} & \multicolumn{1}{r}{$3.2$} & \multicolumn{1}{r}{$3.5$} & \multicolumn{1}{r}{$\mathbf{1.5}$} & \multicolumn{1}{r}{$4.2$} \\
\multicolumn{1}{r}{iEBW}  & \multicolumn{1}{r}{$\mathbf{1.1}$} & \multicolumn{1}{r}{$\mathbf{1.3}$} & \multicolumn{1}{r}{$\mathbf{1.8}$} & \multicolumn{1}{r}{$\mathbf{2.8}$} & \multicolumn{1}{r}{$2.4$} & \multicolumn{1}{r}{$2.9$} & \multicolumn{1}{r}{$\mathbf{1.8}$} & \multicolumn{1}{r}{$\mathbf{2.4}$} & \multicolumn{1}{r}{$3.2$} & \multicolumn{1}{r}{$4.5$} \\
\multicolumn{1}{r}{KCB}  & \multicolumn{1}{r}{$2.6$} & \multicolumn{1}{r}{$2.9$} & \multicolumn{1}{r}{$2.9$} & \multicolumn{1}{r}{$3.7$} & \multicolumn{1}{r}{$2.7$} & \multicolumn{1}{r}{$2.7$} & \multicolumn{1}{r}{$2.8$} & \multicolumn{1}{r}{$3.0$} & \multicolumn{1}{r}{$3.2$} & \multicolumn{1}{r}{$3.2$} \\
\multicolumn{1}{r}{IPW}  & \multicolumn{1}{r}{$6.4$} & \multicolumn{1}{r}{$6.1$} & \multicolumn{1}{r}{$5.0$} & \multicolumn{1}{r}{$4.2$} & \multicolumn{1}{r}{$6.3$} & \multicolumn{1}{r}{$5.7$} & \multicolumn{1}{r}{$5.2$} & \multicolumn{1}{r}{$5.0$} & \multicolumn{1}{r}{$6.9$} & \multicolumn{1}{r}{$4.3$} \\
\multicolumn{1}{r}{CBPS}  & \multicolumn{1}{r}{$5.3$} & \multicolumn{1}{r}{$5.0$} & \multicolumn{1}{r}{$4.8$} & \multicolumn{1}{r}{$4.0$} & \multicolumn{1}{r}{$5.2$} & \multicolumn{1}{r}{$5.4$} & \multicolumn{1}{r}{$5.3$} & \multicolumn{1}{r}{$4.6$} & \multicolumn{1}{r}{$4.8$} & \multicolumn{1}{r}{$3.7$} \\
\multicolumn{1}{r}{Cal}  & \multicolumn{1}{r}{$5.2$} & \multicolumn{1}{r}{$5.3$} & \multicolumn{1}{r}{$3.8$} & \multicolumn{1}{r}{$4.7$} & \multicolumn{1}{r}{$4.6$} & \multicolumn{1}{r}{$4.5$} & \multicolumn{1}{r}{$3.8$} & \multicolumn{1}{r}{$4.7$} & \multicolumn{1}{r}{$4.7$} & \multicolumn{1}{r}{$5.8$} \\
   \bottomrule
\end{tabular}%
}
\end{table}

\begin{table}[ht]
\centering
\resizebox{1\textwidth}{!}{
		\begin{tabular}{crccRcRcRcRcRc}
			\toprule
			& Propensity Model: & \multicolumn{2}{c}{I} & \multicolumn{2}{c}{II} & \multicolumn{2}{c}{III} & \multicolumn{2}{c}{IV} & \multicolumn{2}{c}{V} & \multicolumn{2}{c}{VI} \\ 
			\cmidrule{3-4} \cmidrule{5-6} \cmidrule{7-8} \cmidrule{9-10} \cmidrule{11-12} \cmidrule{13-14}  
			& Method & RMSE & Bias & RMSE & Bias & RMSE & Bias & RMSE & Bias & RMSE & Bias & RMSE & Bias  \\ 
			\midrule
			\multicolumn{1}{r}{$Y$ Model: A} & \multicolumn{1}{r}{Unweighted}  & \multicolumn{1}{r}{$\phantom{-}21.375$} & \multicolumn{1}{r}{$-21.110$} & \multicolumn{1}{r}{$\phantom{-}23.362$} & \multicolumn{1}{r}{$\phantom{-}22.893$} & \multicolumn{1}{r}{$\phantom{0}\phantom{-}3.851$} & \multicolumn{1}{r}{$\phantom{0}\phantom{-}0.032$} & \multicolumn{1}{r}{$\phantom{0}\phantom{-}7.772$} & \multicolumn{1}{r}{$\phantom{0}\phantom{-}6.752$} & \multicolumn{1}{r}{$\phantom{-}35.122$} & \multicolumn{1}{r}{$\phantom{-}34.983$} & \multicolumn{1}{r}{$\phantom{-}12.423$} & \multicolumn{1}{r}{$\phantom{-}11.786$} \\
			 & \multicolumn{1}{r}{EBW}  & \multicolumn{1}{r}{$\phantom{0}\phantom{-}8.450$} & \multicolumn{1}{r}{$\phantom{0}-8.173$} & \multicolumn{1}{r}{$\phantom{0}\phantom{-}6.908$} & \multicolumn{1}{r}{$\phantom{0}\phantom{-}6.714$} & \multicolumn{1}{r}{$\phantom{0}\phantom{-}2.133$} & \multicolumn{1}{r}{$\phantom{0}-0.094$} & \multicolumn{1}{r}{$\phantom{0}\phantom{-}1.521$} & \multicolumn{1}{r}{$\phantom{0}\mathbf{-0.200}$} & \multicolumn{1}{r}{$\phantom{-}13.542$} & \multicolumn{1}{r}{$\phantom{-}13.387$} & \multicolumn{1}{r}{$\phantom{0}\phantom{-}6.673$} & \multicolumn{1}{r}{$\phantom{0}\phantom{-}6.403$} \\
			 & \multicolumn{1}{r}{iEBW}  & \multicolumn{1}{r}{$\phantom{0}\phantom{-}\mathbf{5.239}$} & \multicolumn{1}{r}{$\phantom{0}\mathbf{-5.010}$} & \multicolumn{1}{r}{$\phantom{0}\phantom{-}5.372$} & \multicolumn{1}{r}{$\phantom{0}\phantom{-}5.204$} & \multicolumn{1}{r}{$\phantom{0}\phantom{-}\mathbf{1.790}$} & \multicolumn{1}{r}{$\phantom{0}\phantom{-}\mathbf{0.005}$} & \multicolumn{1}{r}{$\phantom{0}\phantom{-}\mathbf{1.338}$} & \multicolumn{1}{r}{$\phantom{0}-0.496$} & \multicolumn{1}{r}{$\phantom{0}\phantom{-}\mathbf{9.607}$} & \multicolumn{1}{r}{$\phantom{0}\phantom{-}\mathbf{9.475}$} & \multicolumn{1}{r}{$\phantom{0}\phantom{-}\mathbf{5.091}$} & \multicolumn{1}{r}{$\phantom{0}\phantom{-}\mathbf{4.841}$} \\
			 & \multicolumn{1}{r}{KCB}  & \multicolumn{1}{r}{$\phantom{0}\phantom{-}7.305$} & \multicolumn{1}{r}{$\phantom{0}-6.380$} & \multicolumn{1}{r}{$\phantom{0}\phantom{-}\mathbf{4.015}$} & \multicolumn{1}{r}{$\phantom{0}\phantom{-}\mathbf{3.413}$} & \multicolumn{1}{r}{$\phantom{0}\phantom{-}3.025$} & \multicolumn{1}{r}{$\phantom{0}\phantom{-}0.789$} & \multicolumn{1}{r}{$\phantom{0}\phantom{-}1.463$} & \multicolumn{1}{r}{$\phantom{0}-0.446$} & \multicolumn{1}{r}{$\phantom{-}15.210$} & \multicolumn{1}{r}{$\phantom{-}14.643$} & \multicolumn{1}{r}{$\phantom{0}\phantom{-}7.070$} & \multicolumn{1}{r}{$\phantom{0}\phantom{-}6.549$} \\
			 & \multicolumn{1}{r}{IPW}  & \multicolumn{1}{r}{$\phantom{-}21.462$} & \multicolumn{1}{r}{$-21.188$} & \multicolumn{1}{r}{$\phantom{-}23.371$} & \multicolumn{1}{r}{$\phantom{-}22.827$} & \multicolumn{1}{r}{$\phantom{0}\phantom{-}8.083$} & \multicolumn{1}{r}{$\phantom{0}\phantom{-}0.930$} & \multicolumn{1}{r}{$\phantom{0}\phantom{-}7.864$} & \multicolumn{1}{r}{$\phantom{0}\phantom{-}6.766$} & \multicolumn{1}{r}{$\phantom{-}35.114$} & \multicolumn{1}{r}{$\phantom{-}34.971$} & \multicolumn{1}{r}{$\phantom{-}12.731$} & \multicolumn{1}{r}{$\phantom{-}12.028$} \\
			 & \multicolumn{1}{r}{CBPS}  & \multicolumn{1}{r}{$\phantom{-}21.442$} & \multicolumn{1}{r}{$-21.174$} & \multicolumn{1}{r}{$\phantom{-}23.194$} & \multicolumn{1}{r}{$\phantom{-}22.688$} & \multicolumn{1}{r}{$\phantom{0}\phantom{-}4.324$} & \multicolumn{1}{r}{$\phantom{0}\phantom{-}0.612$} & \multicolumn{1}{r}{$\phantom{0}\phantom{-}7.822$} & \multicolumn{1}{r}{$\phantom{0}\phantom{-}6.759$} & \multicolumn{1}{r}{$\phantom{-}35.083$} & \multicolumn{1}{r}{$\phantom{-}34.942$} & \multicolumn{1}{r}{$\phantom{-}12.655$} & \multicolumn{1}{r}{$\phantom{-}11.994$} \\
			 & \multicolumn{1}{r}{Cal}  & \multicolumn{1}{r}{$\phantom{-}21.459$} & \multicolumn{1}{r}{$-21.184$} & \multicolumn{1}{r}{$\phantom{-}23.129$} & \multicolumn{1}{r}{$\phantom{-}22.597$} & \multicolumn{1}{r}{$\phantom{0}\phantom{-}3.567$} & \multicolumn{1}{r}{$\phantom{0}\phantom{-}0.744$} & \multicolumn{1}{r}{$\phantom{0}\phantom{-}7.854$} & \multicolumn{1}{r}{$\phantom{0}\phantom{-}6.764$} & \multicolumn{1}{r}{$\phantom{-}35.033$} & \multicolumn{1}{r}{$\phantom{-}34.892$} & \multicolumn{1}{r}{$\phantom{-}12.791$} & \multicolumn{1}{r}{$\phantom{-}12.112$} \\
			 \midrule
			\multicolumn{1}{r}{$Y$ Model: B} & \multicolumn{1}{r}{Unweighted}  & \multicolumn{1}{r}{$\phantom{-}13.678$} & \multicolumn{1}{r}{$\phantom{0}-0.377$} & \multicolumn{1}{r}{$\phantom{-}30.446$} & \multicolumn{1}{r}{$\phantom{0}\phantom{-}0.861$} & \multicolumn{1}{r}{$\phantom{-}29.859$} & \multicolumn{1}{r}{$-24.281$} & \multicolumn{1}{r}{$\phantom{-}17.670$} & \multicolumn{1}{r}{$\phantom{0}\phantom{-}0.870$} & \multicolumn{1}{r}{$\phantom{-}17.529$} & \multicolumn{1}{r}{$\phantom{0}\phantom{-}0.474$} & \multicolumn{1}{r}{$\phantom{-}18.164$} & \multicolumn{1}{r}{$\phantom{0}-0.855$} \\
			 & \multicolumn{1}{r}{EBW}  & \multicolumn{1}{r}{$\phantom{0}\phantom{-}8.892$} & \multicolumn{1}{r}{$\phantom{0}-0.263$} & \multicolumn{1}{r}{$\phantom{0}\phantom{-}9.596$} & \multicolumn{1}{r}{$\phantom{0}\phantom{-}0.520$} & \multicolumn{1}{r}{$\phantom{-}28.707$} & \multicolumn{1}{r}{$-22.923$} & \multicolumn{1}{r}{$\phantom{-}11.779$} & \multicolumn{1}{r}{$\phantom{0}\phantom{-}0.996$} & \multicolumn{1}{r}{$\phantom{0}\phantom{-}8.679$} & \multicolumn{1}{r}{$\phantom{0}\phantom{-}0.237$} & \multicolumn{1}{r}{$\phantom{-}12.226$} & \multicolumn{1}{r}{$\phantom{0}-0.717$} \\
			 & \multicolumn{1}{r}{iEBW}  & \multicolumn{1}{r}{$\phantom{0}\phantom{-}\mathbf{4.428}$} & \multicolumn{1}{r}{$\phantom{0}-0.198$} & \multicolumn{1}{r}{$\phantom{0}\phantom{-}\mathbf{5.740}$} & \multicolumn{1}{r}{$\phantom{0}\phantom{-}\mathbf{0.419}$} & \multicolumn{1}{r}{$\phantom{-}23.214$} & \multicolumn{1}{r}{$-18.442$} & \multicolumn{1}{r}{$\phantom{0}\phantom{-}\mathbf{8.824}$} & \multicolumn{1}{r}{$\phantom{0}\phantom{-}0.790$} & \multicolumn{1}{r}{$\phantom{0}\phantom{-}4.077$} & \multicolumn{1}{r}{$\phantom{0}\phantom{-}0.109$} & \multicolumn{1}{r}{$\phantom{0}\phantom{-}\mathbf{9.191}$} & \multicolumn{1}{r}{$\phantom{0}-0.376$} \\
			 & \multicolumn{1}{r}{KCB}  & \multicolumn{1}{r}{$\phantom{0}\phantom{-}8.965$} & \multicolumn{1}{r}{$\phantom{0}-0.267$} & \multicolumn{1}{r}{$\phantom{0}\phantom{-}9.334$} & \multicolumn{1}{r}{$\phantom{0}\phantom{-}0.446$} & \multicolumn{1}{r}{$\phantom{-}17.595$} & \multicolumn{1}{r}{$-14.202$} & \multicolumn{1}{r}{$\phantom{0}\phantom{-}9.744$} & \multicolumn{1}{r}{$\phantom{0}\phantom{-}\mathbf{0.567}$} & \multicolumn{1}{r}{$\phantom{0}\phantom{-}8.945$} & \multicolumn{1}{r}{$\phantom{0}\phantom{-}0.278$} & \multicolumn{1}{r}{$\phantom{0}\phantom{-}9.809$} & \multicolumn{1}{r}{$\phantom{0}\mathbf{-0.281}$} \\
			 & \multicolumn{1}{r}{IPW}  & \multicolumn{1}{r}{$\phantom{0}\phantom{-}8.979$} & \multicolumn{1}{r}{$\phantom{0}-0.238$} & \multicolumn{1}{r}{$\phantom{-}12.352$} & \multicolumn{1}{r}{$\phantom{0}\phantom{-}0.817$} & \multicolumn{1}{r}{$\phantom{-}\mathbf{15.395}$} & \multicolumn{1}{r}{$\mathbf{-11.546}$} & \multicolumn{1}{r}{$\phantom{-}17.420$} & \multicolumn{1}{r}{$\phantom{0}\phantom{-}0.972$} & \multicolumn{1}{r}{$\phantom{0}\phantom{-}8.962$} & \multicolumn{1}{r}{$\phantom{0}\phantom{-}0.265$} & \multicolumn{1}{r}{$\phantom{-}72.360$} & \multicolumn{1}{r}{$\phantom{0}-4.526$} \\
			 & \multicolumn{1}{r}{CBPS}  & \multicolumn{1}{r}{$\phantom{0}\phantom{-}8.929$} & \multicolumn{1}{r}{$\phantom{0}-0.241$} & \multicolumn{1}{r}{$\phantom{-}10.603$} & \multicolumn{1}{r}{$\phantom{0}\phantom{-}0.934$} & \multicolumn{1}{r}{$\phantom{-}23.757$} & \multicolumn{1}{r}{$-13.231$} & \multicolumn{1}{r}{$\phantom{-}13.973$} & \multicolumn{1}{r}{$\phantom{0}\phantom{-}0.830$} & \multicolumn{1}{r}{$\phantom{0}\phantom{-}9.074$} & \multicolumn{1}{r}{$\phantom{0}\phantom{-}0.281$} & \multicolumn{1}{r}{$\phantom{-}14.939$} & \multicolumn{1}{r}{$\phantom{0}-0.602$} \\
			 & \multicolumn{1}{r}{Cal}  & \multicolumn{1}{r}{$\phantom{0}\phantom{-}4.643$} & \multicolumn{1}{r}{$\phantom{0}\mathbf{-0.191}$} & \multicolumn{1}{r}{$\phantom{0}\phantom{-}6.762$} & \multicolumn{1}{r}{$\phantom{0}\phantom{-}1.027$} & \multicolumn{1}{r}{$\phantom{-}44.189$} & \multicolumn{1}{r}{$-31.440$} & \multicolumn{1}{r}{$\phantom{-}18.345$} & \multicolumn{1}{r}{$\phantom{0}\phantom{-}1.138$} & \multicolumn{1}{r}{$\phantom{0}\phantom{-}\mathbf{1.702}$} & \multicolumn{1}{r}{$\phantom{0}\phantom{-}\mathbf{0.100}$} & \multicolumn{1}{r}{$\phantom{-}14.375$} & \multicolumn{1}{r}{$\phantom{0}\phantom{-}0.580$} \\
			 \midrule
			\multicolumn{1}{r}{$Y$ Model: C} & \multicolumn{1}{r}{Unweighted}  & \multicolumn{1}{r}{$\phantom{0}\phantom{-}5.406$} & \multicolumn{1}{r}{$\phantom{0}\phantom{-}5.366$} & \multicolumn{1}{r}{$\phantom{0}\phantom{-}5.584$} & \multicolumn{1}{r}{$\phantom{0}-5.203$} & \multicolumn{1}{r}{$\phantom{0}\phantom{-}1.239$} & \multicolumn{1}{r}{$\phantom{0}\phantom{-}0.179$} & \multicolumn{1}{r}{$\phantom{0}\phantom{-}2.493$} & \multicolumn{1}{r}{$\phantom{0}-2.117$} & \multicolumn{1}{r}{$\phantom{0}\phantom{-}6.253$} & \multicolumn{1}{r}{$\phantom{0}-6.099$} & \multicolumn{1}{r}{$\phantom{0}\phantom{-}1.381$} & \multicolumn{1}{r}{$\phantom{0}-0.466$} \\
			 & \multicolumn{1}{r}{EBW}  & \multicolumn{1}{r}{$\phantom{0}\phantom{-}3.666$} & \multicolumn{1}{r}{$\phantom{0}\phantom{-}3.606$} & \multicolumn{1}{r}{$\phantom{0}\phantom{-}1.243$} & \multicolumn{1}{r}{$\phantom{0}-0.802$} & \multicolumn{1}{r}{$\phantom{0}\phantom{-}\mathbf{1.385}$} & \multicolumn{1}{r}{$\phantom{0}\phantom{-}1.013$} & \multicolumn{1}{r}{$\phantom{0}\phantom{-}0.906$} & \multicolumn{1}{r}{$\phantom{0}-0.195$} & \multicolumn{1}{r}{$\phantom{0}\phantom{-}\mathbf{0.993}$} & \multicolumn{1}{r}{$\phantom{0}\mathbf{-0.527}$} & \multicolumn{1}{r}{$\phantom{0}\phantom{-}\mathbf{0.964}$} & \multicolumn{1}{r}{$\phantom{0}\mathbf{-0.003}$} \\
			 & \multicolumn{1}{r}{iEBW}  & \multicolumn{1}{r}{$\phantom{0}\phantom{-}3.784$} & \multicolumn{1}{r}{$\phantom{0}\phantom{-}3.727$} & \multicolumn{1}{r}{$\phantom{0}\phantom{-}\mathbf{1.030}$} & \multicolumn{1}{r}{$\phantom{0}\mathbf{-0.497}$} & \multicolumn{1}{r}{$\phantom{0}\phantom{-}1.991$} & \multicolumn{1}{r}{$\phantom{0}\phantom{-}1.787$} & \multicolumn{1}{r}{$\phantom{0}\phantom{-}\mathbf{0.888}$} & \multicolumn{1}{r}{$\phantom{0}\phantom{-}0.254$} & \multicolumn{1}{r}{$\phantom{0}\phantom{-}1.280$} & \multicolumn{1}{r}{$\phantom{0}\phantom{-}1.107$} & \multicolumn{1}{r}{$\phantom{0}\phantom{-}1.103$} & \multicolumn{1}{r}{$\phantom{0}\phantom{-}0.588$} \\
			 & \multicolumn{1}{r}{KCB}  & \multicolumn{1}{r}{$\phantom{0}\phantom{-}\mathbf{3.313}$} & \multicolumn{1}{r}{$\phantom{0}\phantom{-}\mathbf{3.145}$} & \multicolumn{1}{r}{$\phantom{0}\phantom{-}1.159$} & \multicolumn{1}{r}{$\phantom{0}-0.548$} & \multicolumn{1}{r}{$\phantom{0}\phantom{-}1.732$} & \multicolumn{1}{r}{$\phantom{0}\phantom{-}1.329$} & \multicolumn{1}{r}{$\phantom{0}\phantom{-}0.894$} & \multicolumn{1}{r}{$\phantom{0}\mathbf{-0.078}$} & \multicolumn{1}{r}{$\phantom{0}\phantom{-}1.224$} & \multicolumn{1}{r}{$\phantom{0}-0.786$} & \multicolumn{1}{r}{$\phantom{0}\phantom{-}1.039$} & \multicolumn{1}{r}{$\phantom{0}\phantom{-}0.071$} \\
			 & \multicolumn{1}{r}{IPW}  & \multicolumn{1}{r}{$\phantom{0}\phantom{-}5.406$} & \multicolumn{1}{r}{$\phantom{0}\phantom{-}5.376$} & \multicolumn{1}{r}{$\phantom{0}\phantom{-}5.649$} & \multicolumn{1}{r}{$\phantom{0}-5.214$} & \multicolumn{1}{r}{$\phantom{0}\phantom{-}2.415$} & \multicolumn{1}{r}{$\phantom{0}\phantom{-}1.247$} & \multicolumn{1}{r}{$\phantom{0}\phantom{-}2.524$} & \multicolumn{1}{r}{$\phantom{0}-2.148$} & \multicolumn{1}{r}{$\phantom{0}\phantom{-}6.239$} & \multicolumn{1}{r}{$\phantom{0}-6.082$} & \multicolumn{1}{r}{$\phantom{0}\phantom{-}1.670$} & \multicolumn{1}{r}{$\phantom{0}-1.042$} \\
			 & \multicolumn{1}{r}{CBPS}  & \multicolumn{1}{r}{$\phantom{0}\phantom{-}5.410$} & \multicolumn{1}{r}{$\phantom{0}\phantom{-}5.381$} & \multicolumn{1}{r}{$\phantom{0}\phantom{-}5.530$} & \multicolumn{1}{r}{$\phantom{0}-5.130$} & \multicolumn{1}{r}{$\phantom{0}\phantom{-}2.145$} & \multicolumn{1}{r}{$\phantom{0}\phantom{-}1.692$} & \multicolumn{1}{r}{$\phantom{0}\phantom{-}2.491$} & \multicolumn{1}{r}{$\phantom{0}-2.123$} & \multicolumn{1}{r}{$\phantom{0}\phantom{-}6.215$} & \multicolumn{1}{r}{$\phantom{0}-6.063$} & \multicolumn{1}{r}{$\phantom{0}\phantom{-}1.532$} & \multicolumn{1}{r}{$\phantom{0}-0.850$} \\
			 & \multicolumn{1}{r}{Cal}  & \multicolumn{1}{r}{$\phantom{0}\phantom{-}5.412$} & \multicolumn{1}{r}{$\phantom{0}\phantom{-}5.383$} & \multicolumn{1}{r}{$\phantom{0}\phantom{-}5.509$} & \multicolumn{1}{r}{$\phantom{0}-5.087$} & \multicolumn{1}{r}{$\phantom{0}\phantom{-}1.387$} & \multicolumn{1}{r}{$\phantom{0}\phantom{-}\mathbf{0.731}$} & \multicolumn{1}{r}{$\phantom{0}\phantom{-}2.486$} & \multicolumn{1}{r}{$\phantom{0}-2.106$} & \multicolumn{1}{r}{$\phantom{0}\phantom{-}6.193$} & \multicolumn{1}{r}{$\phantom{0}-6.040$} & \multicolumn{1}{r}{$\phantom{0}\phantom{-}1.638$} & \multicolumn{1}{r}{$\phantom{0}-0.983$} \\
			 \midrule
			\multicolumn{1}{r}{$Y$ Model: D} & \multicolumn{1}{r}{Unweighted}  & \multicolumn{1}{r}{$\phantom{0}\phantom{-}1.193$} & \multicolumn{1}{r}{$\phantom{0}-0.456$} & \multicolumn{1}{r}{$\phantom{0}\phantom{-}1.827$} & \multicolumn{1}{r}{$\phantom{0}\phantom{-}1.077$} & \multicolumn{1}{r}{$\phantom{0}\phantom{-}8.525$} & \multicolumn{1}{r}{$\phantom{0}-8.456$} & \multicolumn{1}{r}{$\phantom{0}\phantom{-}1.226$} & \multicolumn{1}{r}{$\phantom{0}-0.358$} & \multicolumn{1}{r}{$\phantom{0}\phantom{-}1.386$} & \multicolumn{1}{r}{$\phantom{0}\phantom{-}0.708$} & \multicolumn{1}{r}{$\phantom{0}\phantom{-}\mathbf{1.401}$} & \multicolumn{1}{r}{$\phantom{0}\phantom{-}\mathbf{0.724}$} \\
			 & \multicolumn{1}{r}{EBW}  & \multicolumn{1}{r}{$\phantom{0}\phantom{-}0.543$} & \multicolumn{1}{r}{$\phantom{0}\phantom{-}\mathbf{0.039}$} & \multicolumn{1}{r}{$\phantom{0}\phantom{-}0.575$} & \multicolumn{1}{r}{$\phantom{0}\phantom{-}0.192$} & \multicolumn{1}{r}{$\phantom{0}\phantom{-}3.627$} & \multicolumn{1}{r}{$\phantom{0}-3.411$} & \multicolumn{1}{r}{$\phantom{0}\phantom{-}0.577$} & \multicolumn{1}{r}{$\phantom{0}-0.336$} & \multicolumn{1}{r}{$\phantom{0}\phantom{-}0.563$} & \multicolumn{1}{r}{$\phantom{0}-0.186$} & \multicolumn{1}{r}{$\phantom{0}\phantom{-}1.773$} & \multicolumn{1}{r}{$\phantom{0}\phantom{-}1.696$} \\
			 & \multicolumn{1}{r}{iEBW}  & \multicolumn{1}{r}{$\phantom{0}\phantom{-}\mathbf{0.450}$} & \multicolumn{1}{r}{$\phantom{0}\phantom{-}0.066$} & \multicolumn{1}{r}{$\phantom{0}\phantom{-}0.478$} & \multicolumn{1}{r}{$\phantom{0}\phantom{-}0.134$} & \multicolumn{1}{r}{$\phantom{0}\phantom{-}2.908$} & \multicolumn{1}{r}{$\phantom{0}-2.693$} & \multicolumn{1}{r}{$\phantom{0}\phantom{-}0.487$} & \multicolumn{1}{r}{$\phantom{0}-0.264$} & \multicolumn{1}{r}{$\phantom{0}\phantom{-}\mathbf{0.491}$} & \multicolumn{1}{r}{$\phantom{0}\mathbf{-0.180}$} & \multicolumn{1}{r}{$\phantom{0}\phantom{-}1.519$} & \multicolumn{1}{r}{$\phantom{0}\phantom{-}1.445$} \\
			 & \multicolumn{1}{r}{KCB}  & \multicolumn{1}{r}{$\phantom{0}\phantom{-}0.530$} & \multicolumn{1}{r}{$\phantom{0}-0.127$} & \multicolumn{1}{r}{$\phantom{0}\phantom{-}\mathbf{0.467}$} & \multicolumn{1}{r}{$\phantom{0}\phantom{-}\mathbf{0.128}$} & \multicolumn{1}{r}{$\phantom{0}\phantom{-}2.667$} & \multicolumn{1}{r}{$\phantom{0}-2.223$} & \multicolumn{1}{r}{$\phantom{0}\phantom{-}\mathbf{0.430}$} & \multicolumn{1}{r}{$\phantom{0}\mathbf{-0.184}$} & \multicolumn{1}{r}{$\phantom{0}\phantom{-}0.615$} & \multicolumn{1}{r}{$\phantom{0}\phantom{-}0.209$} & \multicolumn{1}{r}{$\phantom{0}\phantom{-}1.611$} & \multicolumn{1}{r}{$\phantom{0}\phantom{-}1.511$} \\
			 & \multicolumn{1}{r}{IPW}  & \multicolumn{1}{r}{$\phantom{0}\phantom{-}0.896$} & \multicolumn{1}{r}{$\phantom{0}-0.505$} & \multicolumn{1}{r}{$\phantom{0}\phantom{-}1.457$} & \multicolumn{1}{r}{$\phantom{0}\phantom{-}1.096$} & \multicolumn{1}{r}{$\phantom{0}\phantom{-}3.144$} & \multicolumn{1}{r}{$\phantom{0}-2.059$} & \multicolumn{1}{r}{$\phantom{0}\phantom{-}0.799$} & \multicolumn{1}{r}{$\phantom{0}-0.359$} & \multicolumn{1}{r}{$\phantom{0}\phantom{-}0.986$} & \multicolumn{1}{r}{$\phantom{0}\phantom{-}0.680$} & \multicolumn{1}{r}{$\phantom{0}\phantom{-}2.325$} & \multicolumn{1}{r}{$\phantom{0}\phantom{-}2.201$} \\
			 & \multicolumn{1}{r}{CBPS}  & \multicolumn{1}{r}{$\phantom{0}\phantom{-}0.978$} & \multicolumn{1}{r}{$\phantom{0}-0.505$} & \multicolumn{1}{r}{$\phantom{0}\phantom{-}1.477$} & \multicolumn{1}{r}{$\phantom{0}\phantom{-}1.061$} & \multicolumn{1}{r}{$\phantom{0}\phantom{-}3.183$} & \multicolumn{1}{r}{$\phantom{0}-2.829$} & \multicolumn{1}{r}{$\phantom{0}\phantom{-}0.893$} & \multicolumn{1}{r}{$\phantom{0}-0.346$} & \multicolumn{1}{r}{$\phantom{0}\phantom{-}1.103$} & \multicolumn{1}{r}{$\phantom{0}\phantom{-}0.687$} & \multicolumn{1}{r}{$\phantom{0}\phantom{-}1.958$} & \multicolumn{1}{r}{$\phantom{0}\phantom{-}1.765$} \\
			 & \multicolumn{1}{r}{Cal}  & \multicolumn{1}{r}{$\phantom{0}\phantom{-}0.880$} & \multicolumn{1}{r}{$\phantom{0}-0.524$} & \multicolumn{1}{r}{$\phantom{0}\phantom{-}1.344$} & \multicolumn{1}{r}{$\phantom{0}\phantom{-}1.065$} & \multicolumn{1}{r}{$\phantom{0}\phantom{-}\mathbf{2.096}$} & \multicolumn{1}{r}{$\phantom{0}\mathbf{-1.378}$} & \multicolumn{1}{r}{$\phantom{0}\phantom{-}0.785$} & \multicolumn{1}{r}{$\phantom{0}-0.354$} & \multicolumn{1}{r}{$\phantom{0}\phantom{-}0.985$} & \multicolumn{1}{r}{$\phantom{0}\phantom{-}0.687$} & \multicolumn{1}{r}{$\phantom{0}\phantom{-}2.303$} & \multicolumn{1}{r}{$\phantom{0}\phantom{-}2.195$} \\
			 \midrule
			\multicolumn{1}{r}{$Y$ Model: E} & \multicolumn{1}{r}{Unweighted}  & \multicolumn{1}{r}{$\phantom{0}\phantom{-}\mathbf{2.463}$} & \multicolumn{1}{r}{$\phantom{0}\phantom{-}\mathbf{0.004}$} & \multicolumn{1}{r}{$\phantom{0}\phantom{-}3.209$} & \multicolumn{1}{r}{$\phantom{0}\mathbf{-0.039}$} & \multicolumn{1}{r}{$\phantom{0}\phantom{-}4.105$} & \multicolumn{1}{r}{$\phantom{0}-3.469$} & \multicolumn{1}{r}{$\phantom{0}\phantom{-}\mathbf{2.013}$} & \multicolumn{1}{r}{$\phantom{0}-0.029$} & \multicolumn{1}{r}{$\phantom{0}\phantom{-}2.466$} & \multicolumn{1}{r}{$\phantom{0}\mathbf{-0.011}$} & \multicolumn{1}{r}{$\phantom{0}\phantom{-}2.458$} & \multicolumn{1}{r}{$\phantom{0}\mathbf{-0.103}$} \\
			 & \multicolumn{1}{r}{EBW}  & \multicolumn{1}{r}{$\phantom{0}\phantom{-}2.764$} & \multicolumn{1}{r}{$\phantom{0}\phantom{-}0.811$} & \multicolumn{1}{r}{$\phantom{0}\phantom{-}\mathbf{2.469}$} & \multicolumn{1}{r}{$\phantom{0}-0.694$} & \multicolumn{1}{r}{$\phantom{0}\phantom{-}\mathbf{2.517}$} & \multicolumn{1}{r}{$\phantom{0}\phantom{-}0.176$} & \multicolumn{1}{r}{$\phantom{0}\phantom{-}2.130$} & \multicolumn{1}{r}{$\phantom{0}\phantom{-}0.118$} & \multicolumn{1}{r}{$\phantom{0}\phantom{-}\mathbf{2.300}$} & \multicolumn{1}{r}{$\phantom{0}-0.079$} & \multicolumn{1}{r}{$\phantom{0}\phantom{-}\mathbf{2.247}$} & \multicolumn{1}{r}{$\phantom{0}\phantom{-}0.462$} \\
			 & \multicolumn{1}{r}{iEBW}  & \multicolumn{1}{r}{$\phantom{0}\phantom{-}2.868$} & \multicolumn{1}{r}{$\phantom{0}\phantom{-}0.975$} & \multicolumn{1}{r}{$\phantom{0}\phantom{-}2.504$} & \multicolumn{1}{r}{$\phantom{0}-0.606$} & \multicolumn{1}{r}{$\phantom{0}\phantom{-}2.776$} & \multicolumn{1}{r}{$\phantom{0}\phantom{-}0.594$} & \multicolumn{1}{r}{$\phantom{0}\phantom{-}2.187$} & \multicolumn{1}{r}{$\phantom{0}\phantom{-}0.188$} & \multicolumn{1}{r}{$\phantom{0}\phantom{-}2.376$} & \multicolumn{1}{r}{$\phantom{0}\phantom{-}0.130$} & \multicolumn{1}{r}{$\phantom{0}\phantom{-}2.375$} & \multicolumn{1}{r}{$\phantom{0}\phantom{-}0.858$} \\
			 & \multicolumn{1}{r}{KCB}  & \multicolumn{1}{r}{$\phantom{0}\phantom{-}2.846$} & \multicolumn{1}{r}{$\phantom{0}\phantom{-}0.757$} & \multicolumn{1}{r}{$\phantom{0}\phantom{-}2.493$} & \multicolumn{1}{r}{$\phantom{0}-0.480$} & \multicolumn{1}{r}{$\phantom{0}\phantom{-}2.812$} & \multicolumn{1}{r}{$\phantom{0}\mathbf{-0.013}$} & \multicolumn{1}{r}{$\phantom{0}\phantom{-}2.215$} & \multicolumn{1}{r}{$\phantom{0}\phantom{-}0.161$} & \multicolumn{1}{r}{$\phantom{0}\phantom{-}2.351$} & \multicolumn{1}{r}{$\phantom{0}\phantom{-}0.244$} & \multicolumn{1}{r}{$\phantom{0}\phantom{-}2.382$} & \multicolumn{1}{r}{$\phantom{0}\phantom{-}0.389$} \\
			 & \multicolumn{1}{r}{IPW}  & \multicolumn{1}{r}{$\phantom{0}\phantom{-}3.112$} & \multicolumn{1}{r}{$\phantom{0}\phantom{-}0.395$} & \multicolumn{1}{r}{$\phantom{0}\phantom{-}3.454$} & \multicolumn{1}{r}{$\phantom{0}-1.957$} & \multicolumn{1}{r}{$\phantom{0}\phantom{-}5.277$} & \multicolumn{1}{r}{$\phantom{0}\phantom{-}0.157$} & \multicolumn{1}{r}{$\phantom{0}\phantom{-}2.928$} & \multicolumn{1}{r}{$\phantom{0}\phantom{-}0.035$} & \multicolumn{1}{r}{$\phantom{0}\phantom{-}3.962$} & \multicolumn{1}{r}{$\phantom{0}-1.895$} & \multicolumn{1}{r}{$\phantom{0}\phantom{-}4.606$} & \multicolumn{1}{r}{$\phantom{0}\phantom{-}0.267$} \\
			 & \multicolumn{1}{r}{CBPS}  & \multicolumn{1}{r}{$\phantom{0}\phantom{-}2.817$} & \multicolumn{1}{r}{$\phantom{0}\phantom{-}0.358$} & \multicolumn{1}{r}{$\phantom{0}\phantom{-}3.002$} & \multicolumn{1}{r}{$\phantom{0}-1.450$} & \multicolumn{1}{r}{$\phantom{0}\phantom{-}3.317$} & \multicolumn{1}{r}{$\phantom{0}-0.517$} & \multicolumn{1}{r}{$\phantom{0}\phantom{-}2.209$} & \multicolumn{1}{r}{$\phantom{0}\mathbf{-0.010}$} & \multicolumn{1}{r}{$\phantom{0}\phantom{-}3.393$} & \multicolumn{1}{r}{$\phantom{0}-1.813$} & \multicolumn{1}{r}{$\phantom{0}\phantom{-}2.473$} & \multicolumn{1}{r}{$\phantom{0}\phantom{-}0.265$} \\
			 & \multicolumn{1}{r}{Cal}  & \multicolumn{1}{r}{$\phantom{0}\phantom{-}3.786$} & \multicolumn{1}{r}{$\phantom{0}\phantom{-}1.279$} & \multicolumn{1}{r}{$\phantom{0}\phantom{-}2.811$} & \multicolumn{1}{r}{$\phantom{0}-1.389$} & \multicolumn{1}{r}{$\phantom{0}\phantom{-}3.482$} & \multicolumn{1}{r}{$\phantom{0}\phantom{-}1.680$} & \multicolumn{1}{r}{$\phantom{0}\phantom{-}2.201$} & \multicolumn{1}{r}{$\phantom{0}\phantom{-}0.265$} & \multicolumn{1}{r}{$\phantom{0}\phantom{-}2.951$} & \multicolumn{1}{r}{$\phantom{0}-0.620$} & \multicolumn{1}{r}{$\phantom{0}\phantom{-}2.336$} & \multicolumn{1}{r}{$\phantom{0}\phantom{-}0.384$} \\
			\bottomrule
		\end{tabular}%
	}
\caption{Displayed are results for $n=250$ and $p=10$ averaged over 1000 independent simulated datasets. The average proportion of those treated in propensity models I, II, III, IV, V, and VI are 0.35, 0.31, 0.50, 0.51, 0.51, and 0.50, respectively.}
\label{tab:simulation_results_n250_p10}
\end{table}

\begin{table}[ht]
\centering
\resizebox{1\textwidth}{!}{
		\begin{tabular}{crccRcRcRcRcRc}
			\toprule
			& Propensity Model: & \multicolumn{2}{c}{I} & \multicolumn{2}{c}{II} & \multicolumn{2}{c}{III} & \multicolumn{2}{c}{IV} & \multicolumn{2}{c}{V} & \multicolumn{2}{c}{VI} \\ 
			\cmidrule{3-4} \cmidrule{5-6} \cmidrule{7-8} \cmidrule{9-10} \cmidrule{11-12} \cmidrule{13-14}  
			& Method & RMSE & Bias & RMSE & Bias & RMSE & Bias & RMSE & Bias & RMSE & Bias & RMSE & Bias  \\ 
			\midrule
			\multicolumn{1}{r}{$Y$ Model: A} & \multicolumn{1}{r}{Unweighted}  & \multicolumn{1}{r}{$\phantom{0}\phantom{-}21.203$} & \multicolumn{1}{r}{$\phantom{0}-20.951$} & \multicolumn{1}{r}{$\phantom{0}\phantom{-}23.159$} & \multicolumn{1}{r}{$\phantom{0}\phantom{-}22.709$} & \multicolumn{1}{r}{$\phantom{00}\phantom{-}3.845$} & \multicolumn{1}{r}{$\phantom{00}\mathbf{-0.036}$} & \multicolumn{1}{r}{$\phantom{00}\phantom{-}7.455$} & \multicolumn{1}{r}{$\phantom{00}\phantom{-}6.419$} & \multicolumn{1}{r}{$\phantom{0}\phantom{-}34.913$} & \multicolumn{1}{r}{$\phantom{0}\phantom{-}34.764$} & \multicolumn{1}{r}{$\phantom{0}\phantom{-}12.252$} & \multicolumn{1}{r}{$\phantom{0}\phantom{-}11.627$} \\
			 & \multicolumn{1}{r}{EBW}  & \multicolumn{1}{r}{$\phantom{0}\phantom{-}15.116$} & \multicolumn{1}{r}{$\phantom{0}-14.852$} & \multicolumn{1}{r}{$\phantom{0}\phantom{-}14.214$} & \multicolumn{1}{r}{$\phantom{0}\phantom{-}13.906$} & \multicolumn{1}{r}{$\phantom{00}\phantom{-}2.789$} & \multicolumn{1}{r}{$\phantom{00}-0.388$} & \multicolumn{1}{r}{$\phantom{00}\phantom{-}3.026$} & \multicolumn{1}{r}{$\phantom{00}\phantom{-}1.630$} & \multicolumn{1}{r}{$\phantom{0}\phantom{-}24.808$} & \multicolumn{1}{r}{$\phantom{0}\phantom{-}24.663$} & \multicolumn{1}{r}{$\phantom{0}\phantom{-}11.005$} & \multicolumn{1}{r}{$\phantom{0}\phantom{-}10.670$} \\
			 & \multicolumn{1}{r}{iEBW}  & \multicolumn{1}{r}{$\phantom{0}\phantom{-}\mathbf{11.547}$} & \multicolumn{1}{r}{$\phantom{0}\mathbf{-11.300}$} & \multicolumn{1}{r}{$\phantom{0}\phantom{-}\mathbf{12.374}$} & \multicolumn{1}{r}{$\phantom{0}\phantom{-}\mathbf{12.093}$} & \multicolumn{1}{r}{$\phantom{00}\phantom{-}\mathbf{2.466}$} & \multicolumn{1}{r}{$\phantom{00}-0.201$} & \multicolumn{1}{r}{$\phantom{00}\phantom{-}\mathbf{2.352}$} & \multicolumn{1}{r}{$\phantom{00}\phantom{-}\mathbf{0.688}$} & \multicolumn{1}{r}{$\phantom{0}\phantom{-}\mathbf{21.365}$} & \multicolumn{1}{r}{$\phantom{0}\phantom{-}\mathbf{21.223}$} & \multicolumn{1}{r}{$\phantom{00}\phantom{-}\mathbf{9.941}$} & \multicolumn{1}{r}{$\phantom{00}\phantom{-}\mathbf{9.642}$} \\
			 & \multicolumn{1}{r}{KCB}  & \multicolumn{1}{r}{$\phantom{0}\phantom{-}15.877$} & \multicolumn{1}{r}{$\phantom{0}-15.534$} & \multicolumn{1}{r}{$\phantom{0}\phantom{-}14.928$} & \multicolumn{1}{r}{$\phantom{0}\phantom{-}14.271$} & \multicolumn{1}{r}{$\phantom{00}\phantom{-}3.036$} & \multicolumn{1}{r}{$\phantom{00}-0.354$} & \multicolumn{1}{r}{$\phantom{00}\phantom{-}3.834$} & \multicolumn{1}{r}{$\phantom{00}\phantom{-}2.348$} & \multicolumn{1}{r}{$\phantom{0}\phantom{-}27.159$} & \multicolumn{1}{r}{$\phantom{0}\phantom{-}26.898$} & \multicolumn{1}{r}{$\phantom{0}\phantom{-}10.764$} & \multicolumn{1}{r}{$\phantom{0}\phantom{-}10.257$} \\
			 & \multicolumn{1}{r}{IPW}  & \multicolumn{1}{r}{$\phantom{0}\phantom{-}21.305$} & \multicolumn{1}{r}{$\phantom{0}-21.014$} & \multicolumn{1}{r}{$\phantom{0}\phantom{-}23.009$} & \multicolumn{1}{r}{$\phantom{0}\phantom{-}22.430$} & \multicolumn{1}{r}{$\phantom{00}\phantom{-}8.154$} & \multicolumn{1}{r}{$\phantom{00}\phantom{-}0.602$} & \multicolumn{1}{r}{$\phantom{00}\phantom{-}7.602$} & \multicolumn{1}{r}{$\phantom{00}\phantom{-}6.382$} & \multicolumn{1}{r}{$\phantom{0}\phantom{-}34.859$} & \multicolumn{1}{r}{$\phantom{0}\phantom{-}34.694$} & \multicolumn{1}{r}{$\phantom{0}\phantom{-}12.644$} & \multicolumn{1}{r}{$\phantom{0}\phantom{-}11.932$} \\
			 & \multicolumn{1}{r}{CBPS}  & \multicolumn{1}{r}{$\phantom{0}\phantom{-}21.327$} & \multicolumn{1}{r}{$\phantom{0}-21.056$} & \multicolumn{1}{r}{$\phantom{0}\phantom{-}22.883$} & \multicolumn{1}{r}{$\phantom{0}\phantom{-}22.385$} & \multicolumn{1}{r}{$\phantom{00}\phantom{-}3.969$} & \multicolumn{1}{r}{$\phantom{00}\phantom{-}0.301$} & \multicolumn{1}{r}{$\phantom{00}\phantom{-}7.538$} & \multicolumn{1}{r}{$\phantom{00}\phantom{-}6.409$} & \multicolumn{1}{r}{$\phantom{0}\phantom{-}34.830$} & \multicolumn{1}{r}{$\phantom{0}\phantom{-}34.675$} & \multicolumn{1}{r}{$\phantom{0}\phantom{-}12.483$} & \multicolumn{1}{r}{$\phantom{0}\phantom{-}11.829$} \\
			 & \multicolumn{1}{r}{Cal}  & \multicolumn{1}{r}{$\phantom{0}\phantom{-}21.324$} & \multicolumn{1}{r}{$\phantom{0}-21.031$} & \multicolumn{1}{r}{$\phantom{0}\phantom{-}22.776$} & \multicolumn{1}{r}{$\phantom{0}\phantom{-}22.183$} & \multicolumn{1}{r}{$\phantom{00}\phantom{-}3.504$} & \multicolumn{1}{r}{$\phantom{00}\phantom{-}0.607$} & \multicolumn{1}{r}{$\phantom{00}\phantom{-}7.596$} & \multicolumn{1}{r}{$\phantom{00}\phantom{-}6.379$} & \multicolumn{1}{r}{$\phantom{0}\phantom{-}34.829$} & \multicolumn{1}{r}{$\phantom{0}\phantom{-}34.669$} & \multicolumn{1}{r}{$\phantom{0}\phantom{-}12.675$} & \multicolumn{1}{r}{$\phantom{0}\phantom{-}11.967$} \\
			 \midrule
			\multicolumn{1}{r}{$Y$ Model: B} & \multicolumn{1}{r}{Unweighted}  & \multicolumn{1}{r}{$\phantom{0}\phantom{-}15.619$} & \multicolumn{1}{r}{$\phantom{00}-0.082$} & \multicolumn{1}{r}{$\phantom{0}\phantom{-}32.211$} & \multicolumn{1}{r}{$\phantom{00}\phantom{-}0.375$} & \multicolumn{1}{r}{$\phantom{0}\phantom{-}30.937$} & \multicolumn{1}{r}{$\phantom{0}-24.117$} & \multicolumn{1}{r}{$\phantom{0}\phantom{-}20.220$} & \multicolumn{1}{r}{$\phantom{00}\phantom{-}0.163$} & \multicolumn{1}{r}{$\phantom{0}\phantom{-}20.323$} & \multicolumn{1}{r}{$\phantom{00}-0.013$} & \multicolumn{1}{r}{$\phantom{0}\phantom{-}18.792$} & \multicolumn{1}{r}{$\phantom{00}-0.765$} \\
			 & \multicolumn{1}{r}{EBW}  & \multicolumn{1}{r}{$\phantom{00}\phantom{-}9.217$} & \multicolumn{1}{r}{$\phantom{00}\phantom{-}\mathbf{0.011}$} & \multicolumn{1}{r}{$\phantom{00}\phantom{-}9.470$} & \multicolumn{1}{r}{$\phantom{00}\phantom{-}\mathbf{0.355}$} & \multicolumn{1}{r}{$\phantom{0}\phantom{-}29.218$} & \multicolumn{1}{r}{$\phantom{0}-22.972$} & \multicolumn{1}{r}{$\phantom{0}\phantom{-}14.260$} & \multicolumn{1}{r}{$\phantom{00}\phantom{-}0.359$} & \multicolumn{1}{r}{$\phantom{00}\phantom{-}8.191$} & \multicolumn{1}{r}{$\phantom{00}-0.199$} & \multicolumn{1}{r}{$\phantom{0}\phantom{-}14.823$} & \multicolumn{1}{r}{$\phantom{00}-0.428$} \\
			 & \multicolumn{1}{r}{iEBW}  & \multicolumn{1}{r}{$\phantom{00}\phantom{-}3.483$} & \multicolumn{1}{r}{$\phantom{00}\phantom{-}0.026$} & \multicolumn{1}{r}{$\phantom{00}\phantom{-}\mathbf{5.725}$} & \multicolumn{1}{r}{$\phantom{00}\phantom{-}0.383$} & \multicolumn{1}{r}{$\phantom{0}\phantom{-}22.282$} & \multicolumn{1}{r}{$\phantom{0}-17.571$} & \multicolumn{1}{r}{$\phantom{0}\phantom{-}11.303$} & \multicolumn{1}{r}{$\phantom{00}\phantom{-}0.389$} & \multicolumn{1}{r}{$\phantom{00}\phantom{-}2.315$} & \multicolumn{1}{r}{$\phantom{00}-0.137$} & \multicolumn{1}{r}{$\phantom{0}\phantom{-}12.119$} & \multicolumn{1}{r}{$\phantom{00}\mathbf{-0.175}$} \\
			 & \multicolumn{1}{r}{KCB}  & \multicolumn{1}{r}{$\phantom{0}\phantom{-}10.093$} & \multicolumn{1}{r}{$\phantom{00}-0.132$} & \multicolumn{1}{r}{$\phantom{0}\phantom{-}11.084$} & \multicolumn{1}{r}{$\phantom{00}\phantom{-}0.775$} & \multicolumn{1}{r}{$\phantom{0}\phantom{-}17.597$} & \multicolumn{1}{r}{$\phantom{0}-14.152$} & \multicolumn{1}{r}{$\phantom{0}\phantom{-}\mathbf{10.572}$} & \multicolumn{1}{r}{$\phantom{00}\phantom{-}0.353$} & \multicolumn{1}{r}{$\phantom{0}\phantom{-}10.082$} & \multicolumn{1}{r}{$\phantom{00}\phantom{-}0.187$} & \multicolumn{1}{r}{$\phantom{0}\phantom{-}\mathbf{11.119}$} & \multicolumn{1}{r}{$\phantom{00}-0.293$} \\
			 & \multicolumn{1}{r}{IPW}  & \multicolumn{1}{r}{$\phantom{00}\phantom{-}9.927$} & \multicolumn{1}{r}{$\phantom{00}-0.097$} & \multicolumn{1}{r}{$\phantom{0}\phantom{-}18.046$} & \multicolumn{1}{r}{$\phantom{00}\phantom{-}0.509$} & \multicolumn{1}{r}{$\phantom{0}\phantom{-}\mathbf{17.506}$} & \multicolumn{1}{r}{$\phantom{0}\mathbf{-11.581}$} & \multicolumn{1}{r}{$\phantom{0}\phantom{-}29.761$} & \multicolumn{1}{r}{$\phantom{00}\phantom{-}0.485$} & \multicolumn{1}{r}{$\phantom{0}\phantom{-}10.180$} & \multicolumn{1}{r}{$\phantom{00}\phantom{-}\mathbf{0.010}$} & \multicolumn{1}{r}{$\phantom{-}128.009$} & \multicolumn{1}{r}{$\phantom{00}-8.725$} \\
			 & \multicolumn{1}{r}{CBPS}  & \multicolumn{1}{r}{$\phantom{0}\phantom{-}10.142$} & \multicolumn{1}{r}{$\phantom{00}-0.096$} & \multicolumn{1}{r}{$\phantom{0}\phantom{-}12.701$} & \multicolumn{1}{r}{$\phantom{00}\phantom{-}0.798$} & \multicolumn{1}{r}{$\phantom{0}\phantom{-}21.586$} & \multicolumn{1}{r}{$\phantom{0}-14.104$} & \multicolumn{1}{r}{$\phantom{0}\phantom{-}16.355$} & \multicolumn{1}{r}{$\phantom{00}\mathbf{-0.025}$} & \multicolumn{1}{r}{$\phantom{0}\phantom{-}12.731$} & \multicolumn{1}{r}{$\phantom{00}\phantom{-}0.085$} & \multicolumn{1}{r}{$\phantom{0}\phantom{-}19.754$} & \multicolumn{1}{r}{$\phantom{00}-0.683$} \\
			 & \multicolumn{1}{r}{Cal}  & \multicolumn{1}{r}{$\phantom{00}\phantom{-}\mathbf{3.075}$} & \multicolumn{1}{r}{$\phantom{00}-0.062$} & \multicolumn{1}{r}{$\phantom{00}\phantom{-}8.836$} & \multicolumn{1}{r}{$\phantom{00}\phantom{-}0.912$} & \multicolumn{1}{r}{$\phantom{0}\phantom{-}39.914$} & \multicolumn{1}{r}{$\phantom{0}-27.307$} & \multicolumn{1}{r}{$\phantom{0}\phantom{-}23.360$} & \multicolumn{1}{r}{$\phantom{00}\phantom{-}0.428$} & \multicolumn{1}{r}{$\phantom{00}\phantom{-}\mathbf{1.333}$} & \multicolumn{1}{r}{$\phantom{00}\phantom{-}0.054$} & \multicolumn{1}{r}{$\phantom{0}\phantom{-}19.856$} & \multicolumn{1}{r}{$\phantom{00}\phantom{-}0.428$} \\
			 \midrule
			\multicolumn{1}{r}{$Y$ Model: C} & \multicolumn{1}{r}{Unweighted}  & \multicolumn{1}{r}{$\phantom{00}\phantom{-}5.404$} & \multicolumn{1}{r}{$\phantom{00}\phantom{-}5.362$} & \multicolumn{1}{r}{$\phantom{00}\phantom{-}5.544$} & \multicolumn{1}{r}{$\phantom{00}-5.171$} & \multicolumn{1}{r}{$\phantom{00}\phantom{-}\mathbf{1.240}$} & \multicolumn{1}{r}{$\phantom{00}\phantom{-}\mathbf{0.246}$} & \multicolumn{1}{r}{$\phantom{00}\phantom{-}2.458$} & \multicolumn{1}{r}{$\phantom{00}-2.060$} & \multicolumn{1}{r}{$\phantom{00}\phantom{-}6.174$} & \multicolumn{1}{r}{$\phantom{00}-6.012$} & \multicolumn{1}{r}{$\phantom{00}\phantom{-}1.367$} & \multicolumn{1}{r}{$\phantom{00}-0.431$} \\
			 & \multicolumn{1}{r}{EBW}  & \multicolumn{1}{r}{$\phantom{00}\phantom{-}\mathbf{4.884}$} & \multicolumn{1}{r}{$\phantom{00}\phantom{-}\mathbf{4.843}$} & \multicolumn{1}{r}{$\phantom{00}\phantom{-}2.687$} & \multicolumn{1}{r}{$\phantom{00}-2.364$} & \multicolumn{1}{r}{$\phantom{00}\phantom{-}1.845$} & \multicolumn{1}{r}{$\phantom{00}\phantom{-}1.605$} & \multicolumn{1}{r}{$\phantom{00}\phantom{-}1.153$} & \multicolumn{1}{r}{$\phantom{00}-0.602$} & \multicolumn{1}{r}{$\phantom{00}\phantom{-}2.456$} & \multicolumn{1}{r}{$\phantom{00}-2.268$} & \multicolumn{1}{r}{$\phantom{00}\phantom{-}1.119$} & \multicolumn{1}{r}{$\phantom{00}-0.366$} \\
			 & \multicolumn{1}{r}{iEBW}  & \multicolumn{1}{r}{$\phantom{00}\phantom{-}4.914$} & \multicolumn{1}{r}{$\phantom{00}\phantom{-}4.876$} & \multicolumn{1}{r}{$\phantom{00}\phantom{-}\mathbf{2.353}$} & \multicolumn{1}{r}{$\phantom{00}\mathbf{-2.016}$} & \multicolumn{1}{r}{$\phantom{00}\phantom{-}2.730$} & \multicolumn{1}{r}{$\phantom{00}\phantom{-}2.601$} & \multicolumn{1}{r}{$\phantom{00}\phantom{-}\mathbf{0.927}$} & \multicolumn{1}{r}{$\phantom{00}\mathbf{-0.075}$} & \multicolumn{1}{r}{$\phantom{00}\phantom{-}\mathbf{1.160}$} & \multicolumn{1}{r}{$\phantom{00}\mathbf{-0.832}$} & \multicolumn{1}{r}{$\phantom{00}\phantom{-}\mathbf{1.016}$} & \multicolumn{1}{r}{$\phantom{00}\phantom{-}0.147$} \\
			 & \multicolumn{1}{r}{KCB}  & \multicolumn{1}{r}{$\phantom{00}\phantom{-}5.024$} & \multicolumn{1}{r}{$\phantom{00}\phantom{-}4.975$} & \multicolumn{1}{r}{$\phantom{00}\phantom{-}3.215$} & \multicolumn{1}{r}{$\phantom{00}-2.649$} & \multicolumn{1}{r}{$\phantom{00}\phantom{-}2.249$} & \multicolumn{1}{r}{$\phantom{00}\phantom{-}2.045$} & \multicolumn{1}{r}{$\phantom{00}\phantom{-}1.349$} & \multicolumn{1}{r}{$\phantom{00}-0.674$} & \multicolumn{1}{r}{$\phantom{00}\phantom{-}3.440$} & \multicolumn{1}{r}{$\phantom{00}-3.143$} & \multicolumn{1}{r}{$\phantom{00}\phantom{-}1.144$} & \multicolumn{1}{r}{$\phantom{00}\mathbf{-0.096}$} \\
			 & \multicolumn{1}{r}{IPW}  & \multicolumn{1}{r}{$\phantom{00}\phantom{-}5.423$} & \multicolumn{1}{r}{$\phantom{00}\phantom{-}5.390$} & \multicolumn{1}{r}{$\phantom{00}\phantom{-}5.547$} & \multicolumn{1}{r}{$\phantom{00}-5.059$} & \multicolumn{1}{r}{$\phantom{00}\phantom{-}2.680$} & \multicolumn{1}{r}{$\phantom{00}\phantom{-}1.433$} & \multicolumn{1}{r}{$\phantom{00}\phantom{-}2.457$} & \multicolumn{1}{r}{$\phantom{00}-2.039$} & \multicolumn{1}{r}{$\phantom{00}\phantom{-}6.107$} & \multicolumn{1}{r}{$\phantom{00}-5.936$} & \multicolumn{1}{r}{$\phantom{00}\phantom{-}1.663$} & \multicolumn{1}{r}{$\phantom{00}-0.988$} \\
			 & \multicolumn{1}{r}{CBPS}  & \multicolumn{1}{r}{$\phantom{00}\phantom{-}5.414$} & \multicolumn{1}{r}{$\phantom{00}\phantom{-}5.383$} & \multicolumn{1}{r}{$\phantom{00}\phantom{-}5.431$} & \multicolumn{1}{r}{$\phantom{00}-5.035$} & \multicolumn{1}{r}{$\phantom{00}\phantom{-}2.311$} & \multicolumn{1}{r}{$\phantom{00}\phantom{-}1.942$} & \multicolumn{1}{r}{$\phantom{00}\phantom{-}2.447$} & \multicolumn{1}{r}{$\phantom{00}-2.044$} & \multicolumn{1}{r}{$\phantom{00}\phantom{-}6.105$} & \multicolumn{1}{r}{$\phantom{00}-5.945$} & \multicolumn{1}{r}{$\phantom{00}\phantom{-}1.508$} & \multicolumn{1}{r}{$\phantom{00}-0.780$} \\
			 & \multicolumn{1}{r}{Cal}  & \multicolumn{1}{r}{$\phantom{00}\phantom{-}5.438$} & \multicolumn{1}{r}{$\phantom{00}\phantom{-}5.405$} & \multicolumn{1}{r}{$\phantom{00}\phantom{-}5.427$} & \multicolumn{1}{r}{$\phantom{00}-4.937$} & \multicolumn{1}{r}{$\phantom{00}\phantom{-}1.567$} & \multicolumn{1}{r}{$\phantom{00}\phantom{-}1.031$} & \multicolumn{1}{r}{$\phantom{00}\phantom{-}2.420$} & \multicolumn{1}{r}{$\phantom{00}-1.997$} & \multicolumn{1}{r}{$\phantom{00}\phantom{-}6.085$} & \multicolumn{1}{r}{$\phantom{00}-5.920$} & \multicolumn{1}{r}{$\phantom{00}\phantom{-}1.621$} & \multicolumn{1}{r}{$\phantom{00}-0.922$} \\
			 \midrule
			\multicolumn{1}{r}{$Y$ Model: D} & \multicolumn{1}{r}{Unweighted}  & \multicolumn{1}{r}{$\phantom{00}\phantom{-}1.273$} & \multicolumn{1}{r}{$\phantom{00}-0.515$} & \multicolumn{1}{r}{$\phantom{00}\phantom{-}1.761$} & \multicolumn{1}{r}{$\phantom{00}\phantom{-}1.057$} & \multicolumn{1}{r}{$\phantom{00}\phantom{-}8.495$} & \multicolumn{1}{r}{$\phantom{00}-8.428$} & \multicolumn{1}{r}{$\phantom{00}\phantom{-}1.245$} & \multicolumn{1}{r}{$\phantom{00}\mathbf{-0.307}$} & \multicolumn{1}{r}{$\phantom{00}\phantom{-}1.362$} & \multicolumn{1}{r}{$\phantom{00}\phantom{-}0.707$} & \multicolumn{1}{r}{$\phantom{00}\phantom{-}\mathbf{1.410}$} & \multicolumn{1}{r}{$\phantom{00}\phantom{-}\mathbf{0.782}$} \\
			 & \multicolumn{1}{r}{EBW}  & \multicolumn{1}{r}{$\phantom{00}\phantom{-}0.692$} & \multicolumn{1}{r}{$\phantom{00}-0.196$} & \multicolumn{1}{r}{$\phantom{00}\phantom{-}0.908$} & \multicolumn{1}{r}{$\phantom{00}\phantom{-}0.534$} & \multicolumn{1}{r}{$\phantom{00}\phantom{-}4.348$} & \multicolumn{1}{r}{$\phantom{00}-4.185$} & \multicolumn{1}{r}{$\phantom{00}\phantom{-}0.818$} & \multicolumn{1}{r}{$\phantom{00}-0.480$} & \multicolumn{1}{r}{$\phantom{00}\phantom{-}0.639$} & \multicolumn{1}{r}{$\phantom{00}\phantom{-}0.080$} & \multicolumn{1}{r}{$\phantom{00}\phantom{-}2.164$} & \multicolumn{1}{r}{$\phantom{00}\phantom{-}2.066$} \\
			 & \multicolumn{1}{r}{iEBW}  & \multicolumn{1}{r}{$\phantom{00}\phantom{-}\mathbf{0.602}$} & \multicolumn{1}{r}{$\phantom{00}\mathbf{-0.108}$} & \multicolumn{1}{r}{$\phantom{00}\phantom{-}\mathbf{0.817}$} & \multicolumn{1}{r}{$\phantom{00}\phantom{-}\mathbf{0.465}$} & \multicolumn{1}{r}{$\phantom{00}\phantom{-}3.438$} & \multicolumn{1}{r}{$\phantom{00}-3.263$} & \multicolumn{1}{r}{$\phantom{00}\phantom{-}\mathbf{0.760}$} & \multicolumn{1}{r}{$\phantom{00}-0.452$} & \multicolumn{1}{r}{$\phantom{00}\phantom{-}\mathbf{0.588}$} & \multicolumn{1}{r}{$\phantom{00}\phantom{-}\mathbf{0.002}$} & \multicolumn{1}{r}{$\phantom{00}\phantom{-}2.096$} & \multicolumn{1}{r}{$\phantom{00}\phantom{-}2.010$} \\
			 & \multicolumn{1}{r}{KCB}  & \multicolumn{1}{r}{$\phantom{00}\phantom{-}0.881$} & \multicolumn{1}{r}{$\phantom{00}-0.363$} & \multicolumn{1}{r}{$\phantom{00}\phantom{-}1.039$} & \multicolumn{1}{r}{$\phantom{00}\phantom{-}0.582$} & \multicolumn{1}{r}{$\phantom{00}\phantom{-}4.993$} & \multicolumn{1}{r}{$\phantom{00}-4.799$} & \multicolumn{1}{r}{$\phantom{00}\phantom{-}0.863$} & \multicolumn{1}{r}{$\phantom{00}-0.369$} & \multicolumn{1}{r}{$\phantom{00}\phantom{-}0.894$} & \multicolumn{1}{r}{$\phantom{00}\phantom{-}0.390$} & \multicolumn{1}{r}{$\phantom{00}\phantom{-}1.691$} & \multicolumn{1}{r}{$\phantom{00}\phantom{-}1.481$} \\
			 & \multicolumn{1}{r}{IPW}  & \multicolumn{1}{r}{$\phantom{00}\phantom{-}0.981$} & \multicolumn{1}{r}{$\phantom{00}-0.524$} & \multicolumn{1}{r}{$\phantom{00}\phantom{-}1.478$} & \multicolumn{1}{r}{$\phantom{00}\phantom{-}1.002$} & \multicolumn{1}{r}{$\phantom{00}\phantom{-}3.497$} & \multicolumn{1}{r}{$\phantom{00}-2.240$} & \multicolumn{1}{r}{$\phantom{00}\phantom{-}0.879$} & \multicolumn{1}{r}{$\phantom{00}-0.323$} & \multicolumn{1}{r}{$\phantom{00}\phantom{-}1.013$} & \multicolumn{1}{r}{$\phantom{00}\phantom{-}0.694$} & \multicolumn{1}{r}{$\phantom{00}\phantom{-}2.326$} & \multicolumn{1}{r}{$\phantom{00}\phantom{-}2.174$} \\
			 & \multicolumn{1}{r}{CBPS}  & \multicolumn{1}{r}{$\phantom{00}\phantom{-}1.039$} & \multicolumn{1}{r}{$\phantom{00}-0.540$} & \multicolumn{1}{r}{$\phantom{00}\phantom{-}1.442$} & \multicolumn{1}{r}{$\phantom{00}\phantom{-}1.027$} & \multicolumn{1}{r}{$\phantom{00}\phantom{-}3.866$} & \multicolumn{1}{r}{$\phantom{00}-3.645$} & \multicolumn{1}{r}{$\phantom{00}\phantom{-}0.922$} & \multicolumn{1}{r}{$\phantom{00}-0.318$} & \multicolumn{1}{r}{$\phantom{00}\phantom{-}1.075$} & \multicolumn{1}{r}{$\phantom{00}\phantom{-}0.682$} & \multicolumn{1}{r}{$\phantom{00}\phantom{-}1.945$} & \multicolumn{1}{r}{$\phantom{00}\phantom{-}1.740$} \\
			 & \multicolumn{1}{r}{Cal}  & \multicolumn{1}{r}{$\phantom{00}\phantom{-}0.916$} & \multicolumn{1}{r}{$\phantom{00}-0.543$} & \multicolumn{1}{r}{$\phantom{00}\phantom{-}1.338$} & \multicolumn{1}{r}{$\phantom{00}\phantom{-}0.996$} & \multicolumn{1}{r}{$\phantom{00}\phantom{-}\mathbf{2.089}$} & \multicolumn{1}{r}{$\phantom{00}\mathbf{-1.424}$} & \multicolumn{1}{r}{$\phantom{00}\phantom{-}0.832$} & \multicolumn{1}{r}{$\phantom{00}-0.317$} & \multicolumn{1}{r}{$\phantom{00}\phantom{-}0.999$} & \multicolumn{1}{r}{$\phantom{00}\phantom{-}0.697$} & \multicolumn{1}{r}{$\phantom{00}\phantom{-}2.303$} & \multicolumn{1}{r}{$\phantom{00}\phantom{-}2.177$} \\
			 \midrule
			\multicolumn{1}{r}{$Y$ Model: E} & \multicolumn{1}{r}{Unweighted}  & \multicolumn{1}{r}{$\phantom{00}\phantom{-}\mathbf{2.464}$} & \multicolumn{1}{r}{$\phantom{00}\mathbf{-0.063}$} & \multicolumn{1}{r}{$\phantom{00}\phantom{-}3.106$} & \multicolumn{1}{r}{$\phantom{00}\mathbf{-0.158}$} & \multicolumn{1}{r}{$\phantom{00}\phantom{-}4.077$} & \multicolumn{1}{r}{$\phantom{00}-3.469$} & \multicolumn{1}{r}{$\phantom{00}\phantom{-}\mathbf{1.968}$} & \multicolumn{1}{r}{$\phantom{00}-0.061$} & \multicolumn{1}{r}{$\phantom{00}\phantom{-}2.343$} & \multicolumn{1}{r}{$\phantom{00}\mathbf{-0.064}$} & \multicolumn{1}{r}{$\phantom{00}\phantom{-}2.407$} & \multicolumn{1}{r}{$\phantom{00}\mathbf{-0.074}$} \\
			 & \multicolumn{1}{r}{EBW}  & \multicolumn{1}{r}{$\phantom{00}\phantom{-}2.738$} & \multicolumn{1}{r}{$\phantom{00}\phantom{-}0.981$} & \multicolumn{1}{r}{$\phantom{00}\phantom{-}\mathbf{2.662}$} & \multicolumn{1}{r}{$\phantom{00}-0.687$} & \multicolumn{1}{r}{$\phantom{00}\phantom{-}\mathbf{2.416}$} & \multicolumn{1}{r}{$\phantom{00}-0.357$} & \multicolumn{1}{r}{$\phantom{00}\phantom{-}2.022$} & \multicolumn{1}{r}{$\phantom{00}\phantom{-}0.126$} & \multicolumn{1}{r}{$\phantom{00}\phantom{-}\mathbf{2.311}$} & \multicolumn{1}{r}{$\phantom{00}-0.185$} & \multicolumn{1}{r}{$\phantom{00}\phantom{-}\mathbf{2.164}$} & \multicolumn{1}{r}{$\phantom{00}\phantom{-}0.435$} \\
			 & \multicolumn{1}{r}{iEBW}  & \multicolumn{1}{r}{$\phantom{00}\phantom{-}2.751$} & \multicolumn{1}{r}{$\phantom{00}\phantom{-}0.958$} & \multicolumn{1}{r}{$\phantom{00}\phantom{-}2.715$} & \multicolumn{1}{r}{$\phantom{00}-0.511$} & \multicolumn{1}{r}{$\phantom{00}\phantom{-}2.607$} & \multicolumn{1}{r}{$\phantom{00}-0.246$} & \multicolumn{1}{r}{$\phantom{00}\phantom{-}2.074$} & \multicolumn{1}{r}{$\phantom{00}\phantom{-}0.167$} & \multicolumn{1}{r}{$\phantom{00}\phantom{-}2.362$} & \multicolumn{1}{r}{$\phantom{00}-0.184$} & \multicolumn{1}{r}{$\phantom{00}\phantom{-}2.182$} & \multicolumn{1}{r}{$\phantom{00}\phantom{-}0.605$} \\
			 & \multicolumn{1}{r}{KCB}  & \multicolumn{1}{r}{$\phantom{00}\phantom{-}2.605$} & \multicolumn{1}{r}{$\phantom{00}\phantom{-}0.262$} & \multicolumn{1}{r}{$\phantom{00}\phantom{-}2.814$} & \multicolumn{1}{r}{$\phantom{00}-0.653$} & \multicolumn{1}{r}{$\phantom{00}\phantom{-}3.141$} & \multicolumn{1}{r}{$\phantom{00}-1.728$} & \multicolumn{1}{r}{$\phantom{00}\phantom{-}2.038$} & \multicolumn{1}{r}{$\phantom{00}\phantom{-}\mathbf{0.038}$} & \multicolumn{1}{r}{$\phantom{00}\phantom{-}2.344$} & \multicolumn{1}{r}{$\phantom{00}-0.568$} & \multicolumn{1}{r}{$\phantom{00}\phantom{-}2.208$} & \multicolumn{1}{r}{$\phantom{00}\phantom{-}0.131$} \\
			 & \multicolumn{1}{r}{IPW}  & \multicolumn{1}{r}{$\phantom{00}\phantom{-}3.648$} & \multicolumn{1}{r}{$\phantom{00}\phantom{-}0.436$} & \multicolumn{1}{r}{$\phantom{00}\phantom{-}4.151$} & \multicolumn{1}{r}{$\phantom{00}-1.970$} & \multicolumn{1}{r}{$\phantom{00}\phantom{-}5.690$} & \multicolumn{1}{r}{$\phantom{00}\mathbf{-0.102}$} & \multicolumn{1}{r}{$\phantom{00}\phantom{-}3.332$} & \multicolumn{1}{r}{$\phantom{00}-0.070$} & \multicolumn{1}{r}{$\phantom{00}\phantom{-}4.362$} & \multicolumn{1}{r}{$\phantom{00}-1.978$} & \multicolumn{1}{r}{$\phantom{00}\phantom{-}4.691$} & \multicolumn{1}{r}{$\phantom{00}\phantom{-}0.330$} \\
			 & \multicolumn{1}{r}{CBPS}  & \multicolumn{1}{r}{$\phantom{00}\phantom{-}2.626$} & \multicolumn{1}{r}{$\phantom{00}\phantom{-}0.339$} & \multicolumn{1}{r}{$\phantom{00}\phantom{-}3.041$} & \multicolumn{1}{r}{$\phantom{00}-1.159$} & \multicolumn{1}{r}{$\phantom{00}\phantom{-}3.260$} & \multicolumn{1}{r}{$\phantom{00}-1.062$} & \multicolumn{1}{r}{$\phantom{00}\phantom{-}2.237$} & \multicolumn{1}{r}{$\phantom{00}-0.044$} & \multicolumn{1}{r}{$\phantom{00}\phantom{-}3.121$} & \multicolumn{1}{r}{$\phantom{00}-1.555$} & \multicolumn{1}{r}{$\phantom{00}\phantom{-}2.329$} & \multicolumn{1}{r}{$\phantom{00}\phantom{-}0.236$} \\
			 & \multicolumn{1}{r}{Cal}  & \multicolumn{1}{r}{$\phantom{00}\phantom{-}3.377$} & \multicolumn{1}{r}{$\phantom{00}\phantom{-}1.038$} & \multicolumn{1}{r}{$\phantom{00}\phantom{-}2.949$} & \multicolumn{1}{r}{$\phantom{00}-1.388$} & \multicolumn{1}{r}{$\phantom{00}\phantom{-}3.416$} & \multicolumn{1}{r}{$\phantom{00}\phantom{-}1.571$} & \multicolumn{1}{r}{$\phantom{00}\phantom{-}2.150$} & \multicolumn{1}{r}{$\phantom{00}\phantom{-}0.227$} & \multicolumn{1}{r}{$\phantom{00}\phantom{-}2.852$} & \multicolumn{1}{r}{$\phantom{00}-0.833$} & \multicolumn{1}{r}{$\phantom{00}\phantom{-}2.310$} & \multicolumn{1}{r}{$\phantom{00}\phantom{-}0.408$} \\
			\bottomrule
		\end{tabular}%
	}
\caption{Displayed are results for $n=250$ and $p=25$  averaged over 1000 independent simulated datasets.}
\label{tab:simulation_results_n250_p25}
\end{table}

\subsection{Details for weighted energy distance toy examples}

In this section we outline the details for the toy examples in Section \ref{sec:weighted_energy_general} of the main text. In the first example, we generate a 1-dimensional covariate of sample size 250, which impacts treatment assignment for a binary via a logistic model under three scenarios: 1) $\text{logit}(\pi(X)) = -1 + X$, 2) $\text{logit}(\pi(X)) = -1 + X + 2X^2/3$, and 3) $\text{logit}(\pi(X)) = -1 + X + 2X^2/3 - X^3/3$. In each scenario, the response is generated as $Y=X+X^3-1/(0.1 + 0.1X^2)+\varepsilon$, where $\varepsilon\distas{} N(0,\sqrt{2})$. For each scenario, we construct inverse probability weights based off of 3 logistic regression models, which consider only a linear term in $X$ (denoted as ``IPW (1)''), a linear plus quadratic term (denoted as ``IPW (2)''), and up to the cubic term (denoted as ``IPW (3)''), respectively. For each set of weights $\bw$, we compute the sum of the energy distances between each treatment group and the combined sample, i.e $\calE(F_{n,0,\bw}, F_{n}) + \calE(F_{n,1,\bw}, F_{n})$ and compute the bias of \eqref{eqn:wate} for for $\tau$ using each set of weights.

In a second toy example, we consider a two dimensional example where the true assignment mechanism depends on first and second moments of the covariates. In particular, we generate treatment assignments from $\text{logit}(\pi(X)) = -1 + X_1 + 0.5X_1^2 - X_2 - 0.5X_2^2$. The response is generated as $Y = X_1 - 1/(0.1 + 0.1X_1^2) - X_2 + 1/(0.1 + 0.1X_2^2) + \varepsilon$. We consider a collection of methods to estimate weights, including logistic regression, the method of \citet{imai2014covariate}, and the method of \citet{chan2016globally}, each with i) just first order moments included for balancing or estimation and additionally ii) all first and second order moments included. The weights of all methods are then used for weighted estimates of $\tau$. We then compare the weighted energy distances and absolute biases of \eqref{eqn:wate} based on these weights in Figure \ref{fig:toy_example}(b) of the main text.

\subsection{Details for value function optimization toy example}

In this section we detail the setup for the example involving estimation of individualized treatment rules (ITRs) via value function optimization. To demonstrate the effectiveness of using energy balancing weights in optimal ITR estimation, we provide an illustrative example under two data-generating scenarios. For both scenarios we generate outcomes as $Y = g(\bX) + \widetilde{A}\Delta(\bX)/2 + \varepsilon$, where $g(\bX)$ are the main effects of $\bX$, $\widetilde{A} = 2A-1$, and $\Delta(\bX)=\mu_{1}(\bX) - \mu_{0}(\bX)$ is the treatment-covariate interaction, $\varepsilon \distas{}N(0,1)$, and $\bbR^{10}\ni\bX\distas{i.i.d.}\text{Unif(-1,1)}$. Both scenarios are motivated by the simulation studies of \citet{zhao2012estimating} but generate $A$ from a logistic regression model with terms depending on up to third order polynomials in a subset of the predictors and $g(\bX)$ contains non-linear terms in the predictors. Scenario 1 uses $g(\bX) = 8 - \sum_{j=1}^3(-1)^{j}\left\{X_j + 10X_j^3 - 1 / (0.1 + 0.1X_j^2)\right\}$, $\Delta(\bX) = X_2 - 0.25X_1^2 - X_4+0.25X_3^2$, and $\text{logit}(\pi(\bX)) = -1 -\sum_{j=1}^3(-1)^{j}\left\{(7/4)X_j + (7/6)X_j^2 + (7/12)X_j^3\right\}$. Scenario 2 uses $g(\bX) = 8 + 0.5 (X_1 +  10 X_1 ^ 3 - 1 / (0.1 + 0.1 X_1 ^ 2))$, $\Delta(\bX) = -1 - X_1 ^ 3 + \exp(X_3 ^ 2 + X_5) + 0.6 X_6 - (X_7 + X_8) ^ 2$, and $\text{logit}(\pi(\bX)) = -1 + (7/4)X_1 + (7/6)X_1^2 + (7/12)X_1^3$.
 We utilize the OWL method to obtain estimates $\hat{d}$, which uses inverse weighting by the propensity score and adds $\lambda_n\lVert d \rVert^2$ to the objective. For OWL, the propensity score is misspecified to only include linear terms in the covariates. We also estimate $d^*$ by minimizing \eqref{eqn:energy_owl} plus $\lambda_n\lVert d \rVert^2$. We denote this as OWL (EBW) for weights given by \eqref{eqn:energy_balancing_weights_ate} and OWL (iEBW) for weights given by \eqref{eqn:improved_energy_balancing_weights_ate}. We simulate 1000 independent datasets and compute the average value function $\widehat{\bbE}[Y(\hat{d})]$ evaluated on a large independent dataset in addition to the missclassification rate in estimating $I(d^*(X)>0)$ on the independent dataset.


\subsection{Details for RHC simulation and an additional simulation}
\label{sec:rhc_simulation_supp_details}

We now define the outcome model used in the simulation using the RHC data from Section \ref{sec:rhc_simulation} the main text. The outcome model is based on outcome model D from Table \ref{tab:simulation_settings_propensity}. Outcome model D depends on 7 covariates, however the outcome model we use in this section uses an application of this model to multiple sets of 7 covariates from the RHC dataset. Define the mean function from outcome model D of Table \ref{tab:simulation_settings_propensity} to be $f_D(\bx^{1:7})$, where $1:7$ indicates that the first through seventh covariates are used in the mean model. We now define the outcome model of our simulation to be
\begin{equation*}
    Y_i = f(\bx_i) + \varepsilon_i \text{ for } i = 1,\dots, 5735,
\end{equation*}
where $f(\bx_i) = \sum_{k = 0}^8f_D(\bx_i^{(7k + 1):(7(k+1))})$ and $\varepsilon$ are i.i.d $N(0,5)$ random variables. Thus, 63 of the 65 covariates have an impact on the response. The design matrix and the treatment assignment vector are fixed throughout all simulations. Since the ordering of the covariates results in a different outcome model, since the 65 covariates are from the RHC dataset, we create new outcome models by uniformly at random permuting the columns of the design matrix. For each column permutation, we replicate the simulation 1000 times and record the RMSE of each method for that permutation. Since the above outcome model used in the main text has a constant treatment effect of zero, we also include an outcome model with a treatment effect that varies with the covariates $\bX$. The heterogeneous treatment effect model is 
\begin{equation*}
    Y_i = f(\bx_i) + A_i\left( f(\bx_i) - \overline{f(\bx_i)} \right) + \varepsilon_i \text{ for } i = 1,\dots, 5735,
\end{equation*}
where $f(\bx_i)$ is defined as above and $\overline{f(\bx_i)} = \sum_{i=1}^n\overline{f(\bx_i)}/n$ and $\varepsilon$ are i.i.d $N(0,5)$ random variables. The interaction between treatment and covariates is centered so that the sample average treatment effect is always 0, but varies significantly with $\bx$.
The median, average, standard devation, and maximum RMSEs over the 100 permutations of covariates for both the constant treatment effect setting and the heterogeneous treatment effect setting are displayed in Table \ref{tab:rch_simulations}. Both EBW and iEBW perform quite well, with iEBW with the lowest RMSEs on average, by median, with the lowest variability from permutation to permutation, and with the smallest worst-case RMSE. 

\begin{table}
\label{tab:rch_simulations}
\caption{ Displayed are the median, mean, standard deviation, and maximum RMSEs for each method across the 100 simulation settings using the RHC data.
}
\vspace{15pt}
\centering
\begin{tabular}{rrrrrrr}
  \toprule
 & Unweighted & CBPS & IPW & Cal & EBW & iEBW \\ 
  \midrule
  & \multicolumn{6}{c}{Constant treatment effect} \\
  \cmidrule{2-7}
  Median RMSE & 4.6293 & 1.713 & 2.1071 & 1.3496 & 0.6848 & 0.5298 \\
  Mean RMSE & 5.8204 & 2.1394 & 2.5354 & 1.7885 & 0.8336 & 0.6894 \\ 
  SD RMSE & 4.2938 & 1.5409 & 1.8716 & 1.4943 & 0.6261 & 0.5247 \\
  Max RMSE & 22.3084 & 6.8395 & 7.7406 & 5.2275 & 2.6797 & 2.3131 \\
  \midrule
  & \multicolumn{6}{c}{Heterogeneous treatment effect} \\
  \cmidrule{2-7}
  Median RMSE & 7.4944 & 3.7863 & 4.3439 & 2.4788 & 1.2268 & 1.0147 \\
  Mean RMSE & 9.423 & 4.0585 & 4.8919 & 3.272 & 1.4285 & 1.2008 \\ 
  SD RMSE & 6.9536 & 2.887 & 3.6563 & 2.8277 & 1.0519 & 0.8788 \\
  Max RMSE & 36.1208 & 12.6281 & 15.1127 & 9.6681 & 4.5404 & 3.7934 \\
   \bottomrule
\end{tabular}
\end{table}

\section{ Data analyses using the MIMIC-III Critical Care Database }
\label{sec:analysis_mimic_supp}

In this section we present data analyses of the remaining two studies based on the MIMIC-III Critical Care Database \citep{mimiciii} as mentioned in Section \ref{sec:analysis_mimic} of the main text. The first subsection presents additional simulation results from the IAC data. The remaining two subsections present treatment effect estimates and balance statistics using each method used in the main text. As in the main text, all approaches that require positing a propensity score model or which moments to balance involve only first order terms in all of the confounders. For both studies, we use the covariates and treatment assignments of the observed data to conduct simulation studies using the same approach used for the simulation based on the RHC data. In essence, with these simulation studies we preserve the treatment assignment mechanism of the observed data and simulate outcomes that involve a high likelihood of confounding under this real-world treatment assignment mechanism. The simulation studies investigate scenarios with a constant treatment effect over $\bX$ and with a treatment effect that varies with $\bX$ but results in a (population and sample) average treatment effect of 0. The outcome models used are the same as described in Section \ref{sec:rhc_simulation_supp_details}. Each dataset has a differing number of confounders, however, the outcome model across all datasets involves 63 covariates impacting the response for any given scenario. As in the setup in Section \ref{sec:rhc_simulation_supp_details}, the columns are permuted 100 times, resulting in 100 separate outcome models with different covariates impacting the response in different ways. 

\subsection{Mechanical power of ventilation data}
\label{sec:analysis_mpv}

We use the MIMIC-III database to study the impact of a large degree of mechanical power of ventilation on outcomes. Our study and the construction of the cohort from the MIMIC-III database is based the original study of \citet{hsu2015association} and is based on the code provided by the authors  located at \url{https://github.com/alistairewj/mechanical-power}. The authors of \citet{hsu2015association} treat mechanical power as a continuous treatment, however, we treat it as binary (whether mechanical power of ventilation of greater than 25 Joules per minute) for the purpose of demonstrating the use of our proposed EBWs. The study contains 
5014 patients, 1298 of whom received a mechanical power of ventilation of greater than 25 Joules per minute, the amount of energy generated by the mechanical ventilator. The outcome is an indicator of in-hospital mortality. In all, the dimension of the design matrix of confounders is 86. 

All methods explored in the main text were applied to adjust for the 86 confounders. Estimated treatment effects and balance statistics are displayed in Table \ref{tab:mpv_estimates_se_and_balance}. The KCB approach yielded constant weights of 1 regardless of the tuning parameter. From the univariate standardized mean differences (SMDs), Cal and CBPS balance marginal means the most effectively, however iEBW balances means of interactions and polynomials the best, with the smallest worst case mean imbalance and the best average imbalance. iEBW balances marginal distributions the most effectively on average and in the worse case scenario, with Cal a close second, followed by EBW and CBPS. iEBW balances bivariate distributions the best on average and in the worse case, followed by Cal and EBW. Among non-EBW approaches, Cal yields the smallest weighted energy distances, which is in alignment with its ability to balance marginal univariate and bivariate distributions for this data. The point estimates from each approach, including the unweighted analysis, suggest that   mechanical power larger than 25 Joules/min harms patients in terms of in-hospital mortality, however iEBW and EBW suggest less harm than do other approaches. All approaches yield 95\% confidence intervals that do not contain 0, except IPW, which has an extraordinarily large standard error compared with other approaches. iEBW and EBW yield the shortest length confidence intervals, suggesting a significant increase in in-hospital mortality from mechanical power greater than 25 Joules/min despite their attenuated estimate of the impact on mortality. These findings align qualitatively with the analysis conducted by \citet{hsu2015association}.

As mentioned, we also use the MPV data to conduct simulation studies, wherein we fix the confounders and treatment assignment and simulate outcomes. 
The median, average, standard deviation, and maximum RMSEs over the 100 permutations of covariates for both the constant treatment effect setting and the heterogeneous treatment effect setting are displayed in Table \ref{tab:mpv_simulations}. We note that the rankings of each method in terms of their RMSEs across the simulation settings align with their weighted energy distances in Table \ref{tab:mpv_estimates_se_and_balance}, with iEBW performing best in terms of median, mean, and worst-case RMSE across all settings for both the constant treatment effect setup and the heterogeneous treatment effect setup, followed by EBW.  

\begin{table}
\caption{\label{tab:mpv_estimates_se_and_balance} Estimates of the ATE and standard errors for the mechanical power data. Standard errors were computed for all methods using the nonparametric bootstrap with 1000 replications. Also displayed are various measures of discrepancy between the distributions of covariates for the IAC and control groups. We also display the mean and max RIMSE statistic for marginal univariate and bivariate CDF differences, as in Figure \ref{fig:root_imse_cdfs_rhc}. In addition, we display summary statistics of SMDs for marginal means and SMDs for all polynomials up to order 5 and pairwise interactions (denoted SMD(2)). 
}
\vspace{15pt}
\centering
\begin{tabular}{crrrrrr}
  \toprule
 & Unwtd & CBPS & IPW & Cal & EBW & iEBW \\ 
  \midrule
$\widehat{\tau}_\bw$ & 0.0405 & 0.0773 & 0.0997 & 0.0868 & 0.0729 & 0.0683 \\ 
  SE$(\widehat{\tau}_\bw)$ & 0.0149 & 0.0243 & 0.1724 & 0.0276 & 0.0198 & 0.0191 \\
  \midrule
  Energy dist \eqref{eqn:energy_balancing_weights_ate} & 42.7399 & 3.8171 & 52.7812 & 2.1667 & 1.4990 & 1.6333 \\
  Energy dist \eqref{eqn:improved_energy_balancing_weights_ate} & 112.0910 & 6.3079 & 102.8196 & 4.1974 & 3.0852 & 2.8396  \\
  Mean RIMSE, 1d & 0.0880 & 0.0137 & 0.0314 & 0.0117 & 0.0125 & \textbf{0.0098} \\
  Max RIMSE, 1d & 0.3985 & 0.0880 & 0.0979 & 0.0717 & 0.0924 & \textbf{0.0692} \\
  Mean RIMSE, 2d & 0.0892 & 0.0151 & 0.0404 & 0.0131 & 0.0131 & \textbf{0.0105} \\
  Max RIMSE, 2d & 0.2617 & 0.0448 & 0.1443 & 0.0343 & 0.0407 & \textbf{0.0297} \\
  Mean $|$SMD$|$ & 0.2221 & 0.0016 & 0.1032 & \textbf{0.0001} & 0.0107 & 0.0069 \\
  Max $|\text{SMD}|$& 1.1430 & 0.0176 & 2.4637 & \textbf{0.0054} & 0.0803 & 0.0524 \\
  Mean $|$SMD(2)$|$ & 0.1756 & 0.0130 & 0.1017 & 0.0121 & 0.0146 & \textbf{0.0104} \\
  Max $|${SMD(2)}$|$& 1.1872 & 0.2463 & 4.8639 & 0.2456 & 0.1456 & \textbf{0.0888} \\
   \bottomrule
\end{tabular}
\end{table}

\begin{table}
\caption{\label{tab:mpv_simulations} Displayed are the median, mean, standard deviation, and maximum RMSEs for each method across the 100 simulation settings using the mechanical power data.
}
\vspace{15pt}
\centering
\begin{tabular}{rrrrrrr}
  \toprule
 & Unwtd & CBPS & IPW & Cal & EBW & iEBW \\ 
  \midrule
  & \multicolumn{6}{c}{Constant treatment effect} \\
  \cmidrule{2-7}
  Median RMSE & 10.7097 & 5.3193 & 22.0340 & 3.8651 & 3.0773 & 2.3889 \\
  Mean RMSE & 13.0120 & 7.1023 & 52.1948 & 5.2540 & 3.6800 & 2.5932 \\ 
  SD RMSE & 9.2960 & 5.6161 & 61.4692 & 4.6933 & 2.7329 & 1.8679 \\
  Max RMSE & 38.6099 & 22.0997 & 247.9993 & 21.9211 & 10.1050 & 7.6787 \\
  \midrule
  & \multicolumn{6}{c}{Heterogeneous treatment effect} \\
  \cmidrule{2-7}
  Median RMSE & 18.6460 & 7.2053 & 23.3952 & 6.6768 & 6.1644 & 5.2506 \\
  Mean RMSE & 22.6549 & 9.1752 & 53.3998 & 9.6951 & 6.9088 & 5.8760 \\ 
  SD RMSE & 16.1860 & 7.2144 & 61.1497 & 9.1299 & 5.0759 & 4.2979 \\
  Max RMSE & 67.2217 & 28.6804 & 247.4224 & 43.2159 & 19.7568 & 16.7650 \\
   \bottomrule
\end{tabular}
\end{table}

\subsection{Transthoracic echocardiography data}
\label{sec:analysis_echo}

We use the MIMIC-III database to analyse a study of the effect of transthoracic echocardiography on 28 day mortality in sepsis patients originally conducted by \citet{feng2018transthoracic}. Our construction of the study cohort from the MIMIC-III database is based on the code provided by the authors of \citet{feng2018transthoracic} located at \url{https://github.com/nus-mornin-lab/echo-mimiciii}. The study contains information on 6361 patients, 3262 of whom received transthoracic echocardiography. The outcome is an indicator of mortality within 28 days of admission to the ICU. In all, the dimension of the design matrix of confounders is 77. 

All methods explored in the main text were applied to adjust for the 77 confounders. Estimated treatment effects and balance statistics are displayed in Table \ref{tab:echo_estimates_se_and_balance}. The KCB approach yielded constant weights of 1 regardless of the tuning parameter. From the univariate standardized mean differences (SMDs), Cal, CBPS, and iEBW balance marginal means the most effectively, however iEBW and EBW balance means of interactions and polynomials the best, with the smallest worst case mean imbalance and the best average imbalance. EBW and iEBW balance marginal distributions the most effectively on average and in the worse case scenario, with Cal a close second, followed by EBW and CBPS. EBW and iEBW balance bivariate distributions the best on average and in the worse case, followed by Cal and CBPS. The point estimates for all methods of the effect of echocardiography all indicate a potential reduction of 28 day mortality, with EBW, iEBW, Cal, and CBPS all suggesting a similar effect and IPW suggesting a stronger effect. 95\% confidence intervals for all methods do not contain zero, except for CBPS which has a larger standard error. EBW and iEBW result in the smallest standard error and thus shortest length confidence interval. These findings align with the original analysis conducted in \citet{feng2018transthoracic}.

We also use the echocardiography data to conduct simulation studies, wherein we fix the confounders and treatment assignment and simulate outcomes. 
The median, average, standard deviation, and maximum RMSEs over the 100 permutations of covariates for both the constant treatment effect setting and the heterogeneous treatment effect setting are displayed in Table \ref{tab:echo_simulations}.
We note that the rankings of each method in terms of their RMSEs across the simulation settings closely align with their weighted energy distances in Table \ref{tab:echo_estimates_se_and_balance}, with iEBW performing best in terms of median, mean, and worst-case RMSE across all settings for both the constant treatment effect setup and the heterogeneous treatment effect setup, followed by EBW. Here, CBPS performs slightly better than Cal, unlike with the RHC, IAC, and MPV datasets.

\begin{table}
\caption{\label{tab:echo_estimates_se_and_balance} Estimates of the ATE and standard errors for the echocardiography data. Standard errors were computed for all methods using the nonparametric bootstrap with 1000 replications. Also displayed are various measures of discrepancy between the distributions of covariates for the IAC and control groups. We also display the mean and max RIMSE statistic for marginal univariate and bivariate CDF differences, as in Figure \ref{fig:root_imse_cdfs_rhc}. In addition, we display summary statistics of SMDs for marginal means and SMDs for all polynomials up to order 5 and pairwise interactions (denoted SMD(2)). 
}
\vspace{15pt}
\centering
\begin{tabular}{crrrrrr}
  \toprule
 & Unwtd & CBPS & IPW & Cal & EBW & iEBW \\ 
  \midrule
$\widehat{\tau}_\bw$ & -0.0064 & -0.0347 & -0.0446 & -0.0323 & -0.0323 & -0.0325 \\ 
  SE$(\widehat{\tau}_\bw)$ & 0.0113 & 0.0206 & 0.0137 & 0.0113 & 0.0091 & 0.0088 \\
  \midrule
  Energy dist \eqref{eqn:energy_balancing_weights_ate} & 5.7673 & 0.4647 & 0.6012 & 0.3258 & 0.2338 & 0.2378 \\
  Energy dist \eqref{eqn:improved_energy_balancing_weights_ate} & 17.2943 & 1.0378 & 1.2784 & 0.8662 & 0.6094 & 0.5999  \\
  Mean RIMSE, 1d & 0.0282 & 0.0095 & 0.0094 & 0.0095 & 0.0074 & \textbf{0.0072} \\
  Max RIMSE, 1d & 0.0944 & 0.0214 & 0.0207 & 0.0211 & 0.0185 & \textbf{0.0183} \\
  Mean RIMSE, 2d & 0.0424 & 0.0070 & 0.0078 & 0.0069 & 0.0051 & \textbf{0.0048} \\
  Max RIMSE, 2d & 0.2683 & 0.0284 & 0.0249 & 0.0273 & 0.0173 & \textbf{0.0140} \\
  Mean $|$SMD$|$ & 0.0764 & 0.0015 & 0.0086 & \textbf{0.0000} & 0.0027 & 0.0020 \\
  Max $|\text{SMD}|$& 0.2773 & 0.0108 & 0.0279 & \textbf{0.0000} & 0.0169 & 0.0125 \\
  Mean $|$SMD(2)$|$ & 0.1044 & 0.0078 & 0.0133 & 0.0071 & 0.0056 & \textbf{0.0046} \\
  Max $|${SMD(2)}$|$& 0.5057 & 0.0801 & 0.1672 & 0.0739 & 0.0452 & \textbf{0.0443} \\
   \bottomrule
\end{tabular}
\end{table}

\begin{table}
\caption{\label{tab:echo_simulations} Displayed are the median, mean, standard deviation, and maximum RMSEs for each method across the 100 simulation settings using the echocardiography data.
}
\vspace{15pt}
\centering
\begin{tabular}{rrrrrrr}
  \toprule
 & Unwtd & CBPS & IPW & Cal & EBW & iEBW \\ 
  \midrule
  & \multicolumn{6}{c}{Constant treatment effect} \\
  \cmidrule{2-7}
  Median RMSE & 4.1580 & 1.3328 & 1.6951 & 1.3463 & 1.2802 & 0.9243 \\
  Mean RMSE & 4.4938 & 1.7167 & 1.9567 & 1.7288 & 1.6174 & 1.2176 \\ 
  SD RMSE & 3.4166 & 1.3789 & 1.5090 & 1.3744 & 1.3003 & 0.9296 \\
  Max RMSE & 14.6757 & 5.7763 & 7.5684 & 5.7088 & 5.2813 & 3.8240 \\
  \midrule
  & \multicolumn{6}{c}{Heterogeneous treatment effect} \\
  \cmidrule{2-7}
  Median RMSE & 6.1837 & 1.7658 & 2.1917 & 2.1114 & 1.8577 & 1.5091 \\
  Mean RMSE & 6.6822 & 2.4121 & 2.6146 & 2.6639 & 2.2572 & 1.6587 \\ 
  SD RMSE & 5.0821 & 1.8060 & 1.9278 & 2.1047 & 1.6669 & 1.1206 \\
  Max RMSE & 21.8241 & 7.8449 & 9.4596 & 8.6978 & 6.9643 & 4.8831 \\
   \bottomrule
\end{tabular}
\end{table}

\pagebreak

\spacingset{1.05}
\bibliographystyle{Chicago}
\bibliography{Bibliography}

\end{document}